\documentclass[reprint,
 superscriptaddress,
amsmath,amssymb,
prb,
]{revtex4-2}

\usepackage{graphicx}
\usepackage{dcolumn}
\usepackage{bm}
\usepackage{xcolor}

\begin{document}
\preprint{APS/123-QED}
\title{Colossal magnetoresistance from spin-polarized polarons in an Ising system}

\author{Ying-Fei Li}
\thanks{These authors contributed equally.}
\affiliation{Stanford Institute for Materials and Energy Sciences, SLAC National Accelerator Laboratory, Menlo Park, 94025, CA, USA}
\affiliation{Department of Applied Physics and Physics, Stanford University, Stanford, 94305, CA, USA}
\affiliation{Geballe Laboratory for Advanced Materials, Stanford University, Stanford, 94305, CA, USA}

\author{Emily M. Been}
\thanks{These authors contributed equally.}
\affiliation{Stanford Institute for Materials and Energy Sciences, SLAC National Accelerator Laboratory, Menlo Park, 94025, CA, USA}
\affiliation{Department of Applied Physics and Physics, Stanford University, Stanford, 94305, CA, USA}
\affiliation{Geballe Laboratory for Advanced Materials, Stanford University, Stanford, 94305, CA, USA}

\author{Sudhaman Balguri}
\affiliation{Departments of Physics, Boston College, 140 Commonwealth Avenue, Chestnut Hill, 02467, MA, USA}

\author{Chun-Jing Jia}
\affiliation{Geballe Laboratory for Advanced Materials, Stanford University, Stanford, 94305, CA, USA}
\affiliation{Department of Physics, University of Florida, Gainesville 32611, FL}

\author{Mira B. Mahenderu}
\affiliation{Departments of Physics, Boston College, 140 Commonwealth Avenue, Chestnut Hill, 02467, MA, USA}

\author{Zhi-Cheng Wang}
\affiliation{Departments of Physics, Boston College, 140 Commonwealth Avenue, Chestnut Hill, 02467, MA, USA}

\author{Yi Cui}
\affiliation{Stanford Institute for Materials and Energy Sciences, SLAC National Accelerator Laboratory, Menlo Park, 94025, CA, USA}
\affiliation{Department of Applied Physics and Physics, Stanford University, Stanford, 94305, CA, USA}
\affiliation{Geballe Laboratory for Advanced Materials, Stanford University, Stanford, 94305, CA, USA}

\author{Su-Di Chen}
\affiliation{Stanford Institute for Materials and Energy Sciences, SLAC National Accelerator Laboratory, Menlo Park, 94025, CA, USA}
\affiliation{Department of Applied Physics and Physics, Stanford University, Stanford, 94305, CA, USA}
\affiliation{Geballe Laboratory for Advanced Materials, Stanford University, Stanford, 94305, CA, USA}
\affiliation{Department of Physics, University of California, Berkeley, California 94720, USA}

\author{Makoto Hashimoto}
\affiliation{Stanford Synchrotron Radiation Lightsource, SLAC National Accelerator Laboratory, Menlo Park, 94025, CA, USA}

\author{Dong-Hui Lu}
\affiliation{Stanford Synchrotron Radiation Lightsource, SLAC National Accelerator Laboratory, Menlo Park, 94025, CA, USA}

\author{Brian Moritz}
\affiliation{Stanford Institute for Materials and Energy Sciences, SLAC National Accelerator Laboratory, Menlo Park, 94025, CA, USA}
\affiliation{Department of Applied Physics and Physics, Stanford University, Stanford, 94305, CA, USA}
\affiliation{Geballe Laboratory for Advanced Materials, Stanford University, Stanford, 94305, CA, USA}

\author{Jan Zaanen}
\affiliation{Institute Lorentz for Theoretical Physics, Leiden University, 2300 RA Leiden, Netherlands}

\author{Fazel Tafti}
\email{fazel.tafti@bc.edu}
\affiliation{Departments of Physics, Boston College, 140 Commonwealth Avenue, Chestnut Hill, 02467, MA, USA}

\author{Thomas P. Devereaux}
\email{tpd@stanford.edu}
\affiliation{Stanford Institute for Materials and Energy Sciences, SLAC National Accelerator Laboratory, Menlo Park, 94025, CA, USA}
\affiliation{Department of Materials Science and Engineering, Stanford University, Stanford, 94305, CA, USA}

\author{Zhi-Xun Shen}
\email{zxshen@stanford.edu}
\affiliation{Stanford Institute for Materials and Energy Sciences, SLAC National Accelerator Laboratory, Menlo Park, 94025, CA, USA}
\affiliation{Department of Applied Physics and Physics, Stanford University, Stanford, 94305, CA, USA}
\affiliation{Geballe Laboratory for Advanced Materials, Stanford University, Stanford, 94305, CA, USA}

\date{\today}

\begin{abstract}
Recent experiments suggest a new paradigm towards novel colossal magnetoresistance (CMR) in a family of materials EuM\textsubscript{2}X\textsubscript{2}(M=Cd, In, Zn; X=P, As), distinct from the traditional avenues involving Kondo-RKKY crossovers, magnetic phase transitions with structural distortions, or topological phase transitions. Here, we use angle-resolved photoemission spectroscopy (ARPES) and density functional theory (DFT) calculations to explore their origin, particularly focusing on EuCd\textsubscript{2}P\textsubscript{2}. While the low-energy spectral weight royally tracks that of the resistivity anomaly near the temperature with maximum magnetoresistance (T\textsubscript{MR}) as expected from transport-spectroscopy correspondence, the spectra are completely incoherent and strongly suppressed with no hint of a Landau quasiparticle. Using systematic material and temperature dependence investigation complemented by theory, we attribute this non-quasiparticle caricature to the strong presence of entangled magnetic and lattice interactions, a characteristic enabled by the $p$-$f$ mixing. Given the known presence of ferromagnetic clusters, this naturally points to the origin of CMR being the scattering of spin-polarized polarons at the boundaries of ferromagnetic clusters. These results are not only illuminating to investigate the strong correlations and topology in EuCd\textsubscript{2}X\textsubscript{2} family, but, in a broader view, exemplify how multiple cooperative interactions can give rise to extraordinary behaviors in condensed matter systems.
\end{abstract}

\maketitle
\begin{figure*}[h]
\centering
\includegraphics[scale=1]{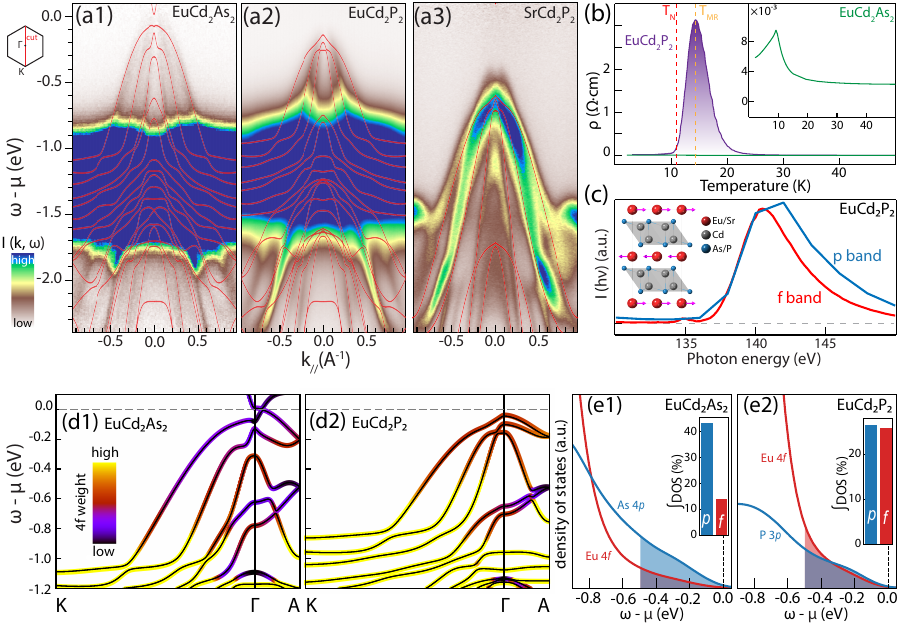}
\caption{\textbf{Overview of the electronic structure: the unique $p$-$f$ mixing low-energy states. (a1-a3)} Angle-resolved photoemission spectroscopy (ARPES) spectra along the $\Gamma$-$K$ cut at 25 K superimposed with DFT-calculated band structures at $k_z=0$ for EuCd\textsubscript{2}As\textsubscript{2}(a1), EuCd\textsubscript{2}P\textsubscript{2}(a2), and SrCd\textsubscript{2}P\textsubscript{2}(a3). \textbf{(b)} Temperature dependence of the resistivity for EuCd\textsubscript{2}As\textsubscript{2} and EuCd\textsubscript{2}P\textsubscript{2}. The inset, scaled by $10^3$,  highlights less pronounced magnetoresistant peak of EuCd\textsubscript{2}As\textsubscript{2} compared to that of the EuCd\textsubscript{2}P\textsubscript{2}. T\textsubscript{N} and T\textsubscript{MR} denote the magnetic phase transition temperature and resistivity maximum temperature, respectively. Detailed transport characterizations are presented in supplementary information (SI) section 4. \textbf{(c)} Spectral weight variations of the localized $4f$ bands and the dispersive $p$ bands near the Eu $4d\rightarrow 4f$ resonant transition. The spectral weights are extracted from the photoemission spectra on EuCd\textsubscript{2}P\textsubscript{2}. Inset exhibits the crystal structure with arrows indicating the A-type antiferromagnetic ground states below T\textsubscript{N} in both EuCd\textsubscript{2}As\textsubscript{2} and EuCd\textsubscript{2}P\textsubscript{2}. \textbf{(d1, d2)} The low-energy DFT band structure along $K$-$\Gamma$-$A$ for EuCd\textsubscript{2}As\textsubscript{2}(d1) and EuCd\textsubscript{2}P\textsubscript{2}(d2). The color map shows Eu $4f$ orbital weight projected onto the bands. \textbf{(e1, e2)} Orbitally-projected density of states (DOS) for EuCd\textsubscript{2}As\textsubscript{2} (e1) and EuCd\textsubscript{2}P\textsubscript{2} (e2) at low energy. Insets show the integrated projected-DOS for pnictogen $p$ orbitals and Eu $4f$ orbitals normalized by the total DOS with the integration window marked as the shaded areas. }
\label{Fig: baseline}
\end{figure*}

The traditional framework for understanding magnetoresistance encompasses a Kondo-Ruderman-Kittel-Kasuya-Yosida (RKKY) transition in heavy fermion systems\cite{kirchner2020colloquium, shaginyan2009energy}, a ferromagnetic-paramagnetic phase transition in concert with structural distortions in manganates\cite{salamon2001physics, uehara1999percolative, mannella2005nodal}, or a topological phase transition in Dirac semimetals \cite{liang2015ultrahigh, shekhar2015extremely, suzuki2019singular, singha2023colossal}. However, the recently discovered colossal magnetoresistance (CMR) in a generic family of materials, particularly in EuCd\textsubscript{2}P\textsubscript{2} (Fig. \ref{Fig: baseline}(b))\cite{wang2021colossal, wang2022anisotropy, homes2023optical, zhang2023electronic, sunko2023spin, krebber2023colossal, usachov2024magnetism}, in the absence of these traditional features, points to a new mechanism. Furthermore, the unconventional CMR effect enriches the physics in the EuM\textsubscript{2}X\textsubscript{2} (M=Cd, In, Zn; X=P, As) family where the strong correlations and topology are being actively explored\cite{ma2019spin, riberolles2021magnetic, XiDai_EIA_axionInsulator}.

Previous explanations of the CMR in Eu\textsubscript{2}Cd\textsubscript{2}P\textsubscript{2} include an insulator-metal transition below $T\textsubscript{MR}$\cite{zhang2023electronic} and the free-electron scattering at the ferromagnetic boundary\cite{sunko2023spin}. Zhang \textit{et al.} proposed a dual-phase model anchored on the observed band reconstruction at low temperatures, and consequently attributed the resistivity increase above $T$\textsubscript{MR} to a nonmagnetic gaped phase and the subsequent decrease below $T$\textsubscript{MR} to an additional $E_F$-crossing band in the antiferromagnetic metallic phase\cite{zhang2023electronic}, in a similar spirit to EuO\cite{torrance1972bound}. However, the proposed scenario is potentially insufficient to explain the following: (i) the resistivity at $T\ll T\textsubscript{MR}$ is similar to that at $T\gg T\textsubscript{MR}$ and still falls outside of the metallic region\cite{zhang2023electronic, wang2021colossal, homes2023optical}, unlike the dramatic change over twelve decades in EuO\cite{torrance1972bound}; (ii) the extracted gap in ARPES spectra is insufficient to account for the resistivity increase above $T\textsubscript{MR}$; (iii) the bad metal behavior in transport at $T\gg T\textsubscript{MR}$ suggests a deviation from a gapped system (details in the supplementary information (SI) section 4); (iv) the disconnect between spectroscopy and transport in that the T\textsubscript{MR} does not align with the temperature at which the band reconstruction feature emerges (details in the supplementary information (SI) section 2.D).

Alternatively, Sunko \textit{et al.} observed the formation of the in-plane ferromagnetic clusters near $T$\textsubscript{MR} in the resonant elastic X-ray scattering and magneto-optical polarimetry\cite{sunko2023spin}. Consequently, they attribute the enhancement of resistivity near T\textsubscript{MR} to the free-electron scattering at the boundary of ferromagnetic clusters. The evidence of ferromagnetic clusters from these photon probes is further supported by detailed analysis on magnetic transport measurements\cite{usachov2024magnetism}; and they will likely play an important role in the physics of the material. Usachocv \textit{et al.} further performed LDA+U calculations of the electronic structure, aiming at addressing the physics within the framework which also only involves electrons scattering at magnetic domains\cite{usachov2024magnetism}. However, this conventional spin-fermion coupling scenario fails to adequately account for the substantial resistivity change either, as this Ising-like picture would typically give a much smaller magnetoresistance\cite{bozorth1946magnetoresistance} than that in EuCd\textsubscript{2}P\textsubscript{2}\cite{wang2021colossal}. Compounding the complexities is the discrepancy in the magnitude of resistivity and their temperature dependence across samples, which, despite comparable magnetoresistance percentage, differs by both three orders of magnitude and the slope in high-temperature regime in two reports\cite{zhang2023electronic, sunko2023spin}. Our research focuses on more metallic samples, consistent with \cite{wang2021colossal, sunko2023spin} (Fig. S20).

In this study, we utilize angle-resolved photoemission spectroscopy (ARPES) and density functional theory (DFT) calculations to decipher these puzzles. We provide spectral results that have close correpondence with the transport, ensuring their relevance to the CMR physics in the bulk material. We then show the spectral lineshapes that indicate the strong presence of polaron formation, which in turn can explain the amplified CMR at $T\textsubscript{MR}$ from scattering at ferromagnetic cluster interfaces, as well as the bad metal behavior at $T\gg T\textsubscript{MR}$. Using resonance soft x-ray photoemission and density functional theory, we show that the origin of the enhanced polaron physics in Eu\textsubscript{2}Cd\textsubscript{2}P\textsubscript{2} versus an other material Eu\textsubscript{2}Cd\textsubscript{2}As\textsubscript{2} in the same family as coming from strong $f$ state mixing into the $p$ bands near the Fermi level. Our finding unified competing pictures for CMR in these materials, and provided microscopy ingredients to engineering future CMR materials. More generally, this result provides further support for the general theme of extreme properties arising from cooperative interactions of different degrees of freedom in complex materials. 

We initiate our discussion by a comprehensive comparison of the ARPES spectra, the band structure calculated by the density functional theory (DFT), and the orbitally-projected density of states in EuCd\textsubscript{2}As\textsubscript{2}, EuCd\textsubscript{2}P\textsubscript{2}, and the non-magnetic counterpart, the SrCd\textsubscript{2}P\textsubscript{2}, respectively (details in SI section 1). While EuCd\textsubscript{2}As\textsubscript{2} may potentially exhibit a topologically non-trivial band structure under a magnetic order with proper symmetry\cite{ma2019spin, jo2020manipulating}, the electronic structure of EuCd\textsubscript{2}P\textsubscript{2} is topologically trivial\cite{ECP_trivial_Cuono_PRB2023} with a discernible gap near the chemical potential $\mu$ (Fig. \ref{Fig: baseline}(a,b)), ruling out the possibilities of a topological phase transition as a potential candidate for CMR. Compared to the SrCd\textsubscript{2}P\textsubscript{2}, the ARPES spectra of EuCd\textsubscript{2}As\textsubscript{2} and EuCd\textsubscript{2}P\textsubscript{2} reveal two main feature: the localized Eu $4f$ states within the -1.5 eV to -1 eV energy range, and the dispersive pnictogen $p$ bands extending towards $\mu$ (Fig. \ref{Fig: baseline}(a1, a2)). Unlike the coexisting itinerant electrons and localized $f$ electrons at low energy in heavy fermion systems\cite{kirchner2020colloquium}, the dispersive $p$ bands solely dominating the low-energy states indicates a departure from the traditional Kondo-RKKY picture in EuCd\textsubscript{2}P\textsubscript{2}.

\begin{figure*}%
\centering
\includegraphics[scale =1]{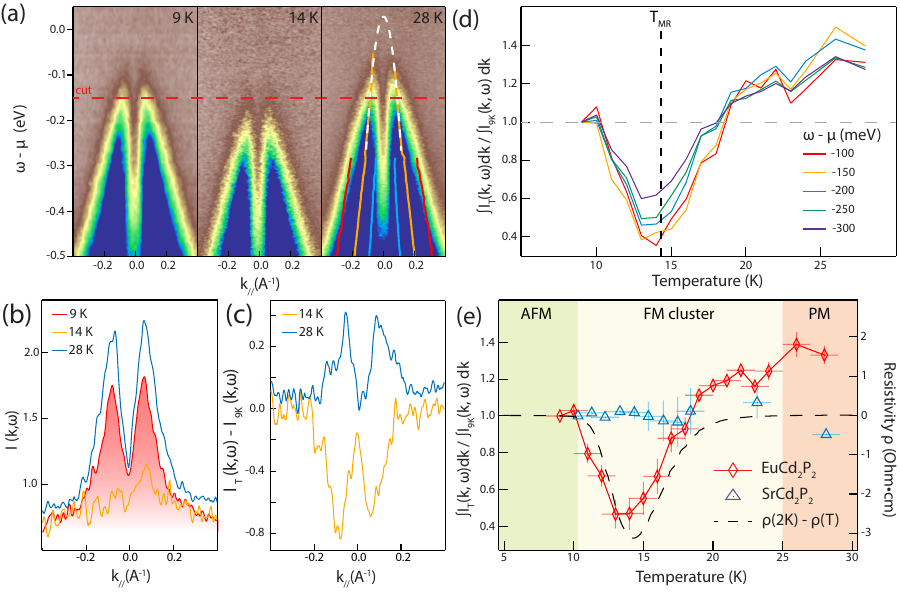}
\caption{\textbf{Spectral weight suppression as the spectroscopic correspondence to the colossal magnetoresistance (CMR). (a)} ARPES spectra of EuCd\textsubscript{2}P\textsubscript{2} below, at, and above T\textsubscript{MR} in the left, middle, and right panels, respectively. The right panel also displays the three bands fitted from MDCs ($\alpha$ in red, $\beta$ in orange, and $\gamma$ in blue). A parabolic fit to the $\beta$ band dispersion highlights its Fermi level-crossing character (white dashed line). \textbf{(b)} MDCs at a binding energy of 150 meV (marked by the red dashed line in (a)) at various temperatures. Spectral weight at T\textsubscript{MR} undergoes pronounced suppression. \textbf{(c)} MDCs in (b) subtracted by the one at 9 K. \textbf{(d)} The temperature dependence of the integrated spectral weight at various binding energies. The red shaded area in (b) indicates the integration area. The spectral weight suppression is more pronounced at lower binding energies. \textbf{(e)} The temperature dependence of integrated spectral weight in EuCd\textsubscript{2}P\textsubscript{2} and SrCd\textsubscript{2}P\textsubscript{2} along with the negative excessive resistivity $\rho$(2K)$-\rho$(T). The spectral weight in EuCd\textsubscript{2}P\textsubscript{2} exhibits a profile akin to anomalous resistivity change and is absent in SrCd\textsubscript{2}P\textsubscript{2}. Error bars of spectral weight indicate the variations at different binding energies in (d). Error bars of temperature indicate the difference between the thermal diode reading and the actual sample temperature. The colored background indicates the anti-ferromagnetism (AFM), short-range ferromagnetic clusters (FM cluster), and paramagnetism (PM) from low to high temperatures.}
\label{Fig: spin_fluctuations}
\end{figure*}

We then focus on the low-energy $p$ states and present a temperature-dependent set of ARPES spectra near $T$\textsubscript{MR}, the corresponding MDCs at a binding energy of 150 meV, and the temperature-varied MDC differential plots in Fig. \ref{Fig: spin_fluctuations}(a,b,c), respectively. At face value, we observe completely incoherent spectra with consistently suppressed low-energy spectral weight, with no resemblance to that of a Landau quasiparticle. Consequently, our analysis is directed toward higher binding energy, and we nevertheless observe a further spectral weight suppression around the $T$\textsubscript{MR}, which becomes increasingly apparent after the background subtraction, spectral weight integration, and normalization in Fig. \ref{Fig: spin_fluctuations}(d,e). As expected from transport-spectroscopy correspondence, the spectral weight mirrors the anomalous resistivity change near $T$\textsubscript{MR} and is absent in the non-magnetic counterpart, the SrCd\textsubscript{2}P\textsubscript{2} (Fig. \ref{Fig: spin_fluctuations}(e)), highlighting its significant link to CMR. Such spectroscopic transport correspondence was not observed in previous photoemission studies\cite{zhang2023electronic, usachov2024magnetism}.

\begin{figure*}
\centering
\includegraphics[scale = .85]{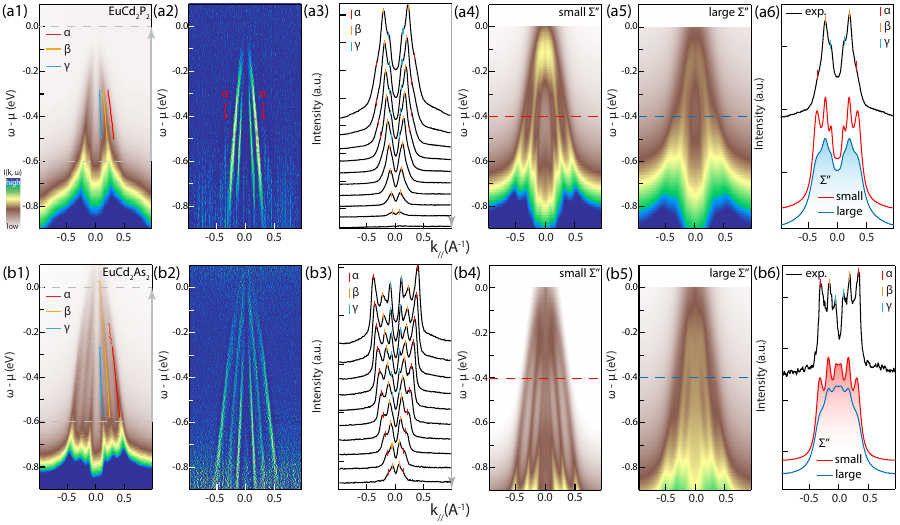}
\caption{\textbf{Low energy spectra in the polaronic limit. (a1, b1)} ARPES spectra of EuCd\textsubscript{2}P\textsubscript{2} (a1) and EuCd\textsubscript{2}As\textsubscript{2} (b1) along the $K$-$\Gamma$-$K$ cut at 25 K. Three pnictogen $p$ bands ($\alpha$, $\beta$, and $\gamma$) are fitted from momentum distribution curves (MDCs). \textbf{(a2, b2)} Second-momentum-derivative of (a1, b1). Red arrows indicate the $\alpha$ band with suppressed spectral weight in EuCd\textsubscript{2}P\textsubscript{2}. \textbf{(a3, b3)} MDCs from the energy window marked by the grey arrow in (a1, b1). MDCs of EuCd\textsubscript{2}P\textsubscript{2} show a single-peak line shape in (a3), contrasting with the multi-peaked MDCs of EuCd\textsubscript{2}As\textsubscript{2} in (b3). \textbf{(a4-a5, b4-b5)} Simulated ARPES spectra with small and large imaginary parts of self-energy ($\Sigma''$) in (a4) and (a5) for EuCd\textsubscript{2}P\textsubscript{2} and in (b4) and (b5) for EuCd\textsubscript{2}As\textsubscript{2}, respectively. \textbf{(a6, b6)} MDCs at a binding energy of -0.4 eV from the simulated ARPES spectra with small (red) and large (blue) $\Sigma''$ juxtaposed with the experimental MDC (black) for EuCd\textsubscript{2}P\textsubscript{2} (a6) and EuCd\textsubscript{2}As\textsubscript{2} (b6), respectively. The MDC with large (small) $\Sigma''$ agrees with the experimental one qualitatively in EuCd\textsubscript{2}P\textsubscript{2} (EuCd\textsubscript{2}As\textsubscript{2}) and is highlighted accordingly. 
}
\label{Fig: polarons}
\end{figure*}

To further discern the essential elements contributing to CMR, the striking difference in transport between EuCd\textsubscript{2}P\textsubscript{2} and EuCd\textsubscript{2}As\textsubscript{2} provides a valuable comparative basis. Hence, we compare the ARPES spectra of EuCd\textsubscript{2}P\textsubscript{2} and EuCd\textsubscript{2}As\textsubscript{2} in Fig. \ref{Fig: polarons}. Consistent with the DFT results, the ARPES spectra exhibit three sets of dispersive bands in both materials, labeled as $\alpha$, $\beta$, and $\gamma$, respectively (Fig. \ref{Fig: polarons}(a1, b1)). Notably, the spectral weight of the $\alpha$ band in EuCd\textsubscript{2}P\textsubscript{2} shows a marked reduction in spectral weight relative to that in EuCd\textsubscript{2}As\textsubscript{2}, a disparity that is more pronounced in the second-momentum-derivatives (Fig. \ref{Fig: polarons}(a2, b2)). This results in a clear divergence between single-peak-like and multi-peak-like momentum distribution curves (MDCs, Fig. \ref{Fig: polarons}(a3, b3)). Unlike the manganates or cuprates where the low-energy spectra are dominated by a single band, the multi-band nature here introduces complexities and necessitates spectral simulation for interpretation. Utilizing the DFT-calculated band structure, we present the simulated ARPES spectra with small and large imaginary parts of self-energy ($\Sigma''$) for EuCd\textsubscript{2}P\textsubscript{2} (Fig. \ref{Fig: polarons}(a4, a5)) and for EuCd\textsubscript{2}As\textsubscript{2} (Fig. \ref{Fig: polarons}(b4, b5)), respectively, after careful treatments of the photoemission matrix elements and the $k_z$ resolution (details in SI section 3). The simulated MDCs are juxtaposed with the experimental results in Fig. \ref{Fig: polarons} (a6, b6), where the single-peak-like (multi-peak-like) scenario signifies strong (weak) interactions in EuCd\textsubscript{2}P\textsubscript{2} (EuCd\textsubscript{2}As\textsubscript{2}). As $\Sigma''$ captures the many-body interactions in the spectral function, the difference in $\Sigma''$ points to more pronounced interactions in EuCd\textsubscript{2}P\textsubscript{2} compared to that in EuCd\textsubscript{2}As\textsubscript{2}. 

We explore the origin of these enhanced interactions. We first exclude the possibility of collective spin excitations, as the single-peak MDCs in EuCd\textsubscript{2}P\textsubscript{2} persist up to 150 K, well above the magnetic transition and deep into the paramagnetic phase (details in SI section 2.C). A more likely candidate for these persistently enhanced interactions is the electron-phonon (e-ph) coupling, further corroborated by several factors: (i) a ``bad metal'' behavior at $T\gg T\textsubscript{MR}$ in transport (Fig. \ref{Fig: baseline}(b))\cite{wang2021colossal}, (ii) a diminished spectral weight at low energy devoid of Landau quasiparticles (Fig. \ref{Fig: spin_fluctuations}(a)), (iii) a broadened spectral dispersion aligned with the DFT results, indicative of strongly polaronic limit\cite{mannella2005nodal, mannella2007polaron, engelsberg1963coupled}, and (iv) the Gaussian-enveloped energy distribution curves due to Frank-Condon broadening (Fig. S19)\cite{shen2004missing}. Consequently, we highlight the polaronic interactions for the itinerant $p$ electrons as a critical factor in the emergence of the CMR. 

A unique $p$-$f$ mixing in the low-energy states could serve as the microscopic origin of the polaronic interactions. We show the orbitally-projected band dispersion and the density of states for EuCd\textsubscript{2}As\textsubscript{2} and EuCd\textsubscript{2}P\textsubscript{2} in Fig. \ref{Fig: baseline}(d1, d2) and (e1, e2), respectively. We note a substantial mixing of polaronic $4f$ states into pnictogen $p$ states near $\mu$ (Fig. \ref{Fig: baseline}(d1, d2)), which is more pronounced in EuCd\textsubscript{2}P\textsubscript{2} (Fig. \ref{Fig: baseline}(e1, e2)) and echoes the comparatively larger polaronic interactions in EuCd\textsubscript{2}P\textsubscript{2} as discussed above. The prominent $p$-$f$ mixing is further supported experimentally by the fact that the spectral weight of the pnictogen $p$ bands follows the same trend as that of the Eu $4f$ bands across the Eu $4d\rightarrow 4f$ resonant transition (Fig. \ref{Fig: baseline}(c), details in SI section 2.A). Remarkably, the $p$-$f$ mixing not only enhances the polaronic interactions on low-energy states in the phonon channel, but also enables the conducting $p$ electrons carrying partial magnetic moments from $f$ electrons, facilitating a direct coupling to the magnetic clusters in the spin channel. As such, we suggest the presence of novel spin-polarized polarons as the conducting electrons in EuCd\textsubscript{2}P\textsubscript{2}. 

With the establishment of spin-polarized polarons, we now discuss the mechanism for the resistivity enhancement near $T$\textsubscript{MR}. Primarily, the CMR in EuCd\textsubscript{2}P\textsubscript{2} originates from the scattering of conducting electrons at the boundaries of ferromagnetic clusters near T\textsubscript{MR} and dissipates in (anti-)ferromagnetic phase (T$\ll$T\textsubscript{MR}) or paramagnetic phase (T$\gg$T\textsubscript{MR}), mirroring the time-reversal symmetry breaking in \cite{sunko2023spin} and aligning with the decoherence of electrons reflected by spectral weight suppression near $T$\textsubscript{MR}(Fig. \ref{Fig: spin_fluctuations}(e)). However, this classical Ising picture involving near-free conducting electrons typically yields $\sim$ 5\% magnetoresistance\cite{bozorth1946magnetoresistance}, insufficient to account for $>$ 5,000\% resistivity changes in EuCd\textsubscript{2}P\textsubscript{2}. To reconcile such dramatic difference, we show the critically helpful role of strong e-ph coupling by considering a tunneling process for spin-polarized polarons across a magnetic cluster boundary. In this process, the tunneling current $I$ in the small voltage limit is given by:
\begin{equation*}
    I = 4eV \sum_{\boldsymbol{k},\boldsymbol{p}} \sum_{\sigma, \sigma'} \left|T_{\boldsymbol{k}, \boldsymbol{p}}^{\sigma, \sigma'}\right|^2 \int \frac{dx}{\pi} A_{\sigma}^L(\boldsymbol{k}, \boldsymbol{x}) A_{\sigma'}^R(\boldsymbol{p}, \boldsymbol{x})\left(\frac{\partial f}{\partial x}\right)
\end{equation*}
, where $e$ is the charge of electron, $V$ is the voltage, $T_{\boldsymbol{k}, \boldsymbol{p}}^{\sigma, \sigma'}$ is the tunneling matrix element between the momentum-spin channel $(\boldsymbol{k}, \sigma)$ and $(\boldsymbol{p}, \sigma')$, and $f$ is the Fermi-Dirac distribution with spatial dependence. $A^{L, R}_{\sigma}$ is the spectral function on either side of the ferromagnetic cluster in the form of
\begin{equation*}
    A_{\boldsymbol{k}, \sigma}^{L, R} = \pi Z \delta(x-\epsilon_{\boldsymbol{k}} \pm J\sigma)
\end{equation*}
, where $Z$ is the quasiparticle residue, $\epsilon_{\boldsymbol{k}}$ is the energy of electrons with momentum $\boldsymbol{k}$, and $J$ is the spin exchange splitting. Neglecting the momentum dependence of quasiparticle residue and tunneling matrix for this localized process, the resistance across the ferromagnetic cluster wall becomes
\begin{equation*}
    \frac{V}{I} \propto \frac{1}{Z^2}\cdot \frac{1}{D}
\end{equation*}
, with the joint density of spin-split states $D = 4\pi |T|^2 \sum_{\sigma, \boldsymbol{k}, \boldsymbol{p}} \delta(\epsilon_{\boldsymbol{k}}\pm \sigma J)\delta(\epsilon_{\boldsymbol{p}}\pm \sigma J)$. Therefore, the CMR in EuCd\textsubscript{2}P\textsubscript{2} can be conceptually decomposed into two factors. First, the formation of ferromagnetic clusters increases the tunneling resistance by a reduction in the joint density of spin-split states $D$ under large spin splitting. Second, the quasiparticle residue $Z$ is significantly diminished by strong e-ph coupling, which amplifies the resistivity change by $1/Z^2$. This scenario highlights the cooperative interactions between the spin and phonon channels that account for pronounced resistivity modulation.

In summary, we propose a novel scenario where the CMR arises from the scattering of the spin-polarized polarons at the interfaces of ferromagnetic cluster in bad metals. In stark contrast with the magnetoresistance from the conventional spin-fermion coupling\cite{bozorth1946magnetoresistance}, it is the critically helpful role of e-ph coupling that dramatically amplifies the magnitude. Such polaronic behavior may be the characteristic of the entire EuM\textsubscript{2}X\textsubscript{2} material family (M=Cd, In, Zn; X=P, As)  with EuCd\textsubscript{2}P\textsubscript{2} being the most extreme case, and offer insights into their strong correlations and topological properties. In a broader view, our findings echo those in the high-temperature superconductivity, where $T_c$ is substantially enhanced by e-ph coupling \cite{he2018rapid, lee2014interfacial}, illustrating the potential for multiple collaborative interactions to catalyze dramatic changes during phase transitions.

\textbf{Experimental methods.} The samples, grown by the flux method\cite{wang2021colossal}, demonstrated good homogeneity within the same batch in transport measurements. The resistivity measurements were thereby measured on a sample that is either the same or within the identical batch for the ARPES experiments. The ARPES measurements were mainly performed with a Scienta DA-30 analyzer with an overall energy resolution of 12 meV at Beamline 5-2, Stanford Synchrotron Radiation Lightsource, SLAC National Accelerator Laboratory. Temperature-dependent measurements are carefully controlled on sample aging by limiting the measurement times below 20 h and maintaining the vacuum better than $2\times10^{-11}$ Torr below 25 K and than $4\times10^{-11}$ Torr above 25 K. Photon-flux-dependent measurements were also performed on the samples to confirm the absence of the charge effect. The Fermi edge of polycrystalline gold was measured after each high statistic spectra on samples to determine the temperature-dependent chemical potential that also fluctuates $< 1$ meV in 1 hour. The samples are pre-polished with large and flat surfaces to overcome the difficulties in sample cleavage. A real space mapping utilizing a small beam spot (~60 $\mu$m $\times$ 40 $\mu$m diameter) at the synchrotron beamline station was carried out before taking photoemission spectra to ensure a consistent dispersion across a surface area greater than ~250 $\mu$m $\times$ 250 $\mu$m.

\noindent \textbf{DFT calculations.} The band structures and magnetic configuration energies were calculated with DFT implemented in the Vienna ab initio simulation package (VASP) \cite{VASP0, VASP1, VASP2, VASP3}, using the projector-augmented wave method \cite{PAW} and the Perdew-Burke-Ernzerhof (PBE) exchange-correlation functional \cite{PBE}. We used the europium pseudopotential with 4$f$ electrons explicitly in the valence, with an additional Hubbard $U=5$ eV and spin-orbit coupling (SOC) to treat their localization properly \cite{LLWang_ECA-Weyl, XiDai_EIA_axionInsulator, wang2021colossal}. The DFT$+U+$SOC calculations were performed using the rotationally invariant DFT+$U$ method\cite{LSDApU_Dudarev}. We chose the A-type antiferromagnetic as the ground state for calculating band structures. Ionic relaxation was carried out on a $k$-point Monkhorst-Pack grid sizes of $17\times 17\times 9$ for 2-Eu supercell and $17\times 17\times 17$ for 1-Sr unit cell, followed by the self-consistent field calculations using a $k$-point grid of $21 \times 21 \times 11$ for the 2-Eu supercell and $21 \times 21 \times 21$ for the single-Sr unit cell.

\textbf{Acknowledgement.} This work is dedicated to the memory of Prof. J. Zaanen, whose insights and guidance were fundamental in shaping this research. The works at Stanford University and SLAC are supported by the U.S. Department of Energy, Office of Science, Office of Basic Energy Sciences, Division of Materials Sciences and Engineering, under Contract No. DE-AC02-76SF00515.
This research used resources of the National Energy Research Scientific Computing Center (NERSC), a U.S. Department of Energy Office of Science User Facility, operated under Contract No. DEAC02-05CH11231. The work at Boston College was funded by the U.S. Department of Energy, Office of Basic Energy Sciences, Division of Physical Behavior of Materials under award DE-SC0023124.

\textbf{Author contributions.} Y.-F.L., E.M.B., F.T., T.P.D., and Z.-X.S. designed the project. Y.-F.L. performed ARPES measurements, analyzed data, and implemented the ARPES spectra simulations. E.M.B. performed the DFT calculations. M.M., Z.C.W., S.B., and F.T. synthesized samples and performed transport measurements. S.-D.C., D.H.L., and M.H. contributed to instrument development. Y.-F.L., E.M.B., T.P.D., and Z.-X.S. wrote the manuscript with inputs from all authors. T.P.D. and Z.-X.S. supervised the project.

\bibliography{bib_main}

\begin{thebibliography}{36}%
\makeatletter
\providecommand \@ifxundefined [1]{%
 \@ifx{#1\undefined}
}%
\providecommand \@ifnum [1]{%
 \ifnum #1\expandafter \@firstoftwo
 \else \expandafter \@secondoftwo
 \fi
}%
\providecommand \@ifx [1]{%
 \ifx #1\expandafter \@firstoftwo
 \else \expandafter \@secondoftwo
 \fi
}%
\providecommand \natexlab [1]{#1}%
\providecommand \enquote  [1]{``#1''}%
\providecommand \bibnamefont  [1]{#1}%
\providecommand \bibfnamefont [1]{#1}%
\providecommand \citenamefont [1]{#1}%
\providecommand \href@noop [0]{\@secondoftwo}%
\providecommand \href [0]{\begingroup \@sanitize@url \@href}%
\providecommand \@href[1]{\@@startlink{#1}\@@href}%
\providecommand \@@href[1]{\endgroup#1\@@endlink}%
\providecommand \@sanitize@url [0]{\catcode `\\12\catcode `\$12\catcode `\&12\catcode `\#12\catcode `\^12\catcode `\_12\catcode `\%12\relax}%
\providecommand \@@startlink[1]{}%
\providecommand \@@endlink[0]{}%
\providecommand \url  [0]{\begingroup\@sanitize@url \@url }%
\providecommand \@url [1]{\endgroup\@href {#1}{\urlprefix }}%
\providecommand \urlprefix  [0]{URL }%
\providecommand \Eprint [0]{\href }%
\providecommand \doibase [0]{https://doi.org/}%
\providecommand \selectlanguage [0]{\@gobble}%
\providecommand \bibinfo  [0]{\@secondoftwo}%
\providecommand \bibfield  [0]{\@secondoftwo}%
\providecommand \translation [1]{[#1]}%
\providecommand \BibitemOpen [0]{}%
\providecommand \bibitemStop [0]{}%
\providecommand \bibitemNoStop [0]{.\EOS\space}%
\providecommand \EOS [0]{\spacefactor3000\relax}%
\providecommand \BibitemShut  [1]{\csname bibitem#1\endcsname}%
\let\auto@bib@innerbib\@empty
\bibitem [{\citenamefont {Kirchner}\ \emph {et~al.}(2020)\citenamefont {Kirchner}, \citenamefont {Paschen}, \citenamefont {Chen}, \citenamefont {Wirth}, \citenamefont {Feng}, \citenamefont {Thompson},\ and\ \citenamefont {Si}}]{kirchner2020colloquium}%
  \BibitemOpen
  \bibfield  {author} {\bibinfo {author} {\bibfnamefont {S.}~\bibnamefont {Kirchner}}, \bibinfo {author} {\bibfnamefont {S.}~\bibnamefont {Paschen}}, \bibinfo {author} {\bibfnamefont {Q.}~\bibnamefont {Chen}}, \bibinfo {author} {\bibfnamefont {S.}~\bibnamefont {Wirth}}, \bibinfo {author} {\bibfnamefont {D.}~\bibnamefont {Feng}}, \bibinfo {author} {\bibfnamefont {J.~D.}\ \bibnamefont {Thompson}},\ and\ \bibinfo {author} {\bibfnamefont {Q.}~\bibnamefont {Si}},\ }\bibfield  {title} {\bibinfo {title} {Colloquium: Heavy-electron quantum criticality and single-particle spectroscopy},\ }\href@noop {} {\bibfield  {journal} {\bibinfo  {journal} {Reviews of Modern Physics}\ }\textbf {\bibinfo {volume} {92}},\ \bibinfo {pages} {011002} (\bibinfo {year} {2020})}\BibitemShut {NoStop}%
\bibitem [{\citenamefont {Shaginyan}\ \emph {et~al.}(2009)\citenamefont {Shaginyan}, \citenamefont {Amusia}, \citenamefont {Msezane}, \citenamefont {Popov},\ and\ \citenamefont {Stephanovich}}]{shaginyan2009energy}%
  \BibitemOpen
  \bibfield  {author} {\bibinfo {author} {\bibfnamefont {V.}~\bibnamefont {Shaginyan}}, \bibinfo {author} {\bibfnamefont {M.~Y.}\ \bibnamefont {Amusia}}, \bibinfo {author} {\bibfnamefont {A.}~\bibnamefont {Msezane}}, \bibinfo {author} {\bibfnamefont {K.}~\bibnamefont {Popov}},\ and\ \bibinfo {author} {\bibfnamefont {V.}~\bibnamefont {Stephanovich}},\ }\bibfield  {title} {\bibinfo {title} {Energy scales and magnetoresistance at a quantum critical point},\ }\href@noop {} {\bibfield  {journal} {\bibinfo  {journal} {Physics Letters A}\ }\textbf {\bibinfo {volume} {373}},\ \bibinfo {pages} {986} (\bibinfo {year} {2009})}\BibitemShut {NoStop}%
\bibitem [{\citenamefont {Salamon}\ and\ \citenamefont {Jaime}(2001)}]{salamon2001physics}%
  \BibitemOpen
  \bibfield  {author} {\bibinfo {author} {\bibfnamefont {M.~B.}\ \bibnamefont {Salamon}}\ and\ \bibinfo {author} {\bibfnamefont {M.}~\bibnamefont {Jaime}},\ }\bibfield  {title} {\bibinfo {title} {The physics of manganites: Structure and transport},\ }\href@noop {} {\bibfield  {journal} {\bibinfo  {journal} {Reviews of modern physics}\ }\textbf {\bibinfo {volume} {73}},\ \bibinfo {pages} {583} (\bibinfo {year} {2001})}\BibitemShut {NoStop}%
\bibitem [{\citenamefont {Uehara}\ \emph {et~al.}(1999)\citenamefont {Uehara}, \citenamefont {Mori}, \citenamefont {Chen},\ and\ \citenamefont {Cheong}}]{uehara1999percolative}%
  \BibitemOpen
  \bibfield  {author} {\bibinfo {author} {\bibfnamefont {M.}~\bibnamefont {Uehara}}, \bibinfo {author} {\bibfnamefont {S.}~\bibnamefont {Mori}}, \bibinfo {author} {\bibfnamefont {C.}~\bibnamefont {Chen}},\ and\ \bibinfo {author} {\bibfnamefont {S.-W.}\ \bibnamefont {Cheong}},\ }\bibfield  {title} {\bibinfo {title} {Percolative phase separation underlies colossal magnetoresistance in mixed-valent manganites},\ }\href@noop {} {\bibfield  {journal} {\bibinfo  {journal} {Nature}\ }\textbf {\bibinfo {volume} {399}},\ \bibinfo {pages} {560} (\bibinfo {year} {1999})}\BibitemShut {NoStop}%
\bibitem [{\citenamefont {Mannella}\ \emph {et~al.}(2005)\citenamefont {Mannella}, \citenamefont {Yang}, \citenamefont {Zhou}, \citenamefont {Zheng}, \citenamefont {Mitchell}, \citenamefont {Zaanen}, \citenamefont {Devereaux}, \citenamefont {Nagaosa}, \citenamefont {Hussain},\ and\ \citenamefont {Shen}}]{mannella2005nodal}%
  \BibitemOpen
  \bibfield  {author} {\bibinfo {author} {\bibfnamefont {N.}~\bibnamefont {Mannella}}, \bibinfo {author} {\bibfnamefont {W.~L.}\ \bibnamefont {Yang}}, \bibinfo {author} {\bibfnamefont {X.~J.}\ \bibnamefont {Zhou}}, \bibinfo {author} {\bibfnamefont {H.}~\bibnamefont {Zheng}}, \bibinfo {author} {\bibfnamefont {J.~F.}\ \bibnamefont {Mitchell}}, \bibinfo {author} {\bibfnamefont {J.}~\bibnamefont {Zaanen}}, \bibinfo {author} {\bibfnamefont {T.~P.}\ \bibnamefont {Devereaux}}, \bibinfo {author} {\bibfnamefont {N.}~\bibnamefont {Nagaosa}}, \bibinfo {author} {\bibfnamefont {Z.}~\bibnamefont {Hussain}},\ and\ \bibinfo {author} {\bibfnamefont {Z.-X.}\ \bibnamefont {Shen}},\ }\bibfield  {title} {\bibinfo {title} {Nodal quasiparticle in pseudogapped colossal magnetoresistive manganites},\ }\href@noop {} {\bibfield  {journal} {\bibinfo  {journal} {Nature}\ }\textbf {\bibinfo {volume} {438}},\ \bibinfo {pages} {474} (\bibinfo {year} {2005})}\BibitemShut {NoStop}%
\bibitem [{\citenamefont {Liang}\ \emph {et~al.}(2015)\citenamefont {Liang}, \citenamefont {Gibson}, \citenamefont {Ali}, \citenamefont {Liu}, \citenamefont {Cava},\ and\ \citenamefont {Ong}}]{liang2015ultrahigh}%
  \BibitemOpen
  \bibfield  {author} {\bibinfo {author} {\bibfnamefont {T.}~\bibnamefont {Liang}}, \bibinfo {author} {\bibfnamefont {Q.}~\bibnamefont {Gibson}}, \bibinfo {author} {\bibfnamefont {M.~N.}\ \bibnamefont {Ali}}, \bibinfo {author} {\bibfnamefont {M.}~\bibnamefont {Liu}}, \bibinfo {author} {\bibfnamefont {R.}~\bibnamefont {Cava}},\ and\ \bibinfo {author} {\bibfnamefont {N.}~\bibnamefont {Ong}},\ }\bibfield  {title} {\bibinfo {title} {Ultrahigh mobility and giant magnetoresistance in the dirac semimetal $\mathrm{Cd}_3\mathrm{As}_2$},\ }\href@noop {} {\bibfield  {journal} {\bibinfo  {journal} {Nature Materials}\ }\textbf {\bibinfo {volume} {14}},\ \bibinfo {pages} {280} (\bibinfo {year} {2015})}\BibitemShut {NoStop}%
\bibitem [{\citenamefont {Shekhar}\ \emph {et~al.}(2015)\citenamefont {Shekhar}, \citenamefont {Nayak}, \citenamefont {Sun}, \citenamefont {Schmidt}, \citenamefont {Nicklas}, \citenamefont {Leermakers}, \citenamefont {Zeitler}, \citenamefont {Skourski}, \citenamefont {Wosnitza}, \citenamefont {Liu} \emph {et~al.}}]{shekhar2015extremely}%
  \BibitemOpen
  \bibfield  {author} {\bibinfo {author} {\bibfnamefont {C.}~\bibnamefont {Shekhar}}, \bibinfo {author} {\bibfnamefont {A.~K.}\ \bibnamefont {Nayak}}, \bibinfo {author} {\bibfnamefont {Y.}~\bibnamefont {Sun}}, \bibinfo {author} {\bibfnamefont {M.}~\bibnamefont {Schmidt}}, \bibinfo {author} {\bibfnamefont {M.}~\bibnamefont {Nicklas}}, \bibinfo {author} {\bibfnamefont {I.}~\bibnamefont {Leermakers}}, \bibinfo {author} {\bibfnamefont {U.}~\bibnamefont {Zeitler}}, \bibinfo {author} {\bibfnamefont {Y.}~\bibnamefont {Skourski}}, \bibinfo {author} {\bibfnamefont {J.}~\bibnamefont {Wosnitza}}, \bibinfo {author} {\bibfnamefont {Z.}~\bibnamefont {Liu}}, \emph {et~al.},\ }\bibfield  {title} {\bibinfo {title} {Extremely large magnetoresistance and ultrahigh mobility in the topological $\mathrm{Weyl}$ semimetal candidate $\mathrm{NbP}$},\ }\href@noop {} {\bibfield  {journal} {\bibinfo  {journal} {Nature Physics}\ }\textbf {\bibinfo {volume} {11}},\ \bibinfo {pages} {645} (\bibinfo {year} {2015})}\BibitemShut {NoStop}%
\bibitem [{\citenamefont {Suzuki}\ \emph {et~al.}(2019)\citenamefont {Suzuki}, \citenamefont {Savary}, \citenamefont {Liu}, \citenamefont {Lynn}, \citenamefont {Balents},\ and\ \citenamefont {Checkelsky}}]{suzuki2019singular}%
  \BibitemOpen
  \bibfield  {author} {\bibinfo {author} {\bibfnamefont {T.}~\bibnamefont {Suzuki}}, \bibinfo {author} {\bibfnamefont {L.}~\bibnamefont {Savary}}, \bibinfo {author} {\bibfnamefont {J.-P.}\ \bibnamefont {Liu}}, \bibinfo {author} {\bibfnamefont {J.~W.}\ \bibnamefont {Lynn}}, \bibinfo {author} {\bibfnamefont {L.}~\bibnamefont {Balents}},\ and\ \bibinfo {author} {\bibfnamefont {J.~G.}\ \bibnamefont {Checkelsky}},\ }\bibfield  {title} {\bibinfo {title} {Singular angular magnetoresistance in a magnetic nodal semimetal},\ }\href@noop {} {\bibfield  {journal} {\bibinfo  {journal} {Science}\ }\textbf {\bibinfo {volume} {365}},\ \bibinfo {pages} {377} (\bibinfo {year} {2019})}\BibitemShut {NoStop}%
\bibitem [{\citenamefont {Singha}\ \emph {et~al.}(2023)\citenamefont {Singha}, \citenamefont {Dalgaard}, \citenamefont {Marchenko}, \citenamefont {Krivenkov}, \citenamefont {Rienks}, \citenamefont {Jovanovic}, \citenamefont {Teicher}, \citenamefont {Hu}, \citenamefont {Salters}, \citenamefont {Lin} \emph {et~al.}}]{singha2023colossal}%
  \BibitemOpen
  \bibfield  {author} {\bibinfo {author} {\bibfnamefont {R.}~\bibnamefont {Singha}}, \bibinfo {author} {\bibfnamefont {K.~J.}\ \bibnamefont {Dalgaard}}, \bibinfo {author} {\bibfnamefont {D.}~\bibnamefont {Marchenko}}, \bibinfo {author} {\bibfnamefont {M.}~\bibnamefont {Krivenkov}}, \bibinfo {author} {\bibfnamefont {E.~D.}\ \bibnamefont {Rienks}}, \bibinfo {author} {\bibfnamefont {M.}~\bibnamefont {Jovanovic}}, \bibinfo {author} {\bibfnamefont {S.~M.}\ \bibnamefont {Teicher}}, \bibinfo {author} {\bibfnamefont {J.}~\bibnamefont {Hu}}, \bibinfo {author} {\bibfnamefont {T.~H.}\ \bibnamefont {Salters}}, \bibinfo {author} {\bibfnamefont {J.}~\bibnamefont {Lin}}, \emph {et~al.},\ }\bibfield  {title} {\bibinfo {title} {Colossal magnetoresistance in the multiple wave vector charge density wave regime of an antiferromagnetic dirac semimetal},\ }\href@noop {} {\bibfield  {journal} {\bibinfo  {journal} {Science Advances}\ }\textbf {\bibinfo {volume} {9}},\ \bibinfo {pages} {eadh0145} (\bibinfo {year} {2023})}\BibitemShut
  {NoStop}%
\bibitem [{\citenamefont {Wang}\ \emph {et~al.}(2021)\citenamefont {Wang}, \citenamefont {Rogers}, \citenamefont {Yao}, \citenamefont {Nichols}, \citenamefont {Atay}, \citenamefont {Xu}, \citenamefont {Franklin}, \citenamefont {Sochnikov}, \citenamefont {Ryan}, \citenamefont {Haskel} \emph {et~al.}}]{wang2021colossal}%
  \BibitemOpen
  \bibfield  {author} {\bibinfo {author} {\bibfnamefont {Z.-C.}\ \bibnamefont {Wang}}, \bibinfo {author} {\bibfnamefont {J.~D.}\ \bibnamefont {Rogers}}, \bibinfo {author} {\bibfnamefont {X.}~\bibnamefont {Yao}}, \bibinfo {author} {\bibfnamefont {R.}~\bibnamefont {Nichols}}, \bibinfo {author} {\bibfnamefont {K.}~\bibnamefont {Atay}}, \bibinfo {author} {\bibfnamefont {B.}~\bibnamefont {Xu}}, \bibinfo {author} {\bibfnamefont {J.}~\bibnamefont {Franklin}}, \bibinfo {author} {\bibfnamefont {I.}~\bibnamefont {Sochnikov}}, \bibinfo {author} {\bibfnamefont {P.~J.}\ \bibnamefont {Ryan}}, \bibinfo {author} {\bibfnamefont {D.}~\bibnamefont {Haskel}}, \emph {et~al.},\ }\bibfield  {title} {\bibinfo {title} {Colossal magnetoresistance without mixed valence in a layered phosphide crystal},\ }\href@noop {} {\bibfield  {journal} {\bibinfo  {journal} {Advanced Materials}\ }\textbf {\bibinfo {volume} {33}},\ \bibinfo {pages} {2005755} (\bibinfo {year} {2021})}\BibitemShut {NoStop}%
\bibitem [{\citenamefont {Wang}\ \emph {et~al.}(2022)\citenamefont {Wang}, \citenamefont {Been}, \citenamefont {Gaudet}, \citenamefont {Alqasseri}, \citenamefont {Fruhling}, \citenamefont {Yao}, \citenamefont {Stuhr}, \citenamefont {Zhu}, \citenamefont {Ren}, \citenamefont {Cui} \emph {et~al.}}]{wang2022anisotropy}%
  \BibitemOpen
  \bibfield  {author} {\bibinfo {author} {\bibfnamefont {Z.-C.}\ \bibnamefont {Wang}}, \bibinfo {author} {\bibfnamefont {E.}~\bibnamefont {Been}}, \bibinfo {author} {\bibfnamefont {J.}~\bibnamefont {Gaudet}}, \bibinfo {author} {\bibfnamefont {G.~M.~A.}\ \bibnamefont {Alqasseri}}, \bibinfo {author} {\bibfnamefont {K.}~\bibnamefont {Fruhling}}, \bibinfo {author} {\bibfnamefont {X.}~\bibnamefont {Yao}}, \bibinfo {author} {\bibfnamefont {U.}~\bibnamefont {Stuhr}}, \bibinfo {author} {\bibfnamefont {Q.}~\bibnamefont {Zhu}}, \bibinfo {author} {\bibfnamefont {Z.}~\bibnamefont {Ren}}, \bibinfo {author} {\bibfnamefont {Y.}~\bibnamefont {Cui}}, \emph {et~al.},\ }\bibfield  {title} {\bibinfo {title} {Anisotropy of the magnetic and transport properties of $\mathrm{EuZn}_2\mathrm{As}_2$},\ }\href@noop {} {\bibfield  {journal} {\bibinfo  {journal} {Physical Review B}\ }\textbf {\bibinfo {volume} {105}},\ \bibinfo {pages} {165122} (\bibinfo {year} {2022})}\BibitemShut {NoStop}%
\bibitem [{\citenamefont {Homes}\ \emph {et~al.}(2023)\citenamefont {Homes}, \citenamefont {Wang}, \citenamefont {Fruhling},\ and\ \citenamefont {Tafti}}]{homes2023optical}%
  \BibitemOpen
  \bibfield  {author} {\bibinfo {author} {\bibfnamefont {C.~C.}\ \bibnamefont {Homes}}, \bibinfo {author} {\bibfnamefont {Z.-C.}\ \bibnamefont {Wang}}, \bibinfo {author} {\bibfnamefont {K.}~\bibnamefont {Fruhling}},\ and\ \bibinfo {author} {\bibfnamefont {F.}~\bibnamefont {Tafti}},\ }\bibfield  {title} {\bibinfo {title} {Optical properties and carrier localization in the layered phosphide $\mathrm{EuCd}_2\mathrm{P}_2$},\ }\href@noop {} {\bibfield  {journal} {\bibinfo  {journal} {Physical Review B}\ }\textbf {\bibinfo {volume} {107}},\ \bibinfo {pages} {045106} (\bibinfo {year} {2023})}\BibitemShut {NoStop}%
\bibitem [{\citenamefont {Zhang}\ \emph {et~al.}(2023)\citenamefont {Zhang}, \citenamefont {Du}, \citenamefont {Zheng}, \citenamefont {Luo}, \citenamefont {Wu}, \citenamefont {Zheng}, \citenamefont {Cui}, \citenamefont {Sun}, \citenamefont {Liu}, \citenamefont {Shen} \emph {et~al.}}]{zhang2023electronic}%
  \BibitemOpen
  \bibfield  {author} {\bibinfo {author} {\bibfnamefont {H.}~\bibnamefont {Zhang}}, \bibinfo {author} {\bibfnamefont {F.}~\bibnamefont {Du}}, \bibinfo {author} {\bibfnamefont {X.}~\bibnamefont {Zheng}}, \bibinfo {author} {\bibfnamefont {S.}~\bibnamefont {Luo}}, \bibinfo {author} {\bibfnamefont {Y.}~\bibnamefont {Wu}}, \bibinfo {author} {\bibfnamefont {H.}~\bibnamefont {Zheng}}, \bibinfo {author} {\bibfnamefont {S.}~\bibnamefont {Cui}}, \bibinfo {author} {\bibfnamefont {Z.}~\bibnamefont {Sun}}, \bibinfo {author} {\bibfnamefont {Z.}~\bibnamefont {Liu}}, \bibinfo {author} {\bibfnamefont {D.}~\bibnamefont {Shen}}, \emph {et~al.},\ }\bibfield  {title} {\bibinfo {title} {Electronic band reconstruction across the insulator-metal transition in colossal magnetoresistive $\mathrm{EuCd}_2\mathrm{P}_2$},\ }\href@noop {} {\bibfield  {journal} {\bibinfo  {journal} {arXiv preprint arXiv:2308.16844}\ } (\bibinfo {year} {2023})}\BibitemShut {NoStop}%
\bibitem [{\citenamefont {Sunko}\ \emph {et~al.}(2023)\citenamefont {Sunko}, \citenamefont {Sun}, \citenamefont {Vranas}, \citenamefont {Homes}, \citenamefont {Lee}, \citenamefont {Donoway}, \citenamefont {Wang}, \citenamefont {Balguri}, \citenamefont {Mahendru}, \citenamefont {Ruiz} \emph {et~al.}}]{sunko2023spin}%
  \BibitemOpen
  \bibfield  {author} {\bibinfo {author} {\bibfnamefont {V.}~\bibnamefont {Sunko}}, \bibinfo {author} {\bibfnamefont {Y.}~\bibnamefont {Sun}}, \bibinfo {author} {\bibfnamefont {M.}~\bibnamefont {Vranas}}, \bibinfo {author} {\bibfnamefont {C.~C.}\ \bibnamefont {Homes}}, \bibinfo {author} {\bibfnamefont {C.}~\bibnamefont {Lee}}, \bibinfo {author} {\bibfnamefont {E.}~\bibnamefont {Donoway}}, \bibinfo {author} {\bibfnamefont {Z.-C.}\ \bibnamefont {Wang}}, \bibinfo {author} {\bibfnamefont {S.}~\bibnamefont {Balguri}}, \bibinfo {author} {\bibfnamefont {M.~B.}\ \bibnamefont {Mahendru}}, \bibinfo {author} {\bibfnamefont {A.}~\bibnamefont {Ruiz}}, \emph {et~al.},\ }\bibfield  {title} {\bibinfo {title} {Spin-carrier coupling induced ferromagnetism and giant resistivity peak in $\mathrm{EuCd}_2\mathrm{P}_2$},\ }\href@noop {} {\bibfield  {journal} {\bibinfo  {journal} {Physical Review B}\ }\textbf {\bibinfo {volume} {107}},\ \bibinfo {pages} {144404} (\bibinfo {year} {2023})}\BibitemShut {NoStop}%
\bibitem [{\citenamefont {Krebber}\ \emph {et~al.}(2023)\citenamefont {Krebber}, \citenamefont {Kopp}, \citenamefont {Garg}, \citenamefont {Kummer}, \citenamefont {Sichelschmidt}, \citenamefont {Schulz}, \citenamefont {Poelchen}, \citenamefont {Mende}, \citenamefont {Virovets}, \citenamefont {Warawa} \emph {et~al.}}]{krebber2023colossal}%
  \BibitemOpen
  \bibfield  {author} {\bibinfo {author} {\bibfnamefont {S.}~\bibnamefont {Krebber}}, \bibinfo {author} {\bibfnamefont {M.}~\bibnamefont {Kopp}}, \bibinfo {author} {\bibfnamefont {C.}~\bibnamefont {Garg}}, \bibinfo {author} {\bibfnamefont {K.}~\bibnamefont {Kummer}}, \bibinfo {author} {\bibfnamefont {J.}~\bibnamefont {Sichelschmidt}}, \bibinfo {author} {\bibfnamefont {S.}~\bibnamefont {Schulz}}, \bibinfo {author} {\bibfnamefont {G.}~\bibnamefont {Poelchen}}, \bibinfo {author} {\bibfnamefont {M.}~\bibnamefont {Mende}}, \bibinfo {author} {\bibfnamefont {A.~V.}\ \bibnamefont {Virovets}}, \bibinfo {author} {\bibfnamefont {K.}~\bibnamefont {Warawa}}, \emph {et~al.},\ }\bibfield  {title} {\bibinfo {title} {Colossal magnetoresistance in $\mathrm{EuZn}_2\mathrm{P}_2$ and its electronic and magnetic structure},\ }\href@noop {} {\bibfield  {journal} {\bibinfo  {journal} {Physical Review B}\ }\textbf {\bibinfo {volume} {108}},\ \bibinfo {pages} {045116} (\bibinfo {year} {2023})}\BibitemShut {NoStop}%
\bibitem [{\citenamefont {Usachov}\ \emph {et~al.}(2024)\citenamefont {Usachov}, \citenamefont {Krebber}, \citenamefont {Bokai}, \citenamefont {Tarasov}, \citenamefont {Kopp}, \citenamefont {Garg}, \citenamefont {Virovets}, \citenamefont {M{\"u}ller}, \citenamefont {Mende}, \citenamefont {Poelchen} \emph {et~al.}}]{usachov2024magnetism}%
  \BibitemOpen
  \bibfield  {author} {\bibinfo {author} {\bibfnamefont {D.~Y.}\ \bibnamefont {Usachov}}, \bibinfo {author} {\bibfnamefont {S.}~\bibnamefont {Krebber}}, \bibinfo {author} {\bibfnamefont {K.~A.}\ \bibnamefont {Bokai}}, \bibinfo {author} {\bibfnamefont {A.~V.}\ \bibnamefont {Tarasov}}, \bibinfo {author} {\bibfnamefont {M.}~\bibnamefont {Kopp}}, \bibinfo {author} {\bibfnamefont {C.}~\bibnamefont {Garg}}, \bibinfo {author} {\bibfnamefont {A.}~\bibnamefont {Virovets}}, \bibinfo {author} {\bibfnamefont {J.}~\bibnamefont {M{\"u}ller}}, \bibinfo {author} {\bibfnamefont {M.}~\bibnamefont {Mende}}, \bibinfo {author} {\bibfnamefont {G.}~\bibnamefont {Poelchen}}, \emph {et~al.},\ }\bibfield  {title} {\bibinfo {title} {Magnetism, heat capacity, and electronic structure of $\mathrm{EuCd}_2\mathrm{P}_2$ in view of its colossal magnetoresistance},\ }\href@noop {} {\bibfield  {journal} {\bibinfo  {journal} {Physical Review B}\ }\textbf {\bibinfo {volume} {109}},\ \bibinfo {pages} {104421} (\bibinfo {year} {2024})}\BibitemShut
  {NoStop}%
\bibitem [{\citenamefont {Ma}\ \emph {et~al.}(2019)\citenamefont {Ma}, \citenamefont {Nie}, \citenamefont {Yi}, \citenamefont {Jandke}, \citenamefont {Shang}, \citenamefont {Yao}, \citenamefont {Naamneh}, \citenamefont {Yan}, \citenamefont {Sun}, \citenamefont {Chikina} \emph {et~al.}}]{ma2019spin}%
  \BibitemOpen
  \bibfield  {author} {\bibinfo {author} {\bibfnamefont {J.-Z.}\ \bibnamefont {Ma}}, \bibinfo {author} {\bibfnamefont {S.}~\bibnamefont {Nie}}, \bibinfo {author} {\bibfnamefont {C.}~\bibnamefont {Yi}}, \bibinfo {author} {\bibfnamefont {J.}~\bibnamefont {Jandke}}, \bibinfo {author} {\bibfnamefont {T.}~\bibnamefont {Shang}}, \bibinfo {author} {\bibfnamefont {M.-Y.}\ \bibnamefont {Yao}}, \bibinfo {author} {\bibfnamefont {M.}~\bibnamefont {Naamneh}}, \bibinfo {author} {\bibfnamefont {L.}~\bibnamefont {Yan}}, \bibinfo {author} {\bibfnamefont {Y.}~\bibnamefont {Sun}}, \bibinfo {author} {\bibfnamefont {A.}~\bibnamefont {Chikina}}, \emph {et~al.},\ }\bibfield  {title} {\bibinfo {title} {Spin fluctuation induced weyl semimetal state in the paramagnetic phase of $\mathrm{EuCd}_2\mathrm{As}_2$},\ }\href@noop {} {\bibfield  {journal} {\bibinfo  {journal} {Science advances}\ }\textbf {\bibinfo {volume} {5}},\ \bibinfo {pages} {eaaw4718} (\bibinfo {year} {2019})}\BibitemShut {NoStop}%
\bibitem [{\citenamefont {Riberolles}\ \emph {et~al.}(2021)\citenamefont {Riberolles}, \citenamefont {Trevisan}, \citenamefont {Kuthanazhi}, \citenamefont {Heitmann}, \citenamefont {Ye}, \citenamefont {Johnston}, \citenamefont {Bud’ko}, \citenamefont {Ryan}, \citenamefont {Canfield}, \citenamefont {Kreyssig} \emph {et~al.}}]{riberolles2021magnetic}%
  \BibitemOpen
  \bibfield  {author} {\bibinfo {author} {\bibfnamefont {S.~X.}\ \bibnamefont {Riberolles}}, \bibinfo {author} {\bibfnamefont {T.~V.}\ \bibnamefont {Trevisan}}, \bibinfo {author} {\bibfnamefont {B.}~\bibnamefont {Kuthanazhi}}, \bibinfo {author} {\bibfnamefont {T.}~\bibnamefont {Heitmann}}, \bibinfo {author} {\bibfnamefont {F.}~\bibnamefont {Ye}}, \bibinfo {author} {\bibfnamefont {D.}~\bibnamefont {Johnston}}, \bibinfo {author} {\bibfnamefont {S.}~\bibnamefont {Bud’ko}}, \bibinfo {author} {\bibfnamefont {D.}~\bibnamefont {Ryan}}, \bibinfo {author} {\bibfnamefont {P.}~\bibnamefont {Canfield}}, \bibinfo {author} {\bibfnamefont {A.}~\bibnamefont {Kreyssig}}, \emph {et~al.},\ }\bibfield  {title} {\bibinfo {title} {Magnetic crystalline-symmetry-protected axion electrodynamics and field-tunable unpinned dirac cones in $\mathrm{EuIn}_2\mathrm{As}_2$},\ }\href@noop {} {\bibfield  {journal} {\bibinfo  {journal} {Nature communications}\ }\textbf {\bibinfo {volume} {12}},\ \bibinfo {pages} {999} (\bibinfo {year}
  {2021})}\BibitemShut {NoStop}%
\bibitem [{\citenamefont {Xu}\ \emph {et~al.}(2019)\citenamefont {Xu}, \citenamefont {Song}, \citenamefont {Wang}, \citenamefont {Weng},\ and\ \citenamefont {Dai}}]{XiDai_EIA_axionInsulator}%
  \BibitemOpen
  \bibfield  {author} {\bibinfo {author} {\bibfnamefont {Y.}~\bibnamefont {Xu}}, \bibinfo {author} {\bibfnamefont {Z.}~\bibnamefont {Song}}, \bibinfo {author} {\bibfnamefont {Z.}~\bibnamefont {Wang}}, \bibinfo {author} {\bibfnamefont {H.}~\bibnamefont {Weng}},\ and\ \bibinfo {author} {\bibfnamefont {X.}~\bibnamefont {Dai}},\ }\bibfield  {title} {\bibinfo {title} {Higher-order topology of the axion insulator $\mathrm{EuIn}_{2}\mathrm{As}_{2}$},\ }\href {https://doi.org/10.1103/PhysRevLett.122.256402} {\bibfield  {journal} {\bibinfo  {journal} {Phys. Rev. Lett.}\ }\textbf {\bibinfo {volume} {122}},\ \bibinfo {pages} {256402} (\bibinfo {year} {2019})}\BibitemShut {NoStop}%
\bibitem [{\citenamefont {Torrance}\ \emph {et~al.}(1972)\citenamefont {Torrance}, \citenamefont {Shafer},\ and\ \citenamefont {McGuire}}]{torrance1972bound}%
  \BibitemOpen
  \bibfield  {author} {\bibinfo {author} {\bibfnamefont {J.}~\bibnamefont {Torrance}}, \bibinfo {author} {\bibfnamefont {M.}~\bibnamefont {Shafer}},\ and\ \bibinfo {author} {\bibfnamefont {T.}~\bibnamefont {McGuire}},\ }\bibfield  {title} {\bibinfo {title} {Bound magnetic polarons and the insulator-metal transition in $\mathrm{EuO}$},\ }\href@noop {} {\bibfield  {journal} {\bibinfo  {journal} {Physical Review Letters}\ }\textbf {\bibinfo {volume} {29}},\ \bibinfo {pages} {1168} (\bibinfo {year} {1972})}\BibitemShut {NoStop}%
\bibitem [{\citenamefont {Bozorth}(1946)}]{bozorth1946magnetoresistance}%
  \BibitemOpen
  \bibfield  {author} {\bibinfo {author} {\bibfnamefont {R.}~\bibnamefont {Bozorth}},\ }\bibfield  {title} {\bibinfo {title} {Magnetoresistance and domain theory of iron-nickel alloys},\ }\href@noop {} {\bibfield  {journal} {\bibinfo  {journal} {Physical Review}\ }\textbf {\bibinfo {volume} {70}},\ \bibinfo {pages} {923} (\bibinfo {year} {1946})}\BibitemShut {NoStop}%
\bibitem [{\citenamefont {Jo}\ \emph {et~al.}(2020)\citenamefont {Jo}, \citenamefont {Kuthanazhi}, \citenamefont {Wu}, \citenamefont {Timmons}, \citenamefont {Kim}, \citenamefont {Zhou}, \citenamefont {Wang}, \citenamefont {Ueland}, \citenamefont {Palasyuk}, \citenamefont {Ryan} \emph {et~al.}}]{jo2020manipulating}%
  \BibitemOpen
  \bibfield  {author} {\bibinfo {author} {\bibfnamefont {N.~H.}\ \bibnamefont {Jo}}, \bibinfo {author} {\bibfnamefont {B.}~\bibnamefont {Kuthanazhi}}, \bibinfo {author} {\bibfnamefont {Y.}~\bibnamefont {Wu}}, \bibinfo {author} {\bibfnamefont {E.}~\bibnamefont {Timmons}}, \bibinfo {author} {\bibfnamefont {T.-H.}\ \bibnamefont {Kim}}, \bibinfo {author} {\bibfnamefont {L.}~\bibnamefont {Zhou}}, \bibinfo {author} {\bibfnamefont {L.-L.}\ \bibnamefont {Wang}}, \bibinfo {author} {\bibfnamefont {B.~G.}\ \bibnamefont {Ueland}}, \bibinfo {author} {\bibfnamefont {A.}~\bibnamefont {Palasyuk}}, \bibinfo {author} {\bibfnamefont {D.~H.}\ \bibnamefont {Ryan}}, \emph {et~al.},\ }\bibfield  {title} {\bibinfo {title} {Manipulating magnetism in the topological semimetal $\mathrm{EuCd}_2\mathrm{As}_2$},\ }\href@noop {} {\bibfield  {journal} {\bibinfo  {journal} {Physical Review B}\ }\textbf {\bibinfo {volume} {101}},\ \bibinfo {pages} {140402} (\bibinfo {year} {2020})}\BibitemShut {NoStop}%
\bibitem [{\citenamefont {Cuono}\ \emph {et~al.}(2023)\citenamefont {Cuono}, \citenamefont {Sattigeri}, \citenamefont {Autieri},\ and\ \citenamefont {Dietl}}]{ECP_trivial_Cuono_PRB2023}%
  \BibitemOpen
  \bibfield  {author} {\bibinfo {author} {\bibfnamefont {G.}~\bibnamefont {Cuono}}, \bibinfo {author} {\bibfnamefont {R.~M.}\ \bibnamefont {Sattigeri}}, \bibinfo {author} {\bibfnamefont {C.}~\bibnamefont {Autieri}},\ and\ \bibinfo {author} {\bibfnamefont {T.}~\bibnamefont {Dietl}},\ }\bibfield  {title} {\bibinfo {title} {Ab initio overestimation of the topological region in eu-based compounds},\ }\href {https://doi.org/10.1103/PhysRevB.108.075150} {\bibfield  {journal} {\bibinfo  {journal} {Phys. Rev. B}\ }\textbf {\bibinfo {volume} {108}},\ \bibinfo {pages} {075150} (\bibinfo {year} {2023})}\BibitemShut {NoStop}%
\bibitem [{\citenamefont {Mannella}\ \emph {et~al.}(2007)\citenamefont {Mannella}, \citenamefont {Yang}, \citenamefont {Tanaka}, \citenamefont {Zhou}, \citenamefont {Zheng}, \citenamefont {Mitchell}, \citenamefont {Zaanen}, \citenamefont {Devereaux}, \citenamefont {Nagaosa}, \citenamefont {Hussain} \emph {et~al.}}]{mannella2007polaron}%
  \BibitemOpen
  \bibfield  {author} {\bibinfo {author} {\bibfnamefont {N.}~\bibnamefont {Mannella}}, \bibinfo {author} {\bibfnamefont {W.}~\bibnamefont {Yang}}, \bibinfo {author} {\bibfnamefont {K.}~\bibnamefont {Tanaka}}, \bibinfo {author} {\bibfnamefont {X.}~\bibnamefont {Zhou}}, \bibinfo {author} {\bibfnamefont {H.}~\bibnamefont {Zheng}}, \bibinfo {author} {\bibfnamefont {J.}~\bibnamefont {Mitchell}}, \bibinfo {author} {\bibfnamefont {J.}~\bibnamefont {Zaanen}}, \bibinfo {author} {\bibfnamefont {T.}~\bibnamefont {Devereaux}}, \bibinfo {author} {\bibfnamefont {N.}~\bibnamefont {Nagaosa}}, \bibinfo {author} {\bibfnamefont {Z.}~\bibnamefont {Hussain}}, \emph {et~al.},\ }\bibfield  {title} {\bibinfo {title} {Polaron coherence condensation as the mechanism for colossal magnetoresistance in layered manganites},\ }\href@noop {} {\bibfield  {journal} {\bibinfo  {journal} {Physical Review B}\ }\textbf {\bibinfo {volume} {76}},\ \bibinfo {pages} {233102} (\bibinfo {year} {2007})}\BibitemShut {NoStop}%
\bibitem [{\citenamefont {Engelsberg}\ and\ \citenamefont {Schrieffer}(1963)}]{engelsberg1963coupled}%
  \BibitemOpen
  \bibfield  {author} {\bibinfo {author} {\bibfnamefont {S.}~\bibnamefont {Engelsberg}}\ and\ \bibinfo {author} {\bibfnamefont {J.}~\bibnamefont {Schrieffer}},\ }\bibfield  {title} {\bibinfo {title} {Coupled electron-phonon system},\ }\href@noop {} {\bibfield  {journal} {\bibinfo  {journal} {Physical Review}\ }\textbf {\bibinfo {volume} {131}},\ \bibinfo {pages} {993} (\bibinfo {year} {1963})}\BibitemShut {NoStop}%
\bibitem [{\citenamefont {Shen}\ \emph {et~al.}(2004)\citenamefont {Shen}, \citenamefont {Ronning}, \citenamefont {Lu}, \citenamefont {Lee}, \citenamefont {Ingle}, \citenamefont {Meevasana}, \citenamefont {Baumberger}, \citenamefont {Damascelli}, \citenamefont {Armitage}, \citenamefont {Miller} \emph {et~al.}}]{shen2004missing}%
  \BibitemOpen
  \bibfield  {author} {\bibinfo {author} {\bibfnamefont {K.}~\bibnamefont {Shen}}, \bibinfo {author} {\bibfnamefont {F.}~\bibnamefont {Ronning}}, \bibinfo {author} {\bibfnamefont {D.}~\bibnamefont {Lu}}, \bibinfo {author} {\bibfnamefont {W.}~\bibnamefont {Lee}}, \bibinfo {author} {\bibfnamefont {N.}~\bibnamefont {Ingle}}, \bibinfo {author} {\bibfnamefont {W.}~\bibnamefont {Meevasana}}, \bibinfo {author} {\bibfnamefont {F.}~\bibnamefont {Baumberger}}, \bibinfo {author} {\bibfnamefont {A.}~\bibnamefont {Damascelli}}, \bibinfo {author} {\bibfnamefont {N.}~\bibnamefont {Armitage}}, \bibinfo {author} {\bibfnamefont {L.}~\bibnamefont {Miller}}, \emph {et~al.},\ }\bibfield  {title} {\bibinfo {title} {Missing quasiparticles and the chemical potential puzzle in the doping evolution of the cuprate superconductors},\ }\href@noop {} {\bibfield  {journal} {\bibinfo  {journal} {Physical review letters}\ }\textbf {\bibinfo {volume} {93}},\ \bibinfo {pages} {267002} (\bibinfo {year} {2004})}\BibitemShut {NoStop}%
\bibitem [{\citenamefont {He}\ \emph {et~al.}(2018)\citenamefont {He}, \citenamefont {Hashimoto}, \citenamefont {Song}, \citenamefont {Chen}, \citenamefont {He}, \citenamefont {Vishik}, \citenamefont {Moritz}, \citenamefont {Lee}, \citenamefont {Nagaosa}, \citenamefont {Zaanen} \emph {et~al.}}]{he2018rapid}%
  \BibitemOpen
  \bibfield  {author} {\bibinfo {author} {\bibfnamefont {Y.}~\bibnamefont {He}}, \bibinfo {author} {\bibfnamefont {M.}~\bibnamefont {Hashimoto}}, \bibinfo {author} {\bibfnamefont {D.}~\bibnamefont {Song}}, \bibinfo {author} {\bibfnamefont {S.-D.}\ \bibnamefont {Chen}}, \bibinfo {author} {\bibfnamefont {J.}~\bibnamefont {He}}, \bibinfo {author} {\bibfnamefont {I.}~\bibnamefont {Vishik}}, \bibinfo {author} {\bibfnamefont {B.}~\bibnamefont {Moritz}}, \bibinfo {author} {\bibfnamefont {D.-H.}\ \bibnamefont {Lee}}, \bibinfo {author} {\bibfnamefont {N.}~\bibnamefont {Nagaosa}}, \bibinfo {author} {\bibfnamefont {J.}~\bibnamefont {Zaanen}}, \emph {et~al.},\ }\bibfield  {title} {\bibinfo {title} {Rapid change of superconductivity and electron-phonon coupling through critical doping in $\mathrm{Bi}$-2212},\ }\href@noop {} {\bibfield  {journal} {\bibinfo  {journal} {Science}\ }\textbf {\bibinfo {volume} {362}},\ \bibinfo {pages} {62} (\bibinfo {year} {2018})}\BibitemShut {NoStop}%
\bibitem [{\citenamefont {Lee}\ \emph {et~al.}(2014)\citenamefont {Lee}, \citenamefont {Schmitt}, \citenamefont {Moore}, \citenamefont {Johnston}, \citenamefont {Cui}, \citenamefont {Li}, \citenamefont {Yi}, \citenamefont {Liu}, \citenamefont {Hashimoto}, \citenamefont {Zhang} \emph {et~al.}}]{lee2014interfacial}%
  \BibitemOpen
  \bibfield  {author} {\bibinfo {author} {\bibfnamefont {J.}~\bibnamefont {Lee}}, \bibinfo {author} {\bibfnamefont {F.}~\bibnamefont {Schmitt}}, \bibinfo {author} {\bibfnamefont {R.}~\bibnamefont {Moore}}, \bibinfo {author} {\bibfnamefont {S.}~\bibnamefont {Johnston}}, \bibinfo {author} {\bibfnamefont {Y.-T.}\ \bibnamefont {Cui}}, \bibinfo {author} {\bibfnamefont {W.}~\bibnamefont {Li}}, \bibinfo {author} {\bibfnamefont {M.}~\bibnamefont {Yi}}, \bibinfo {author} {\bibfnamefont {Z.}~\bibnamefont {Liu}}, \bibinfo {author} {\bibfnamefont {M.}~\bibnamefont {Hashimoto}}, \bibinfo {author} {\bibfnamefont {Y.}~\bibnamefont {Zhang}}, \emph {et~al.},\ }\bibfield  {title} {\bibinfo {title} {Interfacial mode coupling as the origin of the enhancement of $\mathrm{T}_\mathrm{c}$ in fese films on $\mathrm{SrTiO}_3$},\ }\href@noop {} {\bibfield  {journal} {\bibinfo  {journal} {Nature}\ }\textbf {\bibinfo {volume} {515}},\ \bibinfo {pages} {245} (\bibinfo {year} {2014})}\BibitemShut {NoStop}%
\bibitem [{\citenamefont {Kresse}\ and\ \citenamefont {Furthmüller}(1996)}]{VASP0}%
  \BibitemOpen
  \bibfield  {author} {\bibinfo {author} {\bibfnamefont {G.}~\bibnamefont {Kresse}}\ and\ \bibinfo {author} {\bibfnamefont {J.}~\bibnamefont {Furthmüller}},\ }\bibfield  {title} {\bibinfo {title} {Efficiency of ab-initio total energy calculations for metals and semiconductors using a plane-wave basis set},\ }\href {https://doi.org/https://doi.org/10.1016/0927-0256(96)00008-0} {\bibfield  {journal} {\bibinfo  {journal} {Computational Materials Science}\ }\textbf {\bibinfo {volume} {6}},\ \bibinfo {pages} {15} (\bibinfo {year} {1996})}\BibitemShut {NoStop}%
\bibitem [{\citenamefont {Kresse}\ and\ \citenamefont {Furthm\"uller}(1996)}]{VASP1}%
  \BibitemOpen
  \bibfield  {author} {\bibinfo {author} {\bibfnamefont {G.}~\bibnamefont {Kresse}}\ and\ \bibinfo {author} {\bibfnamefont {J.}~\bibnamefont {Furthm\"uller}},\ }\bibfield  {title} {\bibinfo {title} {Efficient iterative schemes for ab initio total-energy calculations using a plane-wave basis set},\ }\href {https://doi.org/10.1103/PhysRevB.54.11169} {\bibfield  {journal} {\bibinfo  {journal} {Phys. Rev. B}\ }\textbf {\bibinfo {volume} {54}},\ \bibinfo {pages} {11169} (\bibinfo {year} {1996})}\BibitemShut {NoStop}%
\bibitem [{\citenamefont {Kresse}\ and\ \citenamefont {Hafner}(1993)}]{VASP2}%
  \BibitemOpen
  \bibfield  {author} {\bibinfo {author} {\bibfnamefont {G.}~\bibnamefont {Kresse}}\ and\ \bibinfo {author} {\bibfnamefont {J.}~\bibnamefont {Hafner}},\ }\bibfield  {title} {\bibinfo {title} {Ab initio molecular dynamics for liquid metals},\ }\href {https://doi.org/10.1103/PhysRevB.47.558} {\bibfield  {journal} {\bibinfo  {journal} {Phys. Rev. B}\ }\textbf {\bibinfo {volume} {47}},\ \bibinfo {pages} {558} (\bibinfo {year} {1993})}\BibitemShut {NoStop}%
\bibitem [{\citenamefont {Kresse}\ and\ \citenamefont {Hafner}(1994)}]{VASP3}%
  \BibitemOpen
  \bibfield  {author} {\bibinfo {author} {\bibfnamefont {G.}~\bibnamefont {Kresse}}\ and\ \bibinfo {author} {\bibfnamefont {J.}~\bibnamefont {Hafner}},\ }\bibfield  {title} {\bibinfo {title} {Ab initio molecular-dynamics simulation of the liquid-metal--amorphous-semiconductor transition in germanium},\ }\href {https://doi.org/10.1103/PhysRevB.49.14251} {\bibfield  {journal} {\bibinfo  {journal} {Phys. Rev. B}\ }\textbf {\bibinfo {volume} {49}},\ \bibinfo {pages} {14251} (\bibinfo {year} {1994})}\BibitemShut {NoStop}%
\bibitem [{\citenamefont {Kresse}\ and\ \citenamefont {Joubert}(1999)}]{PAW}%
  \BibitemOpen
  \bibfield  {author} {\bibinfo {author} {\bibfnamefont {G.}~\bibnamefont {Kresse}}\ and\ \bibinfo {author} {\bibfnamefont {D.}~\bibnamefont {Joubert}},\ }\bibfield  {title} {\bibinfo {title} {From ultrasoft pseudopotentials to the projector augmented-wave method},\ }\href {https://doi.org/10.1103/PhysRevB.59.1758} {\bibfield  {journal} {\bibinfo  {journal} {Phys. Rev. B}\ }\textbf {\bibinfo {volume} {59}},\ \bibinfo {pages} {1758} (\bibinfo {year} {1999})}\BibitemShut {NoStop}%
\bibitem [{\citenamefont {Perdew}\ \emph {et~al.}(1996)\citenamefont {Perdew}, \citenamefont {Burke},\ and\ \citenamefont {Ernzerhof}}]{PBE}%
  \BibitemOpen
  \bibfield  {author} {\bibinfo {author} {\bibfnamefont {J.~P.}\ \bibnamefont {Perdew}}, \bibinfo {author} {\bibfnamefont {K.}~\bibnamefont {Burke}},\ and\ \bibinfo {author} {\bibfnamefont {M.}~\bibnamefont {Ernzerhof}},\ }\bibfield  {title} {\bibinfo {title} {Generalized gradient approximation made simple},\ }\href {https://doi.org/10.1103/PhysRevLett.77.3865} {\bibfield  {journal} {\bibinfo  {journal} {Phys. Rev. Lett.}\ }\textbf {\bibinfo {volume} {77}},\ \bibinfo {pages} {3865} (\bibinfo {year} {1996})}\BibitemShut {NoStop}%
\bibitem [{\citenamefont {Wang}\ \emph {et~al.}(2019)\citenamefont {Wang}, \citenamefont {Jo}, \citenamefont {Kuthanazhi}, \citenamefont {Wu}, \citenamefont {McQueeney}, \citenamefont {Kaminski},\ and\ \citenamefont {Canfield}}]{LLWang_ECA-Weyl}%
  \BibitemOpen
  \bibfield  {author} {\bibinfo {author} {\bibfnamefont {L.-L.}\ \bibnamefont {Wang}}, \bibinfo {author} {\bibfnamefont {N.~H.}\ \bibnamefont {Jo}}, \bibinfo {author} {\bibfnamefont {B.}~\bibnamefont {Kuthanazhi}}, \bibinfo {author} {\bibfnamefont {Y.}~\bibnamefont {Wu}}, \bibinfo {author} {\bibfnamefont {R.~J.}\ \bibnamefont {McQueeney}}, \bibinfo {author} {\bibfnamefont {A.}~\bibnamefont {Kaminski}},\ and\ \bibinfo {author} {\bibfnamefont {P.~C.}\ \bibnamefont {Canfield}},\ }\bibfield  {title} {\bibinfo {title} {Single pair of $\mathrm{W}$eyl fermions in the half-metallic semimetal $\mathrm{EuC}{\mathrm{d}}_{2}\mathrm{A}{\mathrm{s}}_{2}$},\ }\href {https://doi.org/10.1103/PhysRevB.99.245147} {\bibfield  {journal} {\bibinfo  {journal} {Phys. Rev. B}\ }\textbf {\bibinfo {volume} {99}},\ \bibinfo {pages} {245147} (\bibinfo {year} {2019})}\BibitemShut {NoStop}%
\bibitem [{\citenamefont {Dudarev}\ \emph {et~al.}(1998)\citenamefont {Dudarev}, \citenamefont {Botton}, \citenamefont {Savrasov}, \citenamefont {Humphreys},\ and\ \citenamefont {Sutton}}]{LSDApU_Dudarev}%
  \BibitemOpen
  \bibfield  {author} {\bibinfo {author} {\bibfnamefont {S.~L.}\ \bibnamefont {Dudarev}}, \bibinfo {author} {\bibfnamefont {G.~A.}\ \bibnamefont {Botton}}, \bibinfo {author} {\bibfnamefont {S.~Y.}\ \bibnamefont {Savrasov}}, \bibinfo {author} {\bibfnamefont {C.~J.}\ \bibnamefont {Humphreys}},\ and\ \bibinfo {author} {\bibfnamefont {A.~P.}\ \bibnamefont {Sutton}},\ }\bibfield  {title} {\bibinfo {title} {Electron-energy-loss spectra and the structural stability of nickel oxide: An $\mathrm{LSDA}+\mathrm{U}$ study},\ }\href {https://doi.org/10.1103/PhysRevB.57.1505} {\bibfield  {journal} {\bibinfo  {journal} {Phys. Rev. B}\ }\textbf {\bibinfo {volume} {57}},\ \bibinfo {pages} {1505} (\bibinfo {year} {1998})}\BibitemShut {NoStop}%
\end{thebibliography}%


\begin{thebibliography}{18}%
\makeatletter
\providecommand \@ifxundefined [1]{%
 \@ifx{#1\undefined}
}%
\providecommand \@ifnum [1]{%
 \ifnum #1\expandafter \@firstoftwo
 \else \expandafter \@secondoftwo
 \fi
}%
\providecommand \@ifx [1]{%
 \ifx #1\expandafter \@firstoftwo
 \else \expandafter \@secondoftwo
 \fi
}%
\providecommand \natexlab [1]{#1}%
\providecommand \enquote  [1]{``#1''}%
\providecommand \bibnamefont  [1]{#1}%
\providecommand \bibfnamefont [1]{#1}%
\providecommand \citenamefont [1]{#1}%
\providecommand \href@noop [0]{\@secondoftwo}%
\providecommand \href [0]{\begingroup \@sanitize@url \@href}%
\providecommand \@href[1]{\@@startlink{#1}\@@href}%
\providecommand \@@href[1]{\endgroup#1\@@endlink}%
\providecommand \@sanitize@url [0]{\catcode `\\12\catcode `\$12\catcode `\&12\catcode `\#12\catcode `\^12\catcode `\_12\catcode `\%12\relax}%
\providecommand \@@startlink[1]{}%
\providecommand \@@endlink[0]{}%
\providecommand \url  [0]{\begingroup\@sanitize@url \@url }%
\providecommand \@url [1]{\endgroup\@href {#1}{\urlprefix }}%
\providecommand \urlprefix  [0]{URL }%
\providecommand \Eprint [0]{\href }%
\providecommand \doibase [0]{https://doi.org/}%
\providecommand \selectlanguage [0]{\@gobble}%
\providecommand \bibinfo  [0]{\@secondoftwo}%
\providecommand \bibfield  [0]{\@secondoftwo}%
\providecommand \translation [1]{[#1]}%
\providecommand \BibitemOpen [0]{}%
\providecommand \bibitemStop [0]{}%
\providecommand \bibitemNoStop [0]{.\EOS\space}%
\providecommand \EOS [0]{\spacefactor3000\relax}%
\providecommand \BibitemShut  [1]{\csname bibitem#1\endcsname}%
\let\auto@bib@innerbib\@empty
\bibitem [{\citenamefont {Ma}\ \emph {et~al.}(2019)\citenamefont {Ma}, \citenamefont {Nie}, \citenamefont {Yi}, \citenamefont {Jandke}, \citenamefont {Shang}, \citenamefont {Yao}, \citenamefont {Naamneh}, \citenamefont {Yan}, \citenamefont {Sun}, \citenamefont {Chikina} \emph {et~al.}}]{ma2019spin}%
  \BibitemOpen
  \bibfield  {author} {\bibinfo {author} {\bibfnamefont {J.-Z.}\ \bibnamefont {Ma}}, \bibinfo {author} {\bibfnamefont {S.}~\bibnamefont {Nie}}, \bibinfo {author} {\bibfnamefont {C.}~\bibnamefont {Yi}}, \bibinfo {author} {\bibfnamefont {J.}~\bibnamefont {Jandke}}, \bibinfo {author} {\bibfnamefont {T.}~\bibnamefont {Shang}}, \bibinfo {author} {\bibfnamefont {M.-Y.}\ \bibnamefont {Yao}}, \bibinfo {author} {\bibfnamefont {M.}~\bibnamefont {Naamneh}}, \bibinfo {author} {\bibfnamefont {L.}~\bibnamefont {Yan}}, \bibinfo {author} {\bibfnamefont {Y.}~\bibnamefont {Sun}}, \bibinfo {author} {\bibfnamefont {A.}~\bibnamefont {Chikina}}, \emph {et~al.},\ }\bibfield  {title} {\bibinfo {title} {Spin fluctuation induced weyl semimetal state in the paramagnetic phase of $\mathrm{EuCd}_2\mathrm{As}_2$},\ }\href@noop {} {\bibfield  {journal} {\bibinfo  {journal} {Science advances}\ }\textbf {\bibinfo {volume} {5}},\ \bibinfo {pages} {eaaw4718} (\bibinfo {year} {2019})}\BibitemShut {NoStop}%
\bibitem [{\citenamefont {Ma}\ \emph {et~al.}(2020)\citenamefont {Ma}, \citenamefont {Wang}, \citenamefont {Nie}, \citenamefont {Yi}, \citenamefont {Xu}, \citenamefont {Li}, \citenamefont {Jandke}, \citenamefont {Wulfhekel}, \citenamefont {Huang}, \citenamefont {West} \emph {et~al.}}]{ma2020emergence}%
  \BibitemOpen
  \bibfield  {author} {\bibinfo {author} {\bibfnamefont {J.}~\bibnamefont {Ma}}, \bibinfo {author} {\bibfnamefont {H.}~\bibnamefont {Wang}}, \bibinfo {author} {\bibfnamefont {S.}~\bibnamefont {Nie}}, \bibinfo {author} {\bibfnamefont {C.}~\bibnamefont {Yi}}, \bibinfo {author} {\bibfnamefont {Y.}~\bibnamefont {Xu}}, \bibinfo {author} {\bibfnamefont {H.}~\bibnamefont {Li}}, \bibinfo {author} {\bibfnamefont {J.}~\bibnamefont {Jandke}}, \bibinfo {author} {\bibfnamefont {W.}~\bibnamefont {Wulfhekel}}, \bibinfo {author} {\bibfnamefont {Y.}~\bibnamefont {Huang}}, \bibinfo {author} {\bibfnamefont {D.}~\bibnamefont {West}}, \emph {et~al.},\ }\bibfield  {title} {\bibinfo {title} {Emergence of nontrivial low-energy dirac fermions in antiferromagnetic $\mathrm{EuCd}_2\mathrm{As}_2$},\ }\href@noop {} {\bibfield  {journal} {\bibinfo  {journal} {Advanced Materials}\ }\textbf {\bibinfo {volume} {32}},\ \bibinfo {pages} {1907565} (\bibinfo {year} {2020})}\BibitemShut {NoStop}%
\bibitem [{\citenamefont {Cuono}\ \emph {et~al.}(2023)\citenamefont {Cuono}, \citenamefont {Sattigeri}, \citenamefont {Autieri},\ and\ \citenamefont {Dietl}}]{ECP_trivial_Cuono_PRB2023}%
  \BibitemOpen
  \bibfield  {author} {\bibinfo {author} {\bibfnamefont {G.}~\bibnamefont {Cuono}}, \bibinfo {author} {\bibfnamefont {R.~M.}\ \bibnamefont {Sattigeri}}, \bibinfo {author} {\bibfnamefont {C.}~\bibnamefont {Autieri}},\ and\ \bibinfo {author} {\bibfnamefont {T.}~\bibnamefont {Dietl}},\ }\bibfield  {title} {\bibinfo {title} {Ab initio overestimation of the topological region in eu-based compounds},\ }\href {https://doi.org/10.1103/PhysRevB.108.075150} {\bibfield  {journal} {\bibinfo  {journal} {Phys. Rev. B}\ }\textbf {\bibinfo {volume} {108}},\ \bibinfo {pages} {075150} (\bibinfo {year} {2023})}\BibitemShut {NoStop}%
\bibitem [{\citenamefont {Wang}\ \emph {et~al.}(2021)\citenamefont {Wang}, \citenamefont {Rogers}, \citenamefont {Yao}, \citenamefont {Nichols}, \citenamefont {Atay}, \citenamefont {Xu}, \citenamefont {Franklin}, \citenamefont {Sochnikov}, \citenamefont {Ryan}, \citenamefont {Haskel} \emph {et~al.}}]{wang2021colossal}%
  \BibitemOpen
  \bibfield  {author} {\bibinfo {author} {\bibfnamefont {Z.-C.}\ \bibnamefont {Wang}}, \bibinfo {author} {\bibfnamefont {J.~D.}\ \bibnamefont {Rogers}}, \bibinfo {author} {\bibfnamefont {X.}~\bibnamefont {Yao}}, \bibinfo {author} {\bibfnamefont {R.}~\bibnamefont {Nichols}}, \bibinfo {author} {\bibfnamefont {K.}~\bibnamefont {Atay}}, \bibinfo {author} {\bibfnamefont {B.}~\bibnamefont {Xu}}, \bibinfo {author} {\bibfnamefont {J.}~\bibnamefont {Franklin}}, \bibinfo {author} {\bibfnamefont {I.}~\bibnamefont {Sochnikov}}, \bibinfo {author} {\bibfnamefont {P.~J.}\ \bibnamefont {Ryan}}, \bibinfo {author} {\bibfnamefont {D.}~\bibnamefont {Haskel}}, \emph {et~al.},\ }\bibfield  {title} {\bibinfo {title} {Colossal magnetoresistance without mixed valence in a layered phosphide crystal},\ }\href@noop {} {\bibfield  {journal} {\bibinfo  {journal} {Advanced Materials}\ }\textbf {\bibinfo {volume} {33}},\ \bibinfo {pages} {2005755} (\bibinfo {year} {2021})}\BibitemShut {NoStop}%
\bibitem [{\citenamefont {Wang}\ \emph {et~al.}(2022)\citenamefont {Wang}, \citenamefont {Been}, \citenamefont {Gaudet}, \citenamefont {Alqasseri}, \citenamefont {Fruhling}, \citenamefont {Yao}, \citenamefont {Stuhr}, \citenamefont {Zhu}, \citenamefont {Ren}, \citenamefont {Cui} \emph {et~al.}}]{wang2022anisotropy}%
  \BibitemOpen
  \bibfield  {author} {\bibinfo {author} {\bibfnamefont {Z.-C.}\ \bibnamefont {Wang}}, \bibinfo {author} {\bibfnamefont {E.}~\bibnamefont {Been}}, \bibinfo {author} {\bibfnamefont {J.}~\bibnamefont {Gaudet}}, \bibinfo {author} {\bibfnamefont {G.~M.~A.}\ \bibnamefont {Alqasseri}}, \bibinfo {author} {\bibfnamefont {K.}~\bibnamefont {Fruhling}}, \bibinfo {author} {\bibfnamefont {X.}~\bibnamefont {Yao}}, \bibinfo {author} {\bibfnamefont {U.}~\bibnamefont {Stuhr}}, \bibinfo {author} {\bibfnamefont {Q.}~\bibnamefont {Zhu}}, \bibinfo {author} {\bibfnamefont {Z.}~\bibnamefont {Ren}}, \bibinfo {author} {\bibfnamefont {Y.}~\bibnamefont {Cui}}, \emph {et~al.},\ }\bibfield  {title} {\bibinfo {title} {Anisotropy of the magnetic and transport properties of $\mathrm{EuZn}_2\mathrm{As}_2$},\ }\href@noop {} {\bibfield  {journal} {\bibinfo  {journal} {Physical Review B}\ }\textbf {\bibinfo {volume} {105}},\ \bibinfo {pages} {165122} (\bibinfo {year} {2022})}\BibitemShut {NoStop}%
\bibitem [{\citenamefont {Zhang}\ \emph {et~al.}(2023)\citenamefont {Zhang}, \citenamefont {Du}, \citenamefont {Zheng}, \citenamefont {Luo}, \citenamefont {Wu}, \citenamefont {Zheng}, \citenamefont {Cui}, \citenamefont {Sun}, \citenamefont {Liu}, \citenamefont {Shen} \emph {et~al.}}]{zhang2023electronic}%
  \BibitemOpen
  \bibfield  {author} {\bibinfo {author} {\bibfnamefont {H.}~\bibnamefont {Zhang}}, \bibinfo {author} {\bibfnamefont {F.}~\bibnamefont {Du}}, \bibinfo {author} {\bibfnamefont {X.}~\bibnamefont {Zheng}}, \bibinfo {author} {\bibfnamefont {S.}~\bibnamefont {Luo}}, \bibinfo {author} {\bibfnamefont {Y.}~\bibnamefont {Wu}}, \bibinfo {author} {\bibfnamefont {H.}~\bibnamefont {Zheng}}, \bibinfo {author} {\bibfnamefont {S.}~\bibnamefont {Cui}}, \bibinfo {author} {\bibfnamefont {Z.}~\bibnamefont {Sun}}, \bibinfo {author} {\bibfnamefont {Z.}~\bibnamefont {Liu}}, \bibinfo {author} {\bibfnamefont {D.}~\bibnamefont {Shen}}, \emph {et~al.},\ }\bibfield  {title} {\bibinfo {title} {Electronic band reconstruction across the insulator-metal transition in colossal magnetoresistive $\mathrm{EuCd}_2\mathrm{P}_2$},\ }\href@noop {} {\bibfield  {journal} {\bibinfo  {journal} {arXiv preprint arXiv:2308.16844}\ } (\bibinfo {year} {2023})}\BibitemShut {NoStop}%
\bibitem [{\citenamefont {Sunko}\ \emph {et~al.}(2023)\citenamefont {Sunko}, \citenamefont {Sun}, \citenamefont {Vranas}, \citenamefont {Homes}, \citenamefont {Lee}, \citenamefont {Donoway}, \citenamefont {Wang}, \citenamefont {Balguri}, \citenamefont {Mahendru}, \citenamefont {Ruiz}, \citenamefont {Gunn}, \citenamefont {Basak}, \citenamefont {Blanco-Canosa}, \citenamefont {Schierle}, \citenamefont {Weschke}, \citenamefont {Tafti}, \citenamefont {Frano},\ and\ \citenamefont {Orenstein}}]{vSunko2023spin}%
  \BibitemOpen
  \bibfield  {author} {\bibinfo {author} {\bibfnamefont {V.}~\bibnamefont {Sunko}}, \bibinfo {author} {\bibfnamefont {Y.}~\bibnamefont {Sun}}, \bibinfo {author} {\bibfnamefont {M.}~\bibnamefont {Vranas}}, \bibinfo {author} {\bibfnamefont {C.~C.}\ \bibnamefont {Homes}}, \bibinfo {author} {\bibfnamefont {C.}~\bibnamefont {Lee}}, \bibinfo {author} {\bibfnamefont {E.}~\bibnamefont {Donoway}}, \bibinfo {author} {\bibfnamefont {Z.-C.}\ \bibnamefont {Wang}}, \bibinfo {author} {\bibfnamefont {S.}~\bibnamefont {Balguri}}, \bibinfo {author} {\bibfnamefont {M.~B.}\ \bibnamefont {Mahendru}}, \bibinfo {author} {\bibfnamefont {A.}~\bibnamefont {Ruiz}}, \bibinfo {author} {\bibfnamefont {B.}~\bibnamefont {Gunn}}, \bibinfo {author} {\bibfnamefont {R.}~\bibnamefont {Basak}}, \bibinfo {author} {\bibfnamefont {S.}~\bibnamefont {Blanco-Canosa}}, \bibinfo {author} {\bibfnamefont {E.}~\bibnamefont {Schierle}}, \bibinfo {author} {\bibfnamefont {E.}~\bibnamefont {Weschke}}, \bibinfo {author} {\bibfnamefont {F.}~\bibnamefont {Tafti}},
  \bibinfo {author} {\bibfnamefont {A.}~\bibnamefont {Frano}},\ and\ \bibinfo {author} {\bibfnamefont {J.}~\bibnamefont {Orenstein}},\ }\bibfield  {title} {\bibinfo {title} {Spin-carrier coupling induced ferromagnetism and giant resistivity peak in ${\mathrm{eucd}}_{2}{\mathrm{p}}_{2}$},\ }\href {https://doi.org/10.1103/PhysRevB.107.144404} {\bibfield  {journal} {\bibinfo  {journal} {Phys. Rev. B}\ }\textbf {\bibinfo {volume} {107}},\ \bibinfo {pages} {144404} (\bibinfo {year} {2023})}\BibitemShut {NoStop}%
\bibitem [{\citenamefont {Damascelli}\ \emph {et~al.}(2003)\citenamefont {Damascelli}, \citenamefont {Hussain},\ and\ \citenamefont {Shen}}]{damascelli2003angle}%
  \BibitemOpen
  \bibfield  {author} {\bibinfo {author} {\bibfnamefont {A.}~\bibnamefont {Damascelli}}, \bibinfo {author} {\bibfnamefont {Z.}~\bibnamefont {Hussain}},\ and\ \bibinfo {author} {\bibfnamefont {Z.-X.}\ \bibnamefont {Shen}},\ }\bibfield  {title} {\bibinfo {title} {Angle-resolved photoemission studies of the cuprate superconductors},\ }\href@noop {} {\bibfield  {journal} {\bibinfo  {journal} {Reviews of modern physics}\ }\textbf {\bibinfo {volume} {75}},\ \bibinfo {pages} {473} (\bibinfo {year} {2003})}\BibitemShut {NoStop}%
\bibitem [{\citenamefont {Sobota}\ \emph {et~al.}(2021)\citenamefont {Sobota}, \citenamefont {He},\ and\ \citenamefont {Shen}}]{sobota2021angle}%
  \BibitemOpen
  \bibfield  {author} {\bibinfo {author} {\bibfnamefont {J.~A.}\ \bibnamefont {Sobota}}, \bibinfo {author} {\bibfnamefont {Y.}~\bibnamefont {He}},\ and\ \bibinfo {author} {\bibfnamefont {Z.-X.}\ \bibnamefont {Shen}},\ }\bibfield  {title} {\bibinfo {title} {Angle-resolved photoemission studies of quantum materials},\ }\href@noop {} {\bibfield  {journal} {\bibinfo  {journal} {Reviews of Modern Physics}\ }\textbf {\bibinfo {volume} {93}},\ \bibinfo {pages} {025006} (\bibinfo {year} {2021})}\BibitemShut {NoStop}%
\bibitem [{\citenamefont {Yeh}\ and\ \citenamefont {Lindau}(1985)}]{yeh1985atomic}%
  \BibitemOpen
  \bibfield  {author} {\bibinfo {author} {\bibfnamefont {J.}~\bibnamefont {Yeh}}\ and\ \bibinfo {author} {\bibfnamefont {I.}~\bibnamefont {Lindau}},\ }\bibfield  {title} {\bibinfo {title} {Atomic subshell photoionization cross sections and asymmetry parameters: $1\leq z\leq 103$},\ }\href@noop {} {\bibfield  {journal} {\bibinfo  {journal} {Atomic data and nuclear data tables}\ }\textbf {\bibinfo {volume} {32}},\ \bibinfo {pages} {1} (\bibinfo {year} {1985})}\BibitemShut {NoStop}%
\bibitem [{\citenamefont {Moser}(2017)}]{moser2017experimentalist}%
  \BibitemOpen
  \bibfield  {author} {\bibinfo {author} {\bibfnamefont {S.}~\bibnamefont {Moser}},\ }\bibfield  {title} {\bibinfo {title} {An experimentalist's guide to the matrix element in angle resolved photoemission},\ }\href@noop {} {\bibfield  {journal} {\bibinfo  {journal} {Journal of Electron Spectroscopy and Related Phenomena}\ }\textbf {\bibinfo {volume} {214}},\ \bibinfo {pages} {29} (\bibinfo {year} {2017})}\BibitemShut {NoStop}%
\bibitem [{\citenamefont {Wang}\ \emph {et~al.}(2012)\citenamefont {Wang}, \citenamefont {Richard}, \citenamefont {Huang}, \citenamefont {Miao}, \citenamefont {Cevey}, \citenamefont {Xu}, \citenamefont {Sun}, \citenamefont {Qian}, \citenamefont {Xu}, \citenamefont {Shi} \emph {et~al.}}]{wang2012orbital}%
  \BibitemOpen
  \bibfield  {author} {\bibinfo {author} {\bibfnamefont {X.-P.}\ \bibnamefont {Wang}}, \bibinfo {author} {\bibfnamefont {P.}~\bibnamefont {Richard}}, \bibinfo {author} {\bibfnamefont {Y.-B.}\ \bibnamefont {Huang}}, \bibinfo {author} {\bibfnamefont {H.}~\bibnamefont {Miao}}, \bibinfo {author} {\bibfnamefont {L.}~\bibnamefont {Cevey}}, \bibinfo {author} {\bibfnamefont {N.}~\bibnamefont {Xu}}, \bibinfo {author} {\bibfnamefont {Y.-J.}\ \bibnamefont {Sun}}, \bibinfo {author} {\bibfnamefont {T.}~\bibnamefont {Qian}}, \bibinfo {author} {\bibfnamefont {Y.-M.}\ \bibnamefont {Xu}}, \bibinfo {author} {\bibfnamefont {M.}~\bibnamefont {Shi}}, \emph {et~al.},\ }\bibfield  {title} {\bibinfo {title} {Orbital characters determined from fermi surface intensity patterns using angle-resolved photoemission spectroscopy},\ }\href@noop {} {\bibfield  {journal} {\bibinfo  {journal} {Physical Review B}\ }\textbf {\bibinfo {volume} {85}},\ \bibinfo {pages} {214518} (\bibinfo {year} {2012})}\BibitemShut {NoStop}%
\bibitem [{\citenamefont {Day}\ \emph {et~al.}(2019)\citenamefont {Day}, \citenamefont {Zwartsenberg}, \citenamefont {Elfimov},\ and\ \citenamefont {Damascelli}}]{day2019computational}%
  \BibitemOpen
  \bibfield  {author} {\bibinfo {author} {\bibfnamefont {R.~P.}\ \bibnamefont {Day}}, \bibinfo {author} {\bibfnamefont {B.}~\bibnamefont {Zwartsenberg}}, \bibinfo {author} {\bibfnamefont {I.~S.}\ \bibnamefont {Elfimov}},\ and\ \bibinfo {author} {\bibfnamefont {A.}~\bibnamefont {Damascelli}},\ }\bibfield  {title} {\bibinfo {title} {Computational framework chinook for angle-resolved photoemission spectroscopy},\ }\href@noop {} {\bibfield  {journal} {\bibinfo  {journal} {npj Quantum Materials}\ }\textbf {\bibinfo {volume} {4}},\ \bibinfo {pages} {54} (\bibinfo {year} {2019})}\BibitemShut {NoStop}%
\bibitem [{\citenamefont {Goldberg}\ \emph {et~al.}(1981)\citenamefont {Goldberg}, \citenamefont {Fadley},\ and\ \citenamefont {Kono}}]{goldberg1981photoionization}%
  \BibitemOpen
  \bibfield  {author} {\bibinfo {author} {\bibfnamefont {S.}~\bibnamefont {Goldberg}}, \bibinfo {author} {\bibfnamefont {C.}~\bibnamefont {Fadley}},\ and\ \bibinfo {author} {\bibfnamefont {S.}~\bibnamefont {Kono}},\ }\bibfield  {title} {\bibinfo {title} {Photoionization cross-sections for atomic orbitals with random and fixed spatial orientation},\ }\href@noop {} {\bibfield  {journal} {\bibinfo  {journal} {Journal of electron spectroscopy and related phenomena}\ }\textbf {\bibinfo {volume} {21}},\ \bibinfo {pages} {285} (\bibinfo {year} {1981})}\BibitemShut {NoStop}%
\bibitem [{\citenamefont {Li}\ \emph {et~al.}(2023)\citenamefont {Li}, \citenamefont {Chen}, \citenamefont {Garcia-Diez}, \citenamefont {Iraola}, \citenamefont {Pfau}, \citenamefont {Zhu}, \citenamefont {Mao}, \citenamefont {Chen}, \citenamefont {Yi}, \citenamefont {Dai} \emph {et~al.}}]{li2023spectroscopic}%
  \BibitemOpen
  \bibfield  {author} {\bibinfo {author} {\bibfnamefont {Y.-F.}\ \bibnamefont {Li}}, \bibinfo {author} {\bibfnamefont {S.-D.}\ \bibnamefont {Chen}}, \bibinfo {author} {\bibfnamefont {M.}~\bibnamefont {Garcia-Diez}}, \bibinfo {author} {\bibfnamefont {M.}~\bibnamefont {Iraola}}, \bibinfo {author} {\bibfnamefont {H.}~\bibnamefont {Pfau}}, \bibinfo {author} {\bibfnamefont {Y.-L.}\ \bibnamefont {Zhu}}, \bibinfo {author} {\bibfnamefont {Z.-Q.}\ \bibnamefont {Mao}}, \bibinfo {author} {\bibfnamefont {T.}~\bibnamefont {Chen}}, \bibinfo {author} {\bibfnamefont {M.}~\bibnamefont {Yi}}, \bibinfo {author} {\bibfnamefont {P.-C.}\ \bibnamefont {Dai}}, \emph {et~al.},\ }\bibfield  {title} {\bibinfo {title} {Spectroscopic evidence for topological band structure in $\mathrm{FeTe}_{0.55}\mathrm{Se}_{0.45}$},\ }\href@noop {} {\bibfield  {journal} {\bibinfo  {journal} {arXiv preprint arXiv:2307.03861}\ } (\bibinfo {year} {2023})}\BibitemShut {NoStop}%
\bibitem [{\citenamefont {Suga}\ \emph {et~al.}(2021)\citenamefont {Suga}, \citenamefont {Sekiyama},\ and\ \citenamefont {Tusche}}]{suga2021photoelectron}%
  \BibitemOpen
  \bibfield  {author} {\bibinfo {author} {\bibfnamefont {S.}~\bibnamefont {Suga}}, \bibinfo {author} {\bibfnamefont {A.}~\bibnamefont {Sekiyama}},\ and\ \bibinfo {author} {\bibfnamefont {C.}~\bibnamefont {Tusche}},\ }\href@noop {} {\emph {\bibinfo {title} {Photoelectron spectroscopy}}}\ (\bibinfo  {publisher} {Springer},\ \bibinfo {year} {2021})\BibitemShut {NoStop}%
\bibitem [{\citenamefont {Homes}\ \emph {et~al.}(2023)\citenamefont {Homes}, \citenamefont {Wang}, \citenamefont {Fruhling},\ and\ \citenamefont {Tafti}}]{homes2023optical}%
  \BibitemOpen
  \bibfield  {author} {\bibinfo {author} {\bibfnamefont {C.~C.}\ \bibnamefont {Homes}}, \bibinfo {author} {\bibfnamefont {Z.-C.}\ \bibnamefont {Wang}}, \bibinfo {author} {\bibfnamefont {K.}~\bibnamefont {Fruhling}},\ and\ \bibinfo {author} {\bibfnamefont {F.}~\bibnamefont {Tafti}},\ }\bibfield  {title} {\bibinfo {title} {Optical properties and carrier localization in the layered phosphide $\mathrm{EuCd}_2\mathrm{P}_2$},\ }\href@noop {} {\bibfield  {journal} {\bibinfo  {journal} {Physical Review B}\ }\textbf {\bibinfo {volume} {107}},\ \bibinfo {pages} {045106} (\bibinfo {year} {2023})}\BibitemShut {NoStop}%
\bibitem [{\citenamefont {Shen}\ \emph {et~al.}(2004)\citenamefont {Shen}, \citenamefont {Ronning}, \citenamefont {Lu}, \citenamefont {Lee}, \citenamefont {Ingle}, \citenamefont {Meevasana}, \citenamefont {Baumberger}, \citenamefont {Damascelli}, \citenamefont {Armitage}, \citenamefont {Miller} \emph {et~al.}}]{shen2004missing}%
  \BibitemOpen
  \bibfield  {author} {\bibinfo {author} {\bibfnamefont {K.}~\bibnamefont {Shen}}, \bibinfo {author} {\bibfnamefont {F.}~\bibnamefont {Ronning}}, \bibinfo {author} {\bibfnamefont {D.}~\bibnamefont {Lu}}, \bibinfo {author} {\bibfnamefont {W.}~\bibnamefont {Lee}}, \bibinfo {author} {\bibfnamefont {N.}~\bibnamefont {Ingle}}, \bibinfo {author} {\bibfnamefont {W.}~\bibnamefont {Meevasana}}, \bibinfo {author} {\bibfnamefont {F.}~\bibnamefont {Baumberger}}, \bibinfo {author} {\bibfnamefont {A.}~\bibnamefont {Damascelli}}, \bibinfo {author} {\bibfnamefont {N.}~\bibnamefont {Armitage}}, \bibinfo {author} {\bibfnamefont {L.}~\bibnamefont {Miller}}, \emph {et~al.},\ }\bibfield  {title} {\bibinfo {title} {Missing quasiparticles and the chemical potential puzzle in the doping evolution of the cuprate superconductors},\ }\href@noop {} {\bibfield  {journal} {\bibinfo  {journal} {Physical review letters}\ }\textbf {\bibinfo {volume} {93}},\ \bibinfo {pages} {267002} (\bibinfo {year} {2004})}\BibitemShut {NoStop}%
\end{thebibliography}%


\end{document}


\preprint{APS/123-QED}
\title{Supplementary information for colossal magnetoresistance from spin-polarized polarons in an Ising system}

\author{Ying-Fei Li}
\thanks{These authors contributed equally.}
\affiliation{Stanford Institute for Materials and Energy Sciences, SLAC National Accelerator Laboratory, Menlo Park, 94025, CA, USA}
\affiliation{Department of Applied Physics and Physics, Stanford University, Stanford, 94305, CA, USA}
\affiliation{Geballe Laboratory for Advanced Materials, Stanford University, Stanford, 94305, CA, USA}

\author{Emily M. Been}
\thanks{These authors contributed equally.}
\affiliation{Stanford Institute for Materials and Energy Sciences, SLAC National Accelerator Laboratory, Menlo Park, 94025, CA, USA}
\affiliation{Department of Applied Physics and Physics, Stanford University, Stanford, 94305, CA, USA}
\affiliation{Geballe Laboratory for Advanced Materials, Stanford University, Stanford, 94305, CA, USA}

\author{Sudhaman Balguri}
\affiliation{Departments of Physics, Boston College, 140 Commonwealth Avenue, Chestnut Hill, 02467, MA, USA}

\author{Chun-Jing Jia}
\affiliation{Geballe Laboratory for Advanced Materials, Stanford University, Stanford, 94305, CA, USA}
\affiliation{Department of Physics, University of Florida, Gainesville 32611, FL}

\author{Mira B. Mahenderu}
\affiliation{Departments of Physics, Boston College, 140 Commonwealth Avenue, Chestnut Hill, 02467, MA, USA}

\author{Zhi-Cheng Wang}
\affiliation{Departments of Physics, Boston College, 140 Commonwealth Avenue, Chestnut Hill, 02467, MA, USA}

\author{Yi Cui}
\affiliation{Stanford Institute for Materials and Energy Sciences, SLAC National Accelerator Laboratory, Menlo Park, 94025, CA, USA}
\affiliation{Department of Applied Physics and Physics, Stanford University, Stanford, 94305, CA, USA}
\affiliation{Geballe Laboratory for Advanced Materials, Stanford University, Stanford, 94305, CA, USA}

\author{Su-Di Chen}
\affiliation{Stanford Institute for Materials and Energy Sciences, SLAC National Accelerator Laboratory, Menlo Park, 94025, CA, USA}
\affiliation{Department of Applied Physics and Physics, Stanford University, Stanford, 94305, CA, USA}
\affiliation{Geballe Laboratory for Advanced Materials, Stanford University, Stanford, 94305, CA, USA}
\affiliation{Department of Physics, University of California, Berkeley, California 94720, USA}

\author{Makoto Hashimoto}
\affiliation{Stanford Synchrotron Radiation Lightsource, SLAC National Accelerator Laboratory, Menlo Park, 94025, CA, USA}

\author{Dong-Hui Lu}
\affiliation{Stanford Synchrotron Radiation Lightsource, SLAC National Accelerator Laboratory, Menlo Park, 94025, CA, USA}

\author{Brian Moritz}
\affiliation{Stanford Institute for Materials and Energy Sciences, SLAC National Accelerator Laboratory, Menlo Park, 94025, CA, USA}
\affiliation{Department of Applied Physics and Physics, Stanford University, Stanford, 94305, CA, USA}
\affiliation{Geballe Laboratory for Advanced Materials, Stanford University, Stanford, 94305, CA, USA}

\author{Jan Zaanen}
\affiliation{Institute Lorentz for Theoretical Physics, Leiden University, 2300 RA Leiden, Netherlands}

\author{Fazel Tafti}
\email{fazel.tafti@bc.edu}
\affiliation{Departments of Physics, Boston College, 140 Commonwealth Avenue, Chestnut Hill, 02467, MA, USA}

\author{Thomas P. Devereaux}
\email{tpd@stanford.edu}
\affiliation{Stanford Institute for Materials and Energy Sciences, SLAC National Accelerator Laboratory, Menlo Park, 94025, CA, USA}
\affiliation{Department of Materials Science and Engineering, Stanford University, Stanford, 94305, CA, USA}

\author{Zhi-Xun Shen}
\email{zxshen@stanford.edu}
\affiliation{Stanford Institute for Materials and Energy Sciences, SLAC National Accelerator Laboratory, Menlo Park, 94025, CA, USA}
\affiliation{Department of Applied Physics and Physics, Stanford University, Stanford, 94305, CA, USA}
\affiliation{Geballe Laboratory for Advanced Materials, Stanford University, Stanford, 94305, CA, USA}

\date{\today}

\maketitle

\section{Density Functional Theory calculations}
\subsection{Electronic structure investigation}
\begin{figure*}[h]
    \centering
    \includegraphics[width=\textwidth,trim={0 120 0 0},clip]{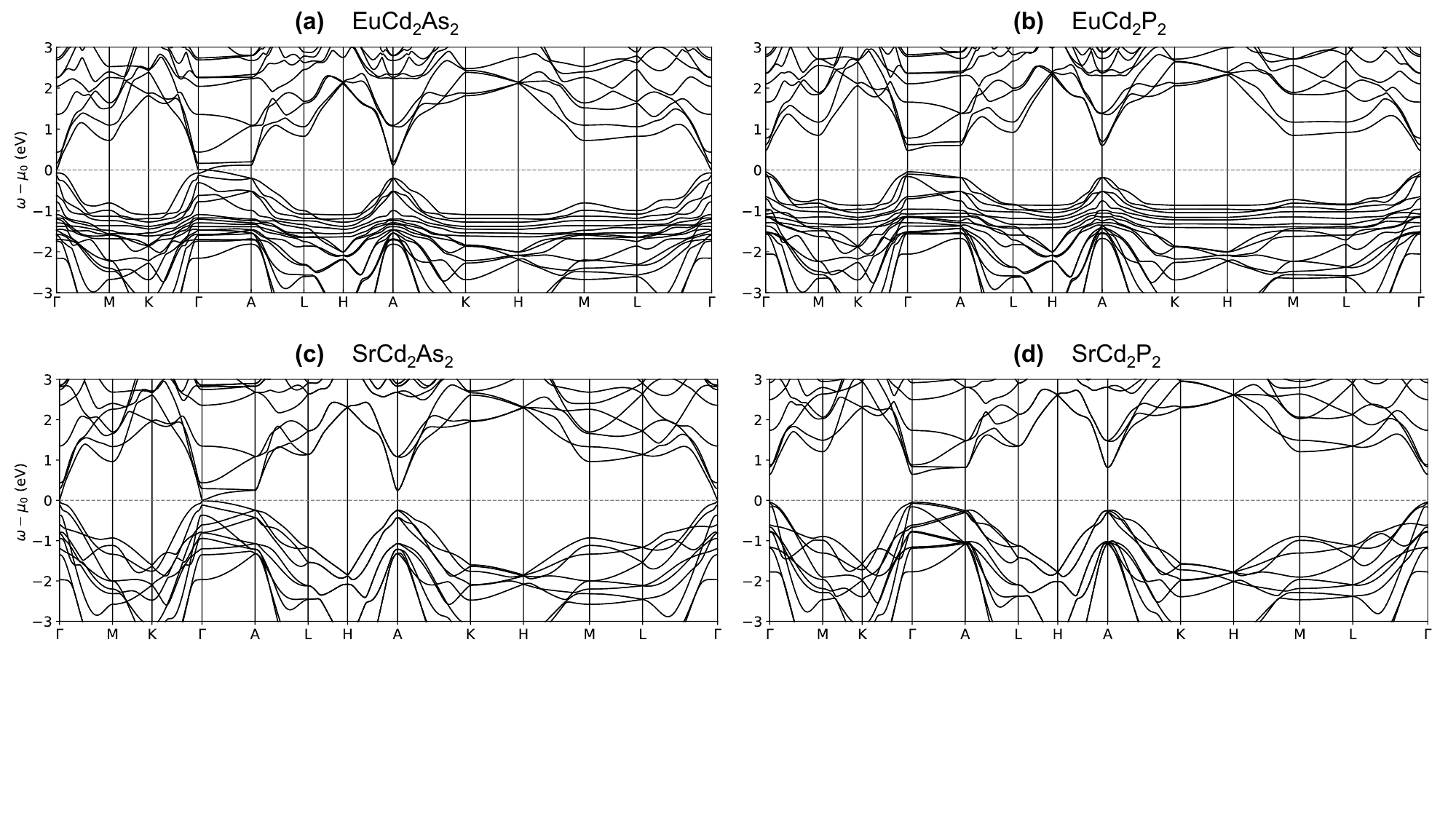}
    \caption{\textbf{The DFT-calculated band structures of the (a) EuCd\textsubscript{2}As\textsubscript{2}, (b) EuCd\textsubscript{2}P\textsubscript{2}, and their non-magnetic counterparts (c) SrCd\textsubscript{2}As\textsubscript{2} and (d) SrCd\textsubscript{2}P\textsubscript{2}.}}
    \label{fig:plain_bands}
\end{figure*}

We delve into the electronic structure of EuCd\textsubscript{2}P\textsubscript{2} and EuCd\textsubscript{2}As\textsubscript{2} through detailed density functional theory (DFT) calculations (Fig. \ref{fig:plain_bands}(a, b)). Compared to their non-magnetic counterparts, SrCd\textsubscript{2}P\textsubscript{2} and SrCd\textsubscript{2}As\textsubscript{2} (Fig. \ref{fig:plain_bands}(c, d)), the dispersive bands in EuCd\textsubscript{2}P\textsubscript{2} and EuCd\textsubscript{2}As\textsubscript{2} exhibit remarkable resemblance, both with additional flat bands within $-1.5$ to $-1$ eV from the binding energy range. Notably, EuCd\textsubscript{2}P\textsubscript{2} lacks any band crossing near the chemical potential ($\mu$) across the entire Brillouin zone, contrasting with the nearly band touching point between Cd $s$ and As $p$ bands in EuCd\textsubscript{2}As\textsubscript{2}. Consequently, despite the proposed proximity to nontrivial topology in EuCd\textsubscript{2}As\textsubscript{2}\cite{ma2019spin, ma2020emergence}, the robust finite gap in EuCd\textsubscript{2}P\textsubscript{2} precludes any contributions to the colossal magnetoresistance (CMR) from low energy topological states \cite{ECP_trivial_Cuono_PRB2023}.

\begin{figure*}[h]
    \centering
    \includegraphics[width=\textwidth,trim={0 110 0 0},clip]{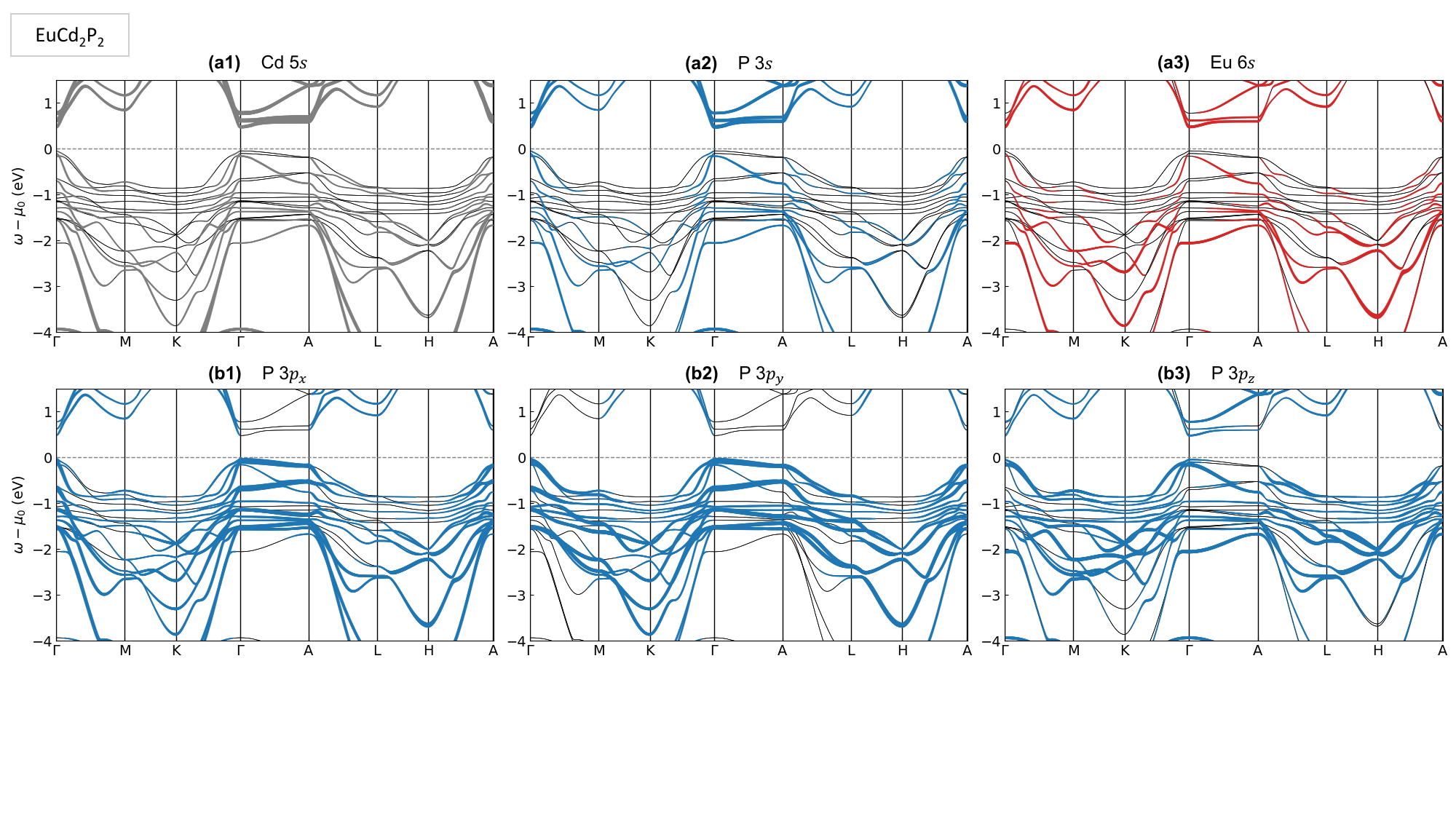}
    \caption{\textbf{The orbital projected band structure for EuCd\textsubscript{2}P\textsubscript{2}.} The P 3$p$ orbitals dominate the low-energy valence band and the $s$ orbitals from Eu, Cd, and P dominate the conduction band near $\mu$.}
    \label{fig:sp_in_ECP}
\end{figure*}

\begin{figure*}[h]
    \centering
    \includegraphics[width=\textwidth,trim={0 110 0 0},clip]{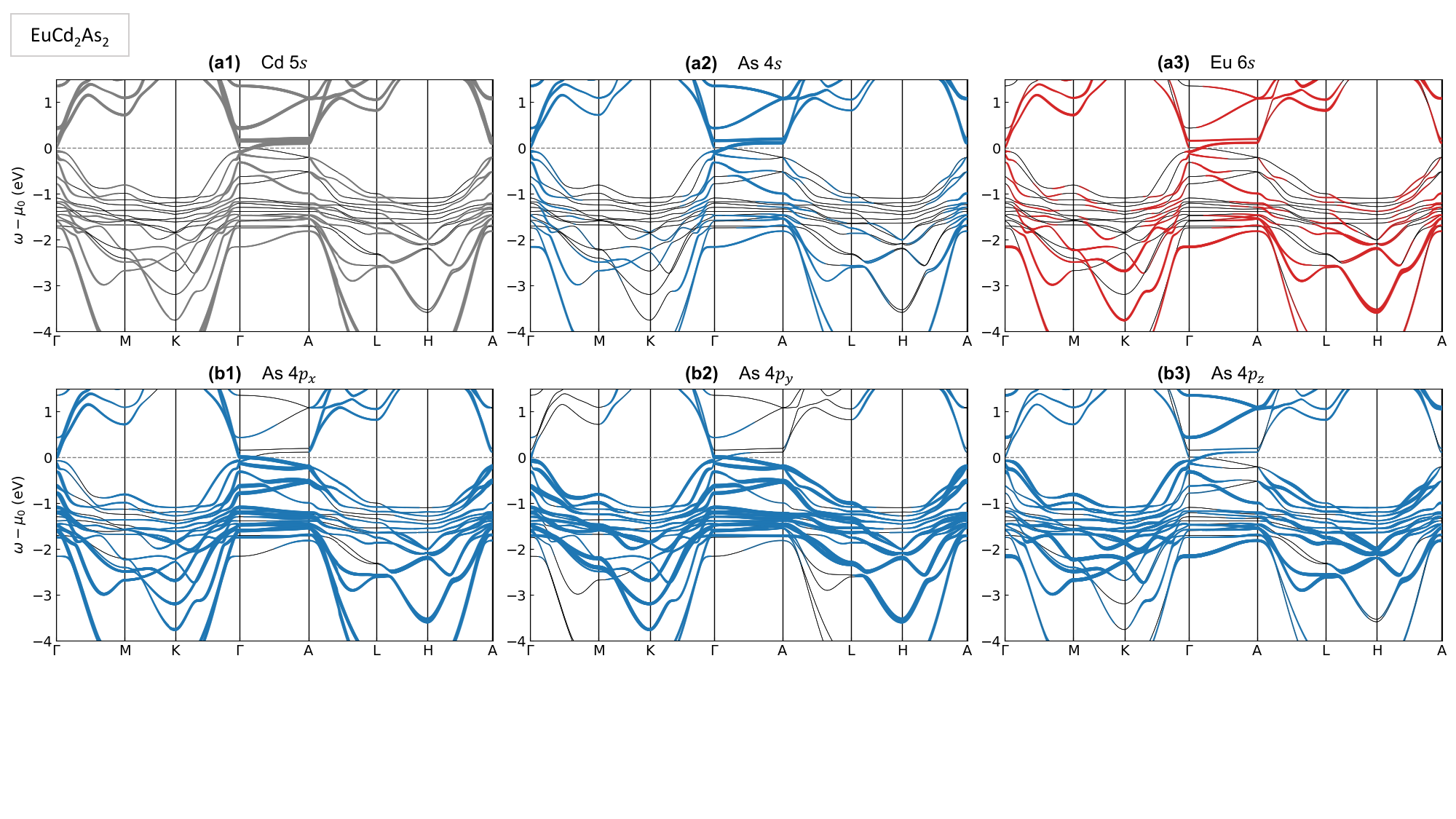}
    \caption{\textbf{The orbital projected band structure for EuCd\textsubscript{2}As\textsubscript{2}.} Similar to EuCd\textsubscript{2}P\textsubscript{2}, the As 4$p$ orbitals dominate the low-energy valence band and the $s$ orbitals from Eu, Cd, and P dominate the conduction band near $\mu$.}
    \label{fig:sp_in_ECA}
\end{figure*}

\begin{figure*}[h]
    \centering
    \includegraphics[width=\textwidth,trim={0 0 120 0},clip]{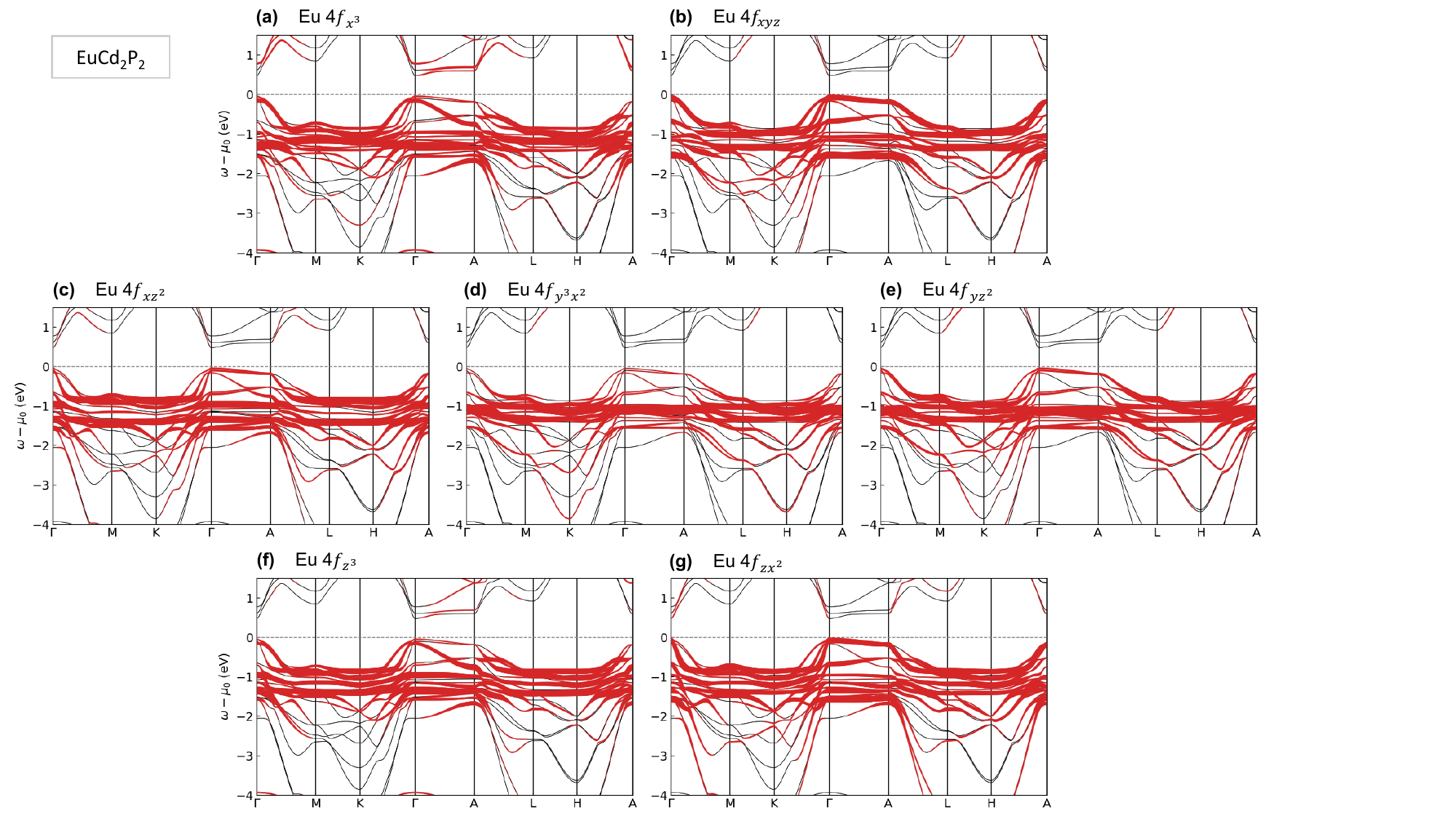}
    \caption{\textbf{The orbital projected band structure for Eu\textsubscript{2}P\textsubscript{2}.} The Eu $4f$ electrons not only dominate the non-dispersive bands within a binding energy window of $-1.5$ eV to $-1$ eV, but also extend expansively towards lower binding energy, with the $f_{xyz}$ and $f_{zx^2}$ orbitals as the prominent contributor.}
    \label{fig:4f_in_ECP}
\end{figure*}

\begin{figure*}[h]
    \centering
    \includegraphics[width=\textwidth,trim={0 0 120 0},clip]{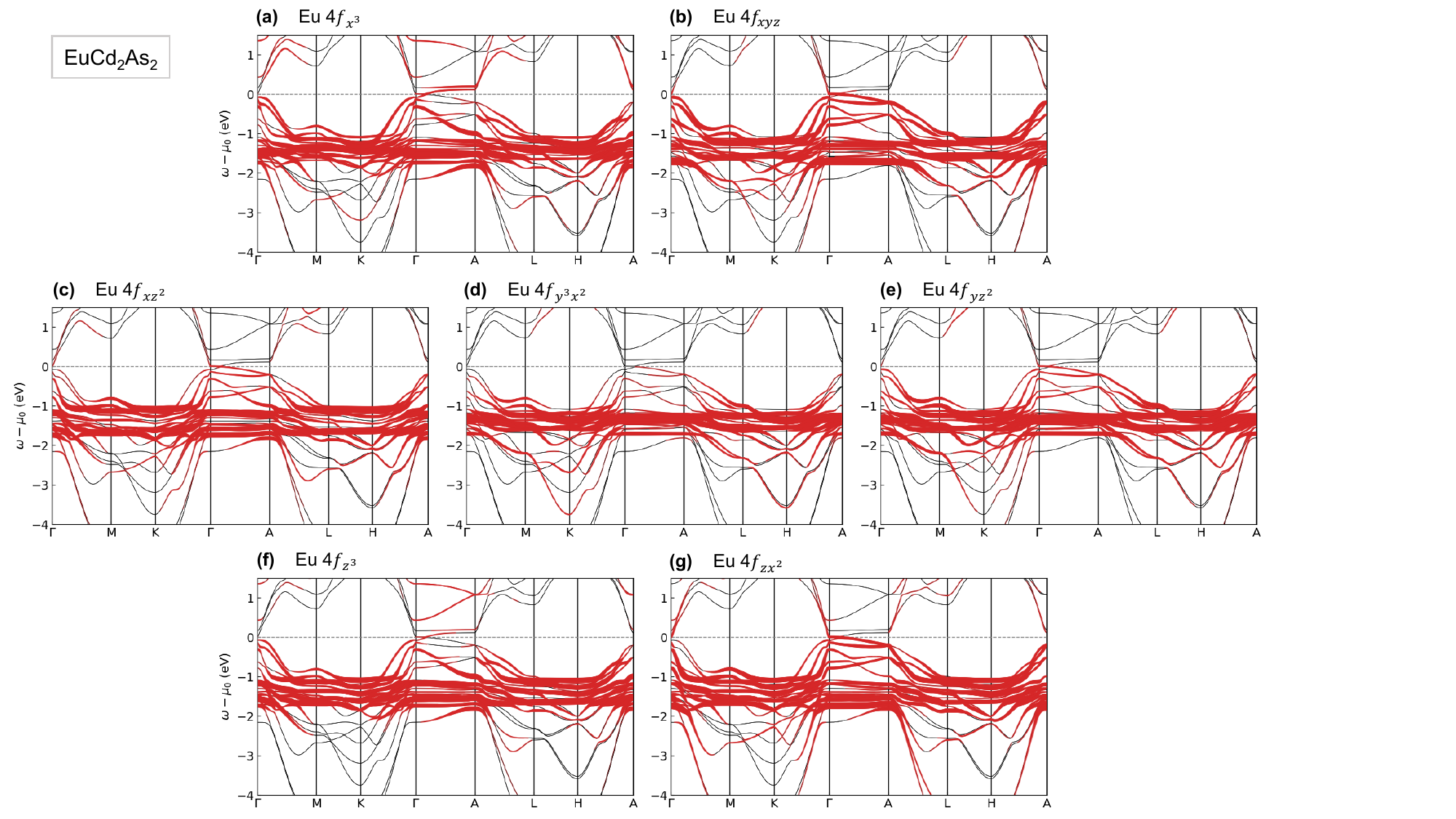}
    \caption{\textbf{The orbital projected band structure for Eu\textsubscript{2}As\textsubscript{2}.} The Eu $4f$ electrons dominate the non-dispersive bands within a binding energy window of $-1.5$ eV to $-1$ eV. Unlike the EuCd\textsubscript{2}P\textsubscript{2} case, they have less mixing into low-energy states.}
    \label{fig:4f_in_ECA}
\end{figure*}

Further scrutiny of the orbital character of the low-energy states reveals that the electronic states below $\mu$ are predominantly composed of pnictogen $p$-electrons, while those above $\mu$ are predominantly Cd $5s$ with contributions from Eu $6s$ and pnictogen $s$-electrons in both EuCd\textsubscript{2}P\textsubscript{2} and EuCd\textsubscript{2}As\textsubscript{2} (Fig. \ref{fig:sp_in_ECP}, Fig. \ref{fig:sp_in_ECA}). Intriguingly, despite the Eu $4f$ bands being deeply embedded at higher binding energies, they show significant hybridization with the low-energy pnictogen $p$ states, particularly for the $f_{xyz}$ and $f_{zx^2}$ orbitals (Fig. \ref{fig:4f_in_ECP}, Fig. \ref{fig:4f_in_ECA}). This substantial $4f$ character within the low-energy states marks a departure from conventional Kondo systems where the localized $4f$-elecrtons and itinerant $p$-electrons form the low-energy landscape independently, and the magnetic response arises from their interactions. Here,  the intermingling $p$-$f$ low-energy states creates a fertile playground for multiple cooperative interactions.

\begin{figure*}[h]
    \centering
    \includegraphics[width=\textwidth]{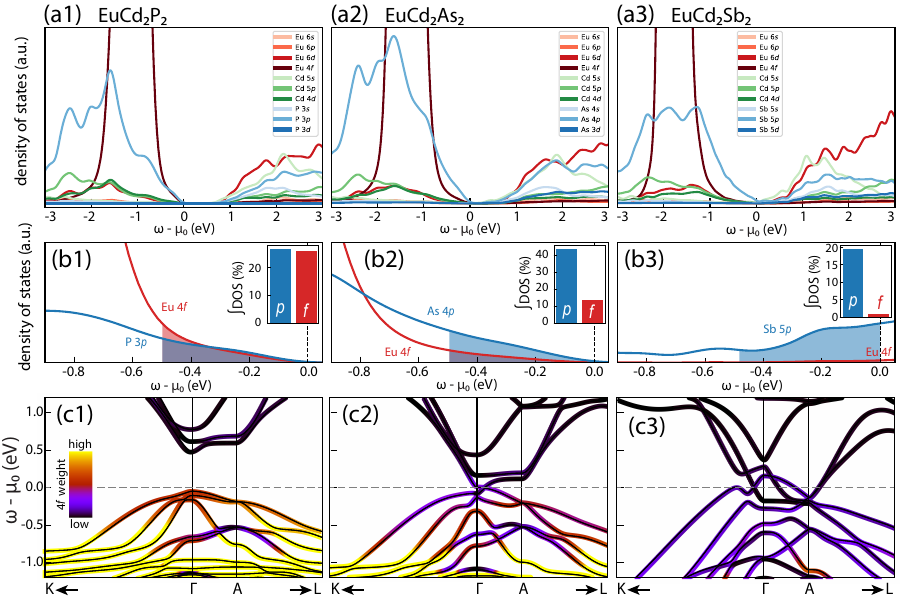}
    \caption{\textbf{The orbitally projected density of states (PDOS) for EuCd\textsubscript{2}P\textsubscript{2}, EuCd\textsubscript{2}As\textsubscript{2}, and EuCd\textsubscript{2}Sb\textsubscript{2}. (a1-a3)} PDOS over a wider range of energies. \textbf{(b1-b3)} Zoom in near $\mu$. The insets shows the integrated PDOS over the shaded area. Notably, EuCd\textsubscript{2}P\textsubscript{2} has much more $f$ weight near $\mu$ than EuCd\textsubscript{2}As\textsubscript{2} or EuCd\textsubscript{2}Sb\textsubscript{2}. (c1-c3) The band structure with projected $f$-character.}
    \label{fig:EXY-PDOS}
\end{figure*}

To quantify the $4f$ orbital contribution at low energies, we integrate the $4f$-projected density of states ranging from -0.5 eV up to $\mu$ (Fig. \ref{fig:EXY-PDOS}). We observe a more pronounced integration of Eu $4f$ states into the low-energy regime in EuCd\textsubscript{2}P\textsubscript{2} compared to EuCd\textsubscript{2}As\textsubscript{2} and EuCd\textsubscript{2}Sb\textsubscript{2}. Notably, the amount of $p$-$f$ mixing is proportional to the magnetoresistance across three materials\cite{wang2021colossal}, echoing the conclusion in the main text that it plays a pivotal factor in magnetoresistance.

Lastly, we examine the $k_z$ dispersion of these compounds to inform a proper selection of photon energy for the photoemission experiments. The band structures of EuCd\textsubscript{2}P\textsubscript{2} and EuCd\textsubscript{2}As\textsubscript{2} along the $K$-$\Gamma$-$K$ direction at various $k_z$ are presented in Fig. \ref{Fig: band kz dependence}. To correlate our spectroscopic findings with the anomalous transport behavior in EuCd\textsubscript{2}P\textsubscript{2}, we select an optimal photon energy of 80 eV, balancing the need for proximity to the $\Gamma$ point in $k_z$ with an effective photoemission cross-section (Fig. \ref{Fig: ECP_hn_dependence}, Fig. \ref{Fig: band kz dependence}).

\subsection{Energetics of magnetic orders}
We investigate the magnetic ground states of EuCd\textsubscript{2}P\textsubscript{2} through DFT calculations. We compute the total energies of various magnetic configurations, as listed in Fig. \ref{tab:graphLabels} and exemplified in Fig. \ref{fig:magnet_ball-stick}, for the corresponding Eu lattice supercell. Analogous to related compounds EuCd\textsubscript{2}As\textsubscript{2} and EuZn\textsubscript{2}As\textsubscript{2}\cite{wang2022anisotropy}, EuCd\textsubscript{2}P\textsubscript{2} exhibits a minimal energy difference between the magnetic states, which is in proximity to the bounds of quantum chemical precision. Detailed analysis reveals that the ferromagnetic configurations are almost degenerate with the antiferromagnetic type-A configuration. This is in contrast to the wider energy separation observed in EuCd\textsubscript{2}As\textsubscript{2} and EuZn\textsubscript{2}As\textsubscript{2}\cite{wang2022anisotropy}. The unique near-degeneracy in EuCd\textsubscript{2}P\textsubscript{2} promotes the formation of in-plane ferromagnetic clusters, which precede the establishment of the antiferromagnetic type-A order in the ground state.

\begin{figure*}[h]
    \centering
    \includegraphics[width=\textwidth,trim={0 170 0 0},clip]{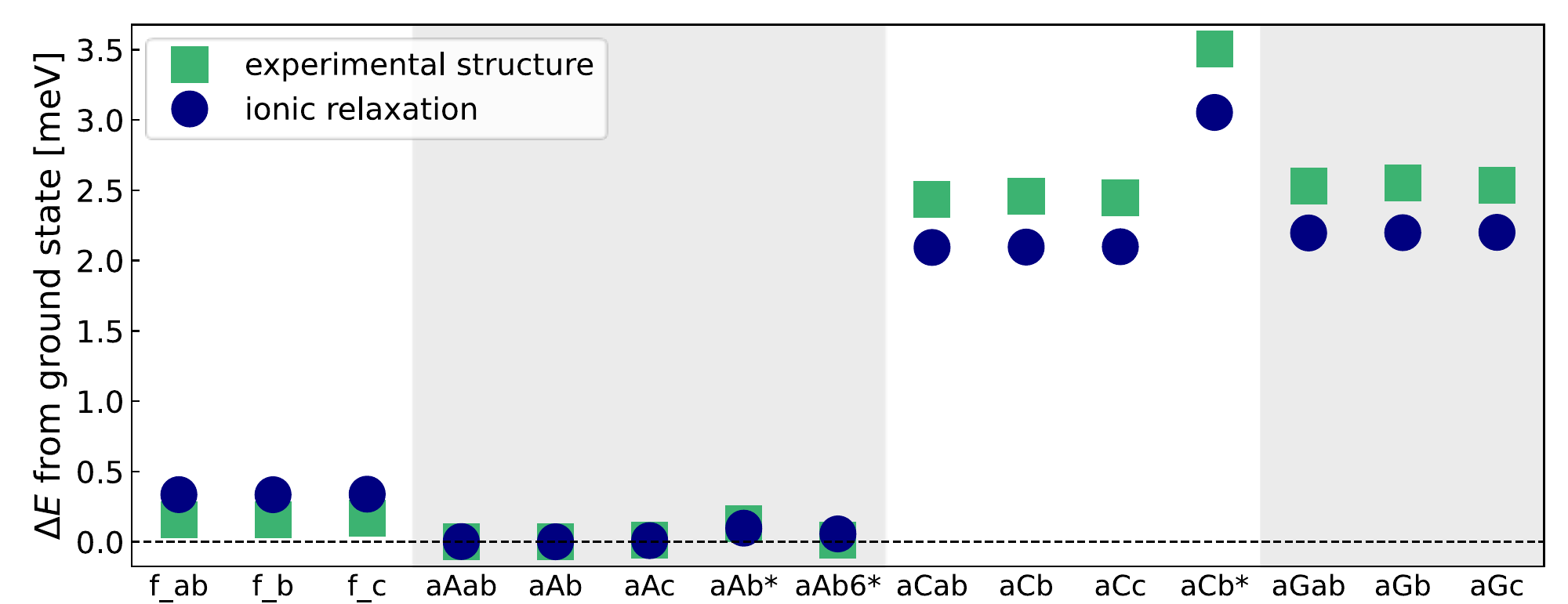}
    \caption{\textbf{Total energies of various magnetic configurations relative to the ground state magnetic order in \ECP.} The square (round) markers show the calculation result from the experimental (ionic relaxation), with lattice parameters listed in Fig. \ref{tab:ECA-EZA-ECP}.}
    \label{fig:ECP-magnetism}
\end{figure*}

\begin{table*}[h]
\caption{\textbf{List of all magnetic configurations considered in the DFT calculations (Fig.~S\ref{fig:ECP-magnetism})}. The asterisks label configurations that rotate the spin by 120$^{\circ}$ in the a-b plane. Details of the lattice supercells are shown in Fig. \ref{fig:magnet_ball-stick}.}
\label{tab:graphLabels}
\begin{tabular}{p{1cm}p{11cm}}
    Label & Magnetic configuration \\
    \hline
    f\_ab   &  Ferromagnetic (FM), spins aligned along [1,1,0] in the $ab$  plane \\
    f\_b   &  FM, spins aligned along $b$ axis \\
    f\_c  & FM, spins aligned along $c$ axis  \\
    aAab   &   Type-A antiferromagnetic (AFM), spins aligned along [1,1,0] in the $ab$ plane \\
    aAb     &   Type-A AFM, spins aligned along $b$ axis  \\
    aAc &  Type-A AFM, spins aligned along $c$ axis   \\
    aAb*   &   Type-A AFM, spins aligned in $ab$ plane with 120$^{\circ}$ rotations along $c$-axis \\
    aAb6*   &   Type-A AFM, spins aligned in $ab$ plane with 60$^{\circ}$ rotations along $c$-axis \\
    aCab    &  Type-C AFM, spins aligned along [1,1,0] in the $ab$ plane \\
    aCb  &  Type-C AFM, spins aligned along $b$ axis  \\
    aCc  &  Type-C AFM, spins aligned along $c$ axis  \\
    aCb*    &  Type-C AFM, spins aligned in $ab$ plane with 120$^{\circ}$ rotations \\
    aGab    &  Type-G AFM, spins aligned along [1,1,0] in the $ab$ plane \\
    aGb    & Type-G AFM, spins aligned along $b$ axis \\
    aGc   &  Type-G AFM, spins aligned along $c$ axis \\
\end{tabular}
\end{table*}

\begin{figure*}[h]
    \centering
    \includegraphics[width=\textwidth,trim={0 0 300 0},clip]{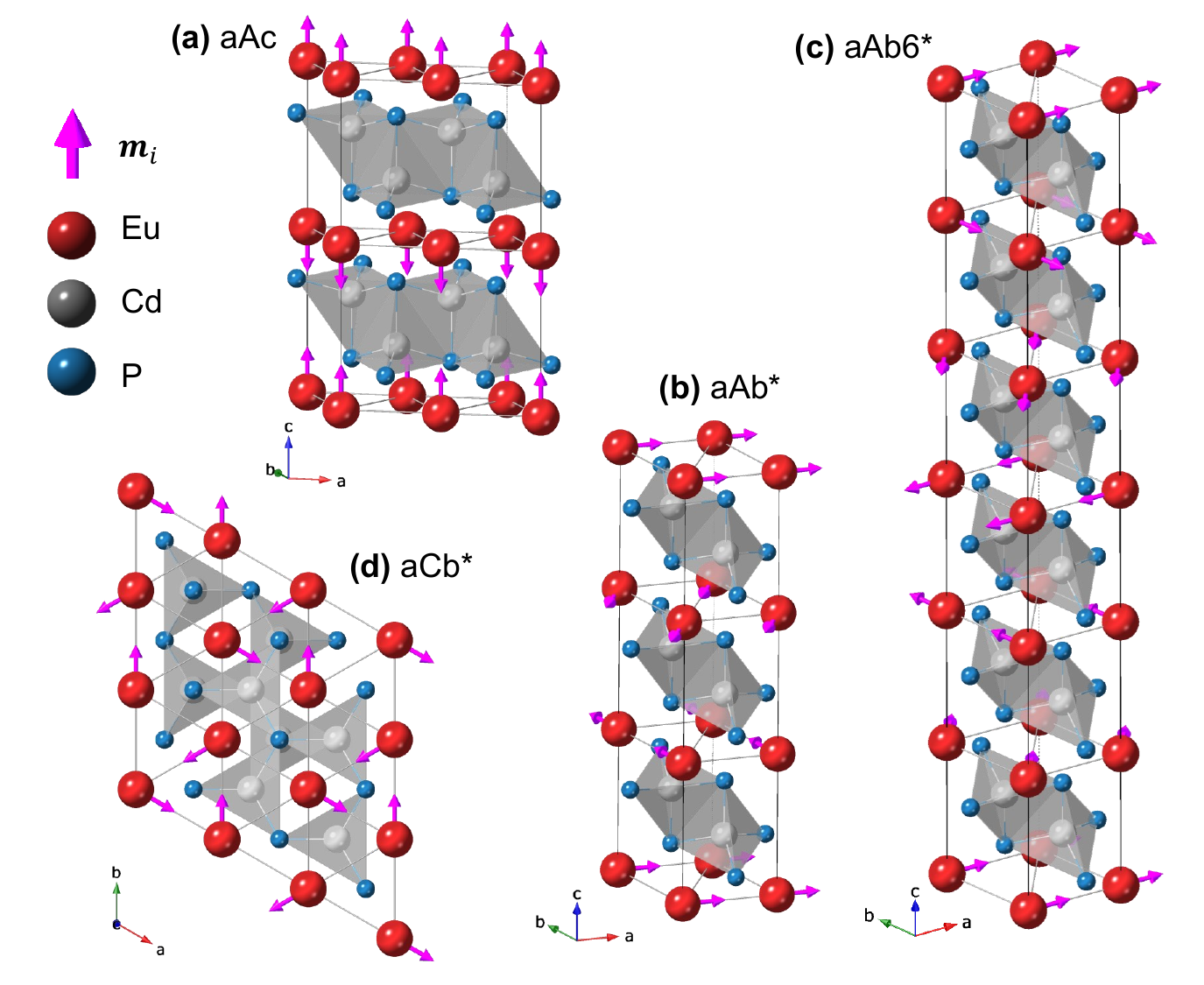}
    \caption{\textbf{Representative long-range magnetic orders.} The magnetic orders without the asterisk were calculated in the 4-Eu supercell with a Monkhorst-Pack $k$-grids of $11\times 21\times 11$ for the self-consistent field calculation and $13\times 17\times 13$ for the ionic relaxation.}
    \label{fig:magnet_ball-stick}
\end{figure*}

\begin{table*}[h]
	\caption{Crystallographic data both from experiment and theory for EuCd\textsubscript{2}As\textsubscript{2} and \ECP. Both belong to the space group $P\bar{3}m1$.
	The Wyckoff sites are Eu~$1a$ (0,0,0), Zn/Cd~$2d$ (1/3,1/3.$z$), and As/P~$2d$ (1/3,2/3,$z$) with full occupancy.}
        \label{tab:ECA-EZA-ECP}

		\begin{tabular}{l|ll|ll}	
                \hline
			  & Experimental &    & Theoretical &   \\
     
			\hline
			Material & \ECA\ & \ECP\ & \ECA\ & \ECP\ \\
   
			\hline
			Lattice parameters & & & & \\
			$a$ (\AA) & 4.440 & 4.325 & 4.507 & 4.369 \\ 
			$c$ (\AA) & 7.328 & 7.177 & 7.229 & 7.229 \\
			$V$ (\AA$^3$)&  125.11 & 116.30 & 127.19& 119.52 \\
			$c/a$ &  1.650 & 1.660 & 1.604 & 1.655 \\

			\hline
			Coordinates ($z$)& & & &  \\
			X & 0.6334 & 0.6357 & 0.6338 & 0.6369 \\
			Y & 0.2459 & 0.2489 & 0.2455 & 0.2486 \\
			\hline
			Bond distances (\AA) & & & & \\
			Eu-Eu~($\times6$) & 4.440 & 4.325 & 4.507 & 4.369 \\
			Eu-X ~($\times6$) & 3.713 & 3.616 & 3.757 & 3.640 \\
			Eu-Y ~($\times6$) & 3.134 & 3.071 & 3.174 & 3.097 \\
			X-X ~ ($\times3$) & 3.224 & 3.167 & 3.270 & 3.207 \\
			X-Y ~ ($\times3$) & 2.712 & 2.631 & 2.751 & 2.655 \\
			X-Y ~ ($\times1$) & 2.840 & 2.777 & 2.874 & 2.807 \\
			\hline
			Bond angles ($^\circ$) & & & & \\
			X-Y ~ ($\times3$) & 109.91 & 110.56 & 109.99 & 110.75 \\
			X-Y ~ ($\times1$) & 109.03 & 108.36 & 108.95 & 108.16 \\
		\end{tabular}
\end{table*}

\section{ARPES experiment}
\subsection{Eu $4d\rightarrow 4f$ resonant transition}
The electronic states predominantly composed of Eu $4f$ character will exhibit a significant increase in the photoemission spectral weight across the Eu $4d\rightarrow 4f$ transition. This enhancement serves as a qualitative indicator of the hybridization of Eu $4f$ states.

\begin{figure*}[h]
    \centering
    \includegraphics[width=\textwidth]{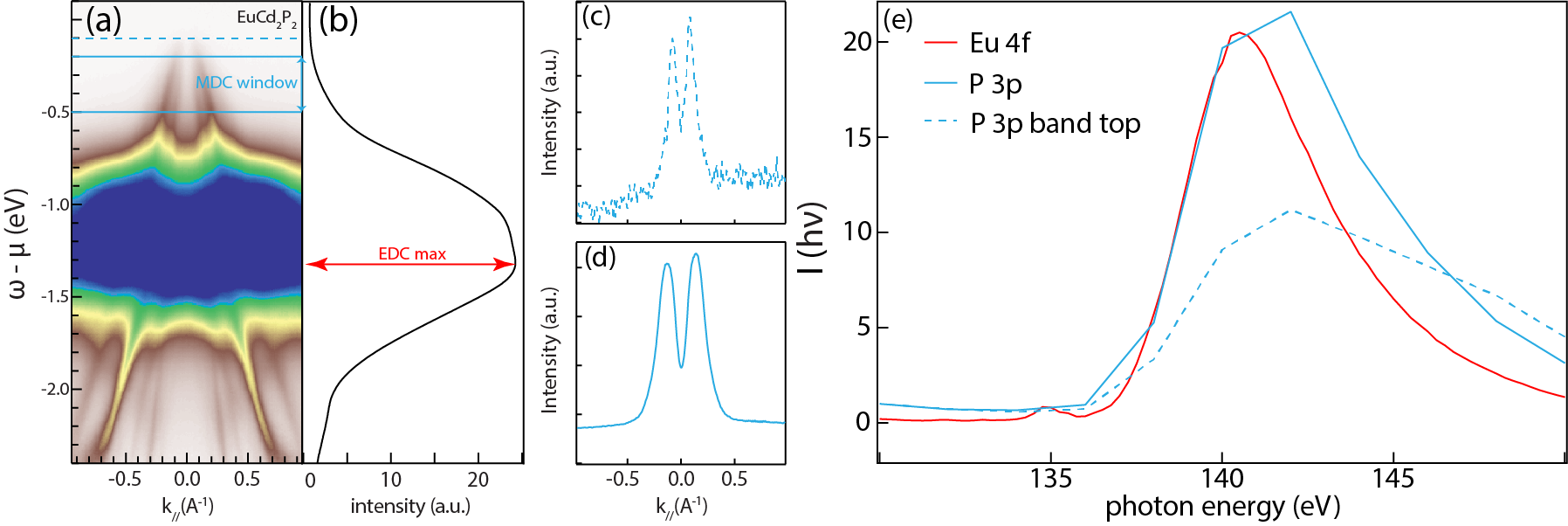}
    \caption{\textbf{ARPES spectral weights across Eu $4d\rightarrow 4f$ resonant transition. (a)} Angle-resolved photoemission spectroscopy (ARPES) spectra of EuCd\textsubscript{2}P\textsubscript{2} at 25 K. \textbf{(b)} The energy distribution curve (EDC) averaged over the entire momentum space in (a). \textbf{(c)} The momentum distribution curve (MDC) at a binding energy cut of $-100$ meV as marked by the blue dashed line in (a). \textbf{(d)} The MDC at a binding energy window marked by the blue box in (a) to encompass the $p$ bands.. \textbf{(e)} Photon energy dependence of the spectral weights across the Eu $4d\rightarrow 4f$ resonant transition from different regions. The photon flux monotonically decreases with increasing photon energy in the photon energy range, as calibrated by the photoemission spectral weight on polycrystalline gold.}
    \label{Fig: Eu_resounance}
\end{figure*}

We assessed the spectral weight of the Eu $4f$ bands within a binding energy range of $-1.5$ eV from $-1$ eV by pinpointing the peak maximum in the energy distribution curves (EDCs, Fig. \ref{Fig: Eu_resounance}(b), red arrow) averaged over the entire momentum space depicted in Fig. \ref{Fig: Eu_resounance}(a). The spectral weights peak in the vicinity of the Eu $4d\rightarrow 4f$ transition at a photon energy of approximately 140 eV (Fig. \ref{Fig: Eu_resounance}(e), red line). Subsequently, we quantified the spectral weights of the dispersive $p$ bands by integrating the momentum distribution curves (MDCs, Fig. \ref{Fig: Eu_resounance}(d)), averaged over an energy window encompassing the $p$ bands (Fig. \ref{Fig: Eu_resounance}(a), blue box). The spectral weight profile of the dispersive $p$ bands closely parallels that of the Eu $4f$ bands, providing compelling evidence of $p$-$f$ hybridization. This hybridization, which extends to very low energies, is further corroborated by monitoring the spectral weight of the MDCs at a binding energy of $-100$ meV (Fig. \ref{Fig: Eu_resounance}(c)), These MDC spectral weights also exhibit a similar non-monotonic trend. Through these observations, we establish that the low-energy Eu $4f$ bands engage in hybridization with the $p$ bands, contributing to the anomalous transport properties observed in EuCd\textsubscript{2}P\textsubscript{2}.

\subsection{Photon-flux dependence to confirm the absence of the charge-up effect}
\begin{figure*}[h]
    \centering
    \includegraphics[width=\textwidth]{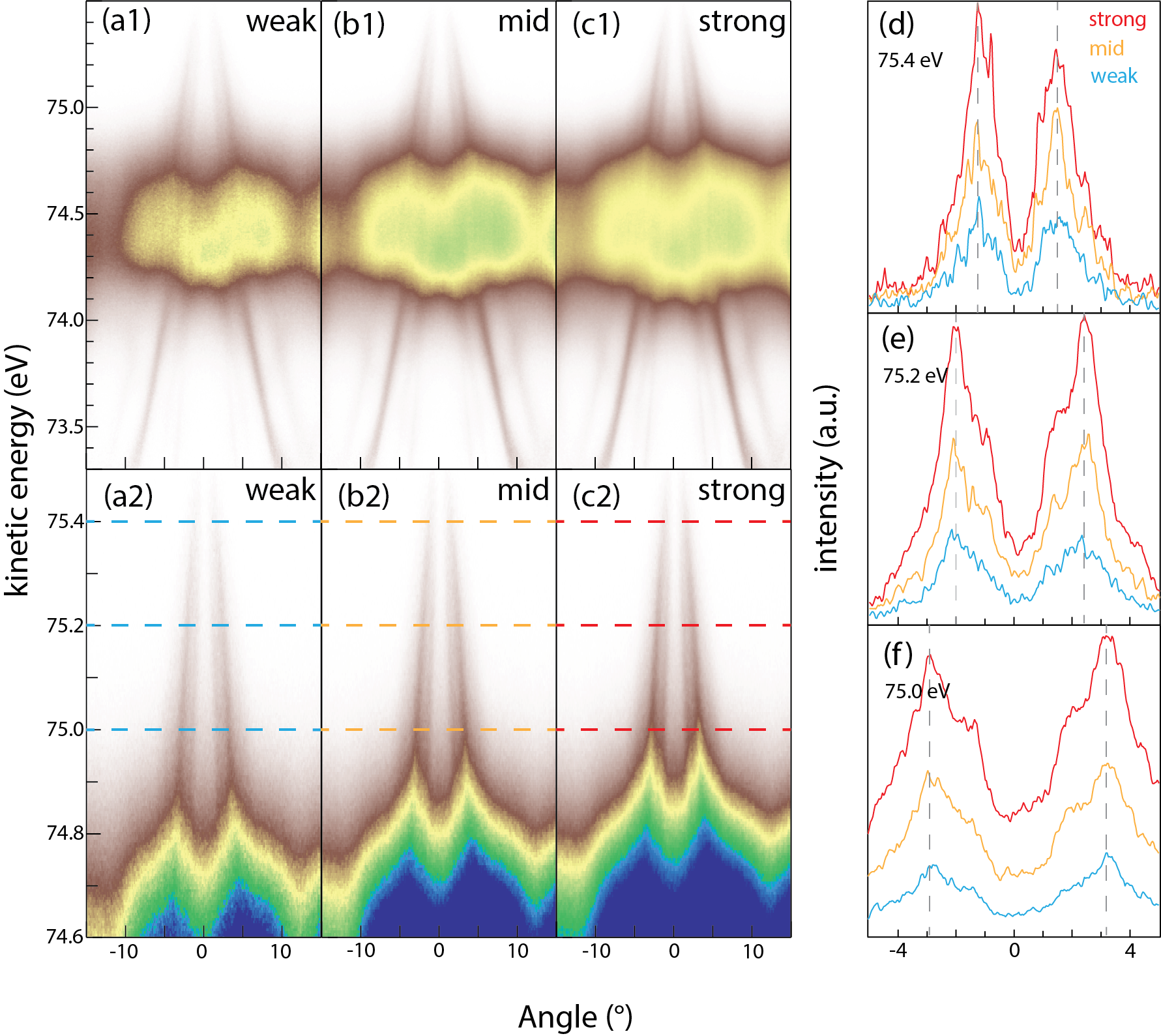}
    \caption{\textbf{Photon flux dependence of ARPES spectra. (a-c)} The ARPES spectra of EuCd\textsubscript{2}P\textsubscript{2} taken at 14 K under weak (a, $I_0=1.4579$ nA), medium (b, $I_0=2.2468$ nA), and strong (c, $I_0=3.022$ nA) incident photon flux with $h\nu=80$ eV, respectively. \textbf{(d-f)} The MDC cut at a kinetic energy of 75.4 eV (d), 75.2 eV (f), respectively. The alignment of the MDC peak positions confirms the absence of the charge-up effect.}
    \label{Fig: charge_up_effect}
\end{figure*}

We show the photon flux dependence of spectral weight at the temperature of maximum resistivity to rule out the charge-up effect as a trivial cause for spectral weight suppression in Fig. 3(e). Photon flux levels are quantified using a photocurrent diode, with strong, medium, and weak photon fluxes corresponding to $I_0 = 3.022/2.2468/1.4579$ nA, respectively. In Fig. \ref{Fig: charge_up_effect}(d-f), the consistent dispersion across different photon fluxes, confirms the absence of the charge-up effect.

\subsection{Temperature dependence of the ARPES spectra}
\begin{figure*}[h]
    \centering
    \includegraphics[width=\textwidth]{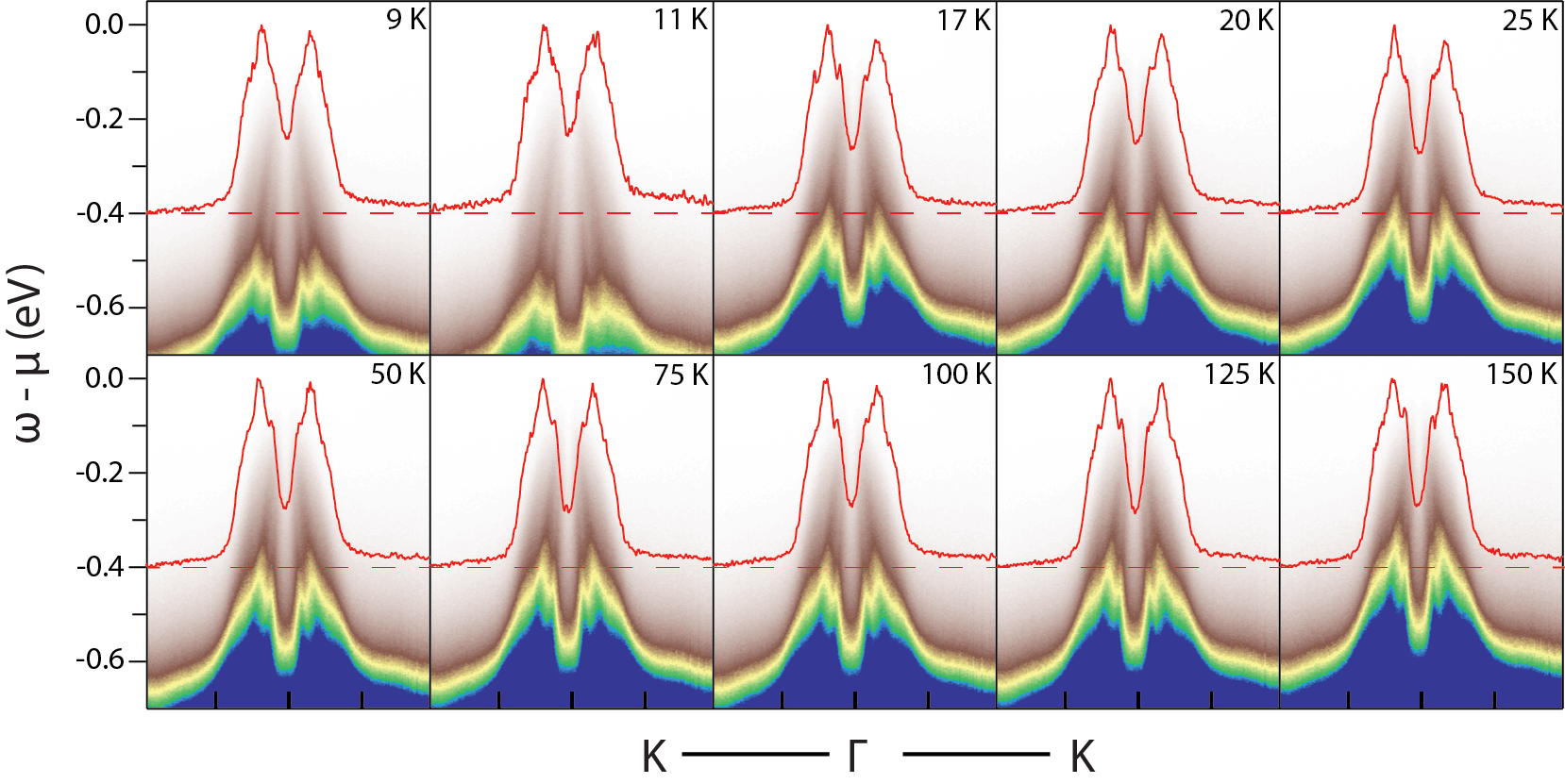}
    \caption{\textbf{Temperature dependence of EuCd\textsubscript{2}P\textsubscript{2} taken with 80 eV $p$-polarized photons.} The MDC cut at the binding energy of 0.4 eV (red dashed line) is superimposed on each spectrum. All cuts go through $K$-$\Gamma$-$K$.}
    \label{Fig: ECP_wide_T_dependence}
\end{figure*}

The temperature-dependent ARPES spectra of \ECP\ are displayed in Fig. \ref{Fig: ECP_wide_T_dependence}, with MDCs at the binding energy of -0.4 eV overlaid in red lines. The consistency of the single-peak MDC up to 150 K, well beyond the temperature of the long-range magnetic order, suggests that the enhanced interaction observed is not due to spin interactions.

\begin{figure*}[h]
    \centering
    \includegraphics[width=\textwidth]{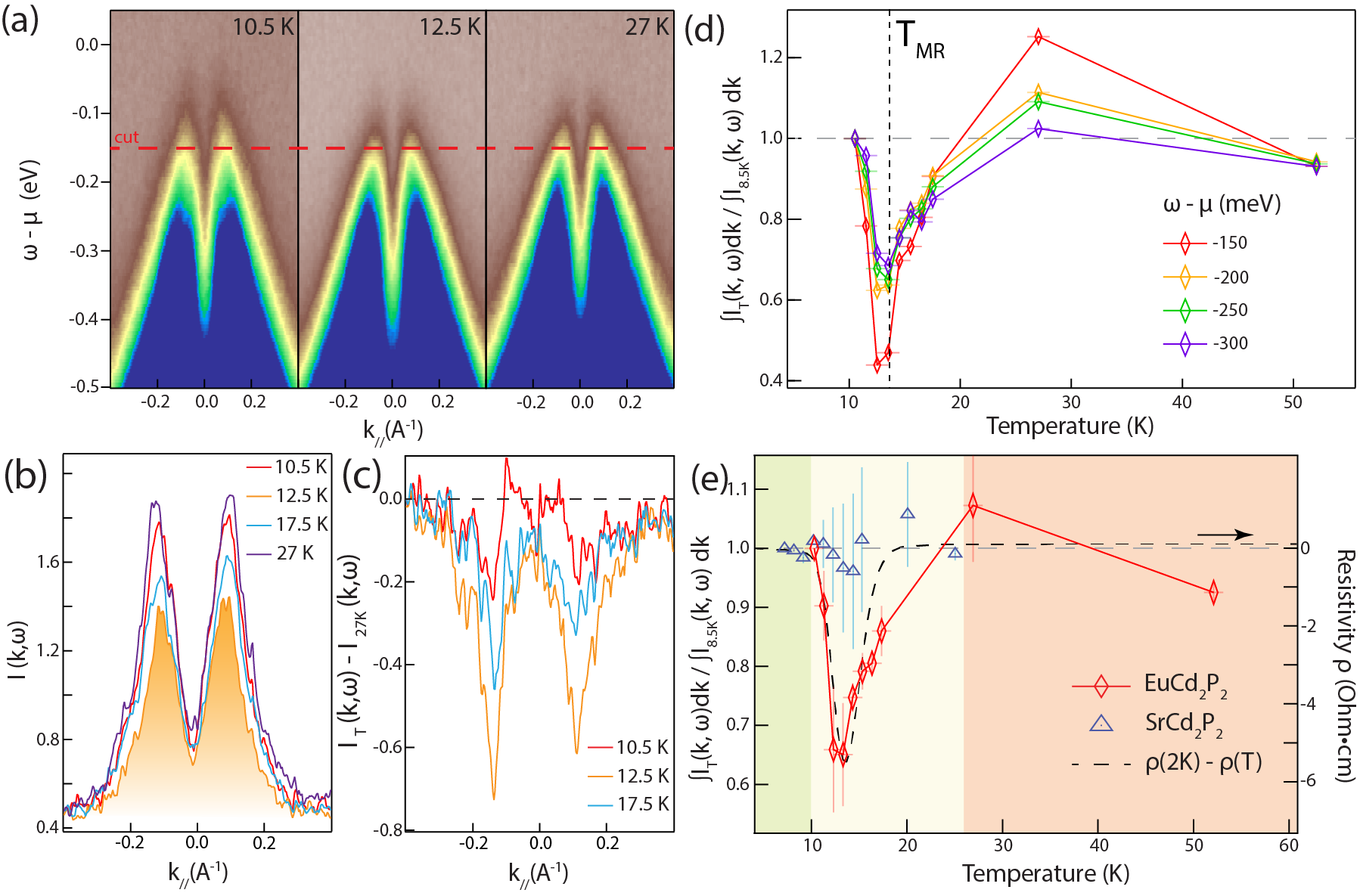}
    \caption{\textbf{Spectral weight suppression as the spectroscopic correspondence to the colossal magnetoresistance (CMR). (a)} ARPES spectra of EuCd\textsubscript{2}P\textsubscript{2} below, at, and above T\textsubscript{MR} in the left, middle, and right panels, respectively. The right panel also displays the three bands fitted from MDCs ($\alpha$ in red, $\beta$ in orange, and $\gamma$ in blue). A parabolic fit to the $\beta$ band dispersion highlights its Fermi level-crossing character (white dashed line). \textbf{(b)} MDCs at a binding energy of 150 meV (marked by the red dashed line in (a)) at various temperatures. Spectral weight at T\textsubscript{MR} undergoes pronounced suppression. \textbf{(c)} MDCs in (b) subtracted by the one at 9 K. \textbf{(d)} The temperature dependence of the integrated spectral weight at various binding energies. The red shaded area in (b) indicates the integration area. The spectral weight suppression is more pronounced at lower binding energies. \textbf{(e)} The temperature dependence of integrated spectral weight in EuCd\textsubscript{2}P\textsubscript{2} and SrCd\textsubscript{2}P\textsubscript{2} along with the negative excessive resistivity $\rho$(2K)$-\rho$(T). The spectral weight in EuCd\textsubscript{2}P\textsubscript{2} exhibits a profile akin to anomalous resistivity change and is absent in SrCd\textsubscript{2}P\textsubscript{2}. Error bars of spectral weight indicate the variations at different binding energies in (d). Error bars of temperature indicate the difference between the thermal diode reading and the actual sample temperature. The colored background indicates the anti-ferromagnetism, ferromagnetism, and paramagnetism from low to high temperatures.}
    \label{fig: spin_fluctuation_reproduce}
\end{figure*}

In Fig. \ref{fig: spin_fluctuation_reproduce}, we present an additional dataset analogous to that of Fig. 2, thereby illustrating the reproducibility of the spectral weight suppression near T\textsubscript{MR}.

\subsection{Band reconstruction at low photon energies}
\begin{figure*}[h]
    \centering
    \includegraphics[width=\textwidth]{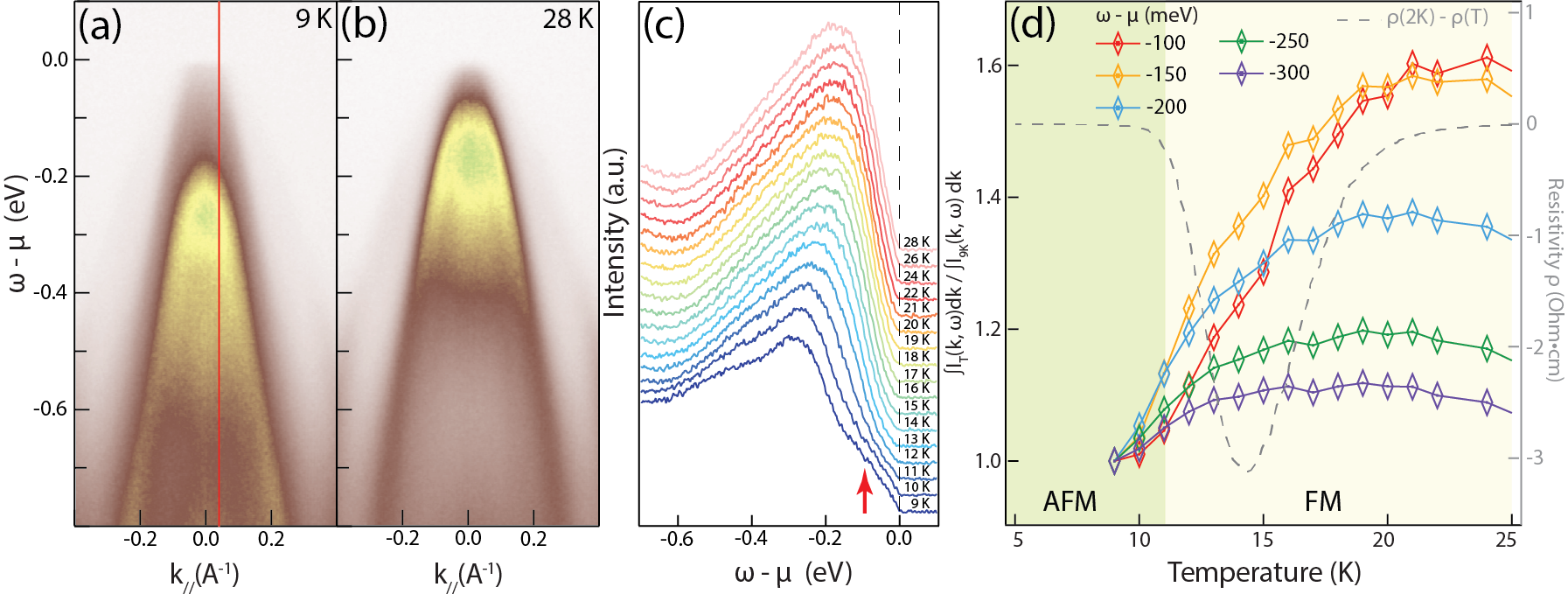}
    \caption{\textbf{Band reconstruction in ferromagnetic state. (a, b)} The ARPES spectra cut along $\Gamma$-$K$ taken with $h\nu=21.2$ eV $p$-polarized photons at (a) 9 K and (b) 28 K. \textbf{(c)} The temperature-dependent EDCs along the cut marked by the red vertical line in (a). The red arrow highlights the emergence of additional low-energy features at 9 K. \textbf{(d)} The temperature-dependent momentum-integrated spectral weight at different binding energies. The grey dashed line shows the negative excessive resistivity $\rho(T)-\rho(2K)$.}
    \label{Fig: band_reconstruction}
\end{figure*}

We present a band reconstruction similar to that reported by Zhang \textit{et al.}\cite{zhang2023electronic}, despite pronounced discrepancies in sample resistivity. Our observations are twofold:\\
\begin{itemize}
    \item We show the ARPES spectra cut along $\Gamma$-$K$ taken with $h\nu=21.2$ eV $p$-polarized photons at 9 K and 28 K in Fig. \ref{Fig: band_reconstruction}(a) and (b), and the corresponding temperature-dependent EDCs in Fig. \ref{Fig: band_reconstruction}(c), respectively. At low temperatures, we observe both a shift of EDC peak towards higher binding energy and the emergence of an additional low-energy feature, in agreement with the reports in \cite{zhang2023electronic};\\
    \item We show the momentum-integrated spectral weight as a function of temperature at various binding energies in Fig. \ref{Fig: band_reconstruction}(d). The monotonic decrease in spectral weight at low temperatures suggests the band splitting \cite{zhang2023electronic} and indicates the emergence of in-plane ferromagnetism around 20 K, aligning with the findings in \cite{vSunko2023spin}.\\ 
\end{itemize}
Notably, the temperature associated with the peak in resistivity (T\textsubscript{MR}$= 14$ K) does not coincide with either the appearance of the low-energy feature ($\sim$10 K) or the onset of in-plane ferromagnetism ($\sim 20$ K). This suggests that the band reconstruction is not directly related to the CMR. It is also noteworthy that the band reconstruction is exclusively observed by 21.2 eV photons. This specificity probably arises from the $k_z$ dependence (Fig. \ref{Fig: band kz dependence}(d)) and points to a need for further investigation.

\subsection{Photon energy dependence of the ARPES spectra}
\begin{figure*}[h]
    \centering
    \includegraphics[width=\textwidth]{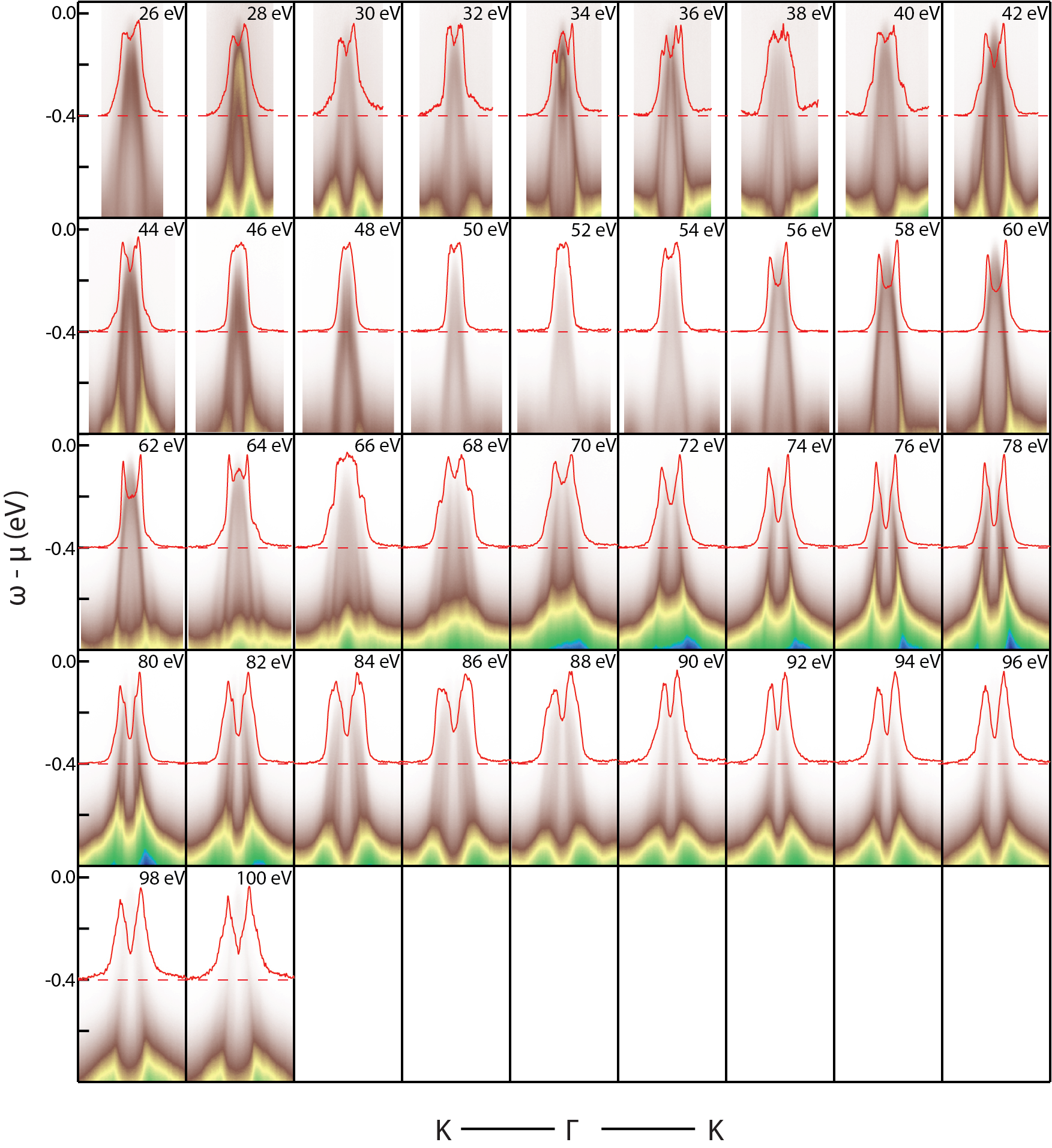}
    \caption{\textbf{Photon-energy dependence of the \ECP\ spectra.} The MDC cut at a binding energy of 0.4 eV (red dashed line) is overlaid on each spectrum. All cuts go through $K$-$\Gamma$-$K$. }
    \label{Fig: ECP_hn_dependence}
\end{figure*}

The variation of ARPES spectra varying with photon energy is presented in Fig. \ref{Fig: ECP_hn_dependence}, accompanied by red lines representing the MDCs at the binding energy of $-0.4$ eV. The consistent appearance of the single-peak MDC across different photon energies rules out the possibility of trivial explanations from an unfavorable $k_z$ or photoemission cross-section, thereby underscoring the intrinsic nature of the observed single-peak MDCs.

\begin{figure*}[h]
    \centering
    \includegraphics[width=\textwidth]{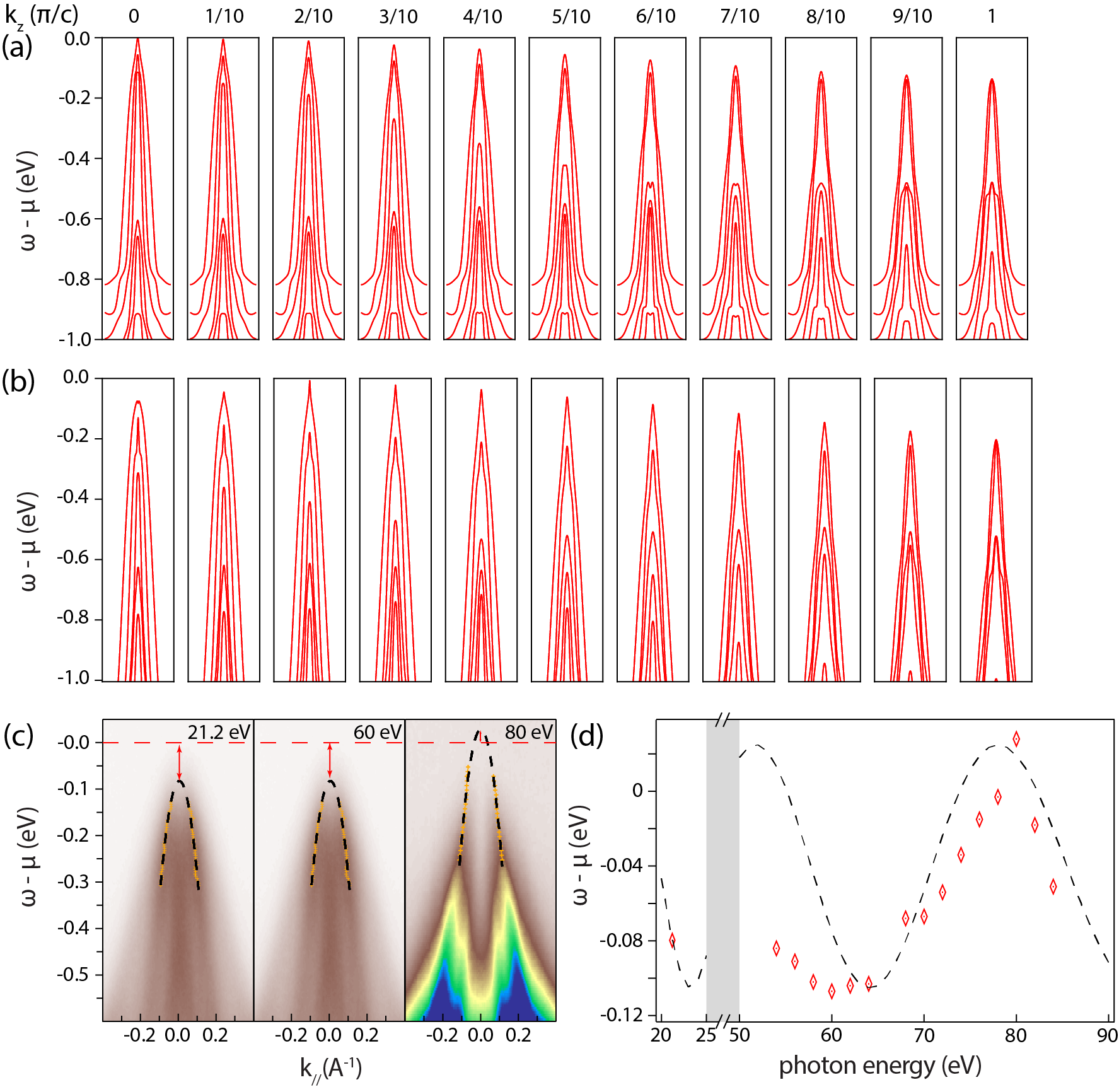}
    \caption{\textbf{$k_z$ dispersion of the ARPES spectra. (a, b)} Calculated band dispersion along $K$-$\Gamma$-$K$ direction at various $k_z$ for (a) EuCd$_2$P$_2$ and (b) EuCd$_2$As$_2$. In both compounds, the valence band tops exhibit dispersions into higher binding energy along $\Gamma$-$Z$. \textbf{(c)} The parabolic fit to the low-energy band dispersion exemplified at 21.2 eV (left), 60 eV (mid), and 80 eV (right), respectively. The black dashed lines represent the parabolic fit to the band dispersion, marked by the orange crosses and derived from the multi-peaked MDC fits. Red arrows indicate the binding energies of parabolic band tops. \textbf{(d)} The binding energies of valence band top indicated by the red markers). The black dashed lines corresponds to the fit described in Fig. \ref{eq: photon_energy_to_kz}.}
    \label{Fig: band kz dependence}
\end{figure*}

The data here also facilitates an estimation of the relationship between photon energy and $k_z$, anchored by the observation that the binding energies of valence band top at $k_{\parallel}=0$ disperses into higher band energies along $\Gamma$-$Z$. We conduct a parabolic fit to the low-energy band dispersion at the photon energies that allows a reliable fit, as exemplified in Fig. \ref{Fig: band kz dependence}(c), and extracted the binding energies of valence band top ($\omega_{\text{VBT}}$, Fig. \ref{Fig: band kz dependence}(d)). The binding energies are fitted to the model defined by Fig. \ref{eq: photon_energy_to_kz},
\begin{widetext}
\begin{equation}\label{eq: photon_energy_to_kz}
    \begin{cases}
        \hbar k_{\perp, \text{solid}} & = \sqrt{2m_e(h\nu-\Phi-\omega_{\text{VBT}}-V_0)}\approx \sqrt{2m_e(h\nu-\Phi-V_0)}\\
        \omega_{\text{VBT}} & = A\cos (c\cdot k_{\perp, \text{solid}}) + B
    \end{cases}
\end{equation}
\end{widetext}
where $k_{\perp, \text{solid}}$ is the momentum of electrons in the solid orthogonal to $k_{\parallel}$, $m_e$ is the mass of an electron, $h\nu$ is the photon energy, $\Phi$ is the work function, $\omega_{\text{VBT}}$ is the binding energy of valence band top, $V_0$ is the inner potential, $c=7.1771\AA$ is the lattice constant along $c$-axis, and $A=65 \text{ meV}, ~ B=-40 \text{ meV}, ~ V_0=5.0 \text{ eV}$ are the fitting parameters. The resulting fits are represented by the black dashed line in Fig. \ref{Fig: band kz dependence}, indicating the proximity to the $\Gamma$ point at 80 eV and to the $Z$ point at 21.2 eV, respectively.

\section{ARPES spectra simulation}
\subsection{Formulation}
In the context of a direct photoemission process within the three-step model, the excitation of electrons by incident photons can be formally described by the transition probability $w_{fi}$ of an $N$-electron initial to final state under Fermi's golden rule\cite{damascelli2003angle, sobota2021angle}:
\begin{equation*}
    w_{fi} = \frac{2\pi}{\hbar} |\bra{\Psi_f^N}H_{\text{int.}}\ket{\Psi_i^N}|^2 \delta(\omega_f^N - \omega_i^N - h\nu),
\end{equation*} 
where $\ket{\Psi_i^N}$ is the $N$-electron initial state with energy $\omega_f^N$ and $\ket{\Psi_f^N}$ is the $N$-electron final state with energy $\omega_i^N$. $H_{\text{int.}}$ is the perturbative Hamiltonian describing the interaction between electrons and photons. Under the dipole approximation, the transition probability becomes,
\begin{equation*}
    w_{fi} \propto \sum_{f,i}|M_{f,i}^{\boldsymbol{k}}|^2\sum_m \braket{\Psi_m^{N-1}|\Psi_i^{N-1}} \delta(\epsilon_f + \omega_m^{N-1} - \omega_i^N -h\nu), 
\end{equation*}
where $M_{f,i}^{\boldsymbol{k}}$ is the photoemission matrix element, $\ket{\Psi_i^{N-1}}$ is the wave function of the initial $N-1$ electrons, $\ket{\Psi_m^{N-1}}$ is the wave function of the final excited states of $N-1$ electrons with energy $\omega_m^{N-1}$ indexed by $m$, and $\omega_i^N$ is the energy of $N$-electron initial state. With a single-electron removal function, the photoemission intensity can be written as \cite{damascelli2003angle, sobota2021angle},
\begin{equation}\label{eq: ARPES intensity}
    I(\boldsymbol{k}, \omega) \propto |M_{f,i}^{\boldsymbol{k}}|^2 f(\omega)A(\boldsymbol{k}, \omega),
\end{equation}
where $f(\omega)$ is the Fermi-Dirac distribution and $A(\boldsymbol{k},\omega)$ is the spectral function. As such, we calculate the matrix element and spectral function explicitly.

\subsection{Matrix element calculation}
The matrix element can be explicitly written as\cite{yeh1985atomic, moser2017experimentalist, wang2012orbital, day2019computational, goldberg1981photoionization, li2023spectroscopic},
\begin{equation}\label{eq: matrix element definition}
    M_{f,i}^{\boldsymbol{k}} = \sum_x \alpha_{i,x} \bra{\Phi_{\text{E}_\text{kin}, \boldsymbol{k}}}\boldsymbol{\hat{\epsilon}\cdot r} \ket{\Phi_{nlx}}
\end{equation}
where $\Phi_{\text{E}_\text{kin}, \boldsymbol{k}}$ is the one-electron final state with kinetic energy $E_{\text{kin}}$ and momentum $k$ and $\Phi_{nlx}$ is the one-electron atomic wave function characterized by quantum numbers $n,~l,~x$ (to elaborate in eq. \ref{eq: real spherical harmonic to complex}). $\alpha_{i,x}$ is the coefficient of the initial state $i$ under the atomic-orbital basis of valence electrons indexed by $x$ and is obtained from the DFT calculations, i.e. $\ket{\Psi_i} = \sum_x \alpha_{i,x}\ket{\Phi_{nlx}}$. $\boldsymbol{\hat{\epsilon}\cdot r}$ is the photon-electron interaction under dipole approximation, with  $\hat{\boldsymbol{\epsilon}}$ is the unit electric field vector of the incident photon, and $\boldsymbol{r}$ is the position vector.

$\Phi_{nlx}$ can be further decomposed as a product of the radial part and the angular part,
\begin{widetext}
\begin{equation}\label{eq: scenario 1 initial state}
    \braket{\boldsymbol{r}|\Phi_{nlx}} = R_{n,l}(r) \sum_{m=-l}^l n(m) Y_{l,m}(\theta,\phi) := \frac{P_{n,l}(r)}{r} \sum_{m=-l}^l n(m) Y_{l,m}(\theta,\phi), 
\end{equation}
\end{widetext}
where $R_{n,l}(r)$ is the radial part of the wave function, and $Y_{l,m}(\theta,\phi)$ is the complex spherical harmonics. $x$ is the index for the initial atomic-orbital states in real spherical harmonics and $n(m)$ is the coefficient that transforms real spherical harmonics $Y_{l,m}^{(R)}(\theta, \phi)$ basis into the complex spherical harmonics $Y_{l,m}(\theta, \phi)$ basis by the relation:
\begin{equation}\label{eq: real spherical harmonic to complex}
    Y_{l,m}^{(R)} = \begin{cases}
        \frac{i}{\sqrt{2}}(Y_{l,m} - (-1)^mY_{l, -m}), &\quad \text{if } m<0\\
        Y_{l,0}, &\quad \text{if } m =0\\
        \frac{1}{\sqrt{2}}(Y_{l,-m} + (-1)^mY_{l,m}), &\quad \text{if } m>0
    \end{cases}
\end{equation}
The radial part of wave function, $P_{n,l}(r):= \frac{R_{n,l}(r)}{r}$ is obtained by solving the one-electron Schrodinger equation\cite{goldberg1981photoionization}
\begin{equation*}\label{eq: one-electron ischrodinger equation}
    \left[\frac{d^2}{dr^2} + V(r) + \epsilon_{n,l} - \frac{l(l+1)}{r^2}\right] P_{n,l}(r) = 0,
\end{equation*}
where $V(r)$ is the effective central-field potential, and $\epsilon_{n,l}$ is the binding energy. Here, we take the form,
\begin{equation}\label{eq: effective central potential}
    V(r) = V^H(r) + V^{\text{ex.}}(r),
\end{equation}
where $V^H(r)$ is the standard Hartree potential and $V^{\text{ex.}}(r)$ is the free-electron exchange potential in the Slater form:
\begin{equation}\label{eq: exchange potential}
    V^{\text{ex.}}(r) = -6\left[\left(\frac{3}{8\pi}\right) \rho(r)\right]^{1/3},
\end{equation}
where $\rho(r)$ is the charge density at radius $r$.

\begin{figure*}[h]
    \centering
    \includegraphics[width=\textwidth]{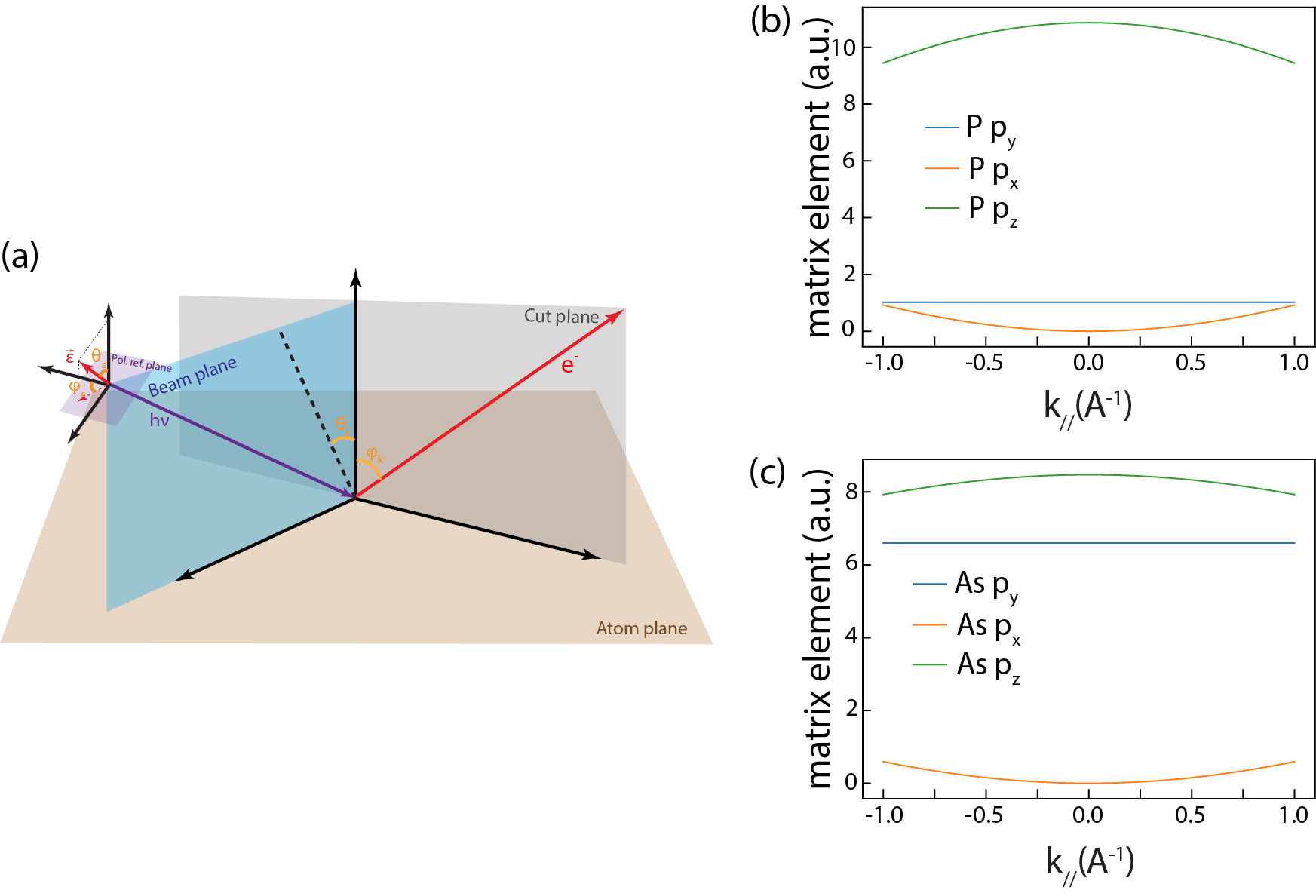}
    \caption{\textbf{Matrix element calculation. (a)} Schematic of the coordinate system for photoemission experiments. The brown, blue, purple, and grey planes are the atom plane, beam plane, polarization reference plane, and cut plane, respectively. The direction of the electric field of the incident photon $\boldsymbol{\varepsilon}$ is defined by $\theta_\varepsilon, \varphi_\varepsilon$ in the spherical coordinate system marked in the figure. The direction of photoemitted electron is similarly defined by $\theta_{\boldsymbol{k}}, \varphi_{\boldsymbol{k}}$. \textbf{(b)} The calculated matrix element for P $p$ orbitals under $p$-polarized photons at $h\nu=80$ eV. \textbf{(c)} The calculated matrix element for As $p$ orbitals under $p$-polarized photons at $h\nu=80$ eV.} 
    \label{Fig: matrix_element}
\end{figure*}

We take the final state as the continuum (scattering) state of the one-electron wave function under the same potential in eq. \ref{eq: effective central potential}. Specifically, we write the final state as a partial-wave expansion,
\begin{widetext}
\begin{equation}\label{eq: scenario 1 final state}
    \braket{\boldsymbol{r}|\Phi_{E_{\text{kin}, \boldsymbol{k}}}} = 4\pi \sum_{l',m'} (i)^{l'} \exp(-i\delta_{l'}) Y_{l',m'}^*(\theta_k,\phi_k) Y_{l',m'}(\theta,\phi) R_{E_\text{kin},l'} (r),
\end{equation}
\end{widetext}
, where $\theta_k, \phi_k$ is the direction of the photo-excited electron in the spherical coordinate system defined in Fig. \ref{Fig: matrix_element} and $Y_{l,m}^*$ is the complex conjugate of $Y_{lm}$ with relation $ Y_{l,m}^*= (-1)^mY_{l,-m}$. $R_{E_{\text{kin}}, l'}(r)$ (equivalently, $P_{E_{\text{kin}}, l'}/r$) is solved by the one-electron Schrodinger's equation:
\begin{equation*}\label{eq: one-electron schrodinger equation scattering state}
    \left[\frac{d^2}{dr^2} + V(r) + E_{\text{kin}} - \frac{l(l+1)}{r^2}\right] P_{E_{\text{kin}},l}(r) = 0,
\end{equation*}
where $E_{\text{kin}} = h\nu - \epsilon_{n,l}$ and $V(r)$ is the same as in eq. \ref{eq: effective central potential}. $P_{E_{\text{kin}},l}(r)$ are normalized by the asymptotic behavior at infinity\cite{goldberg1981photoionization}:
\begin{widetext}
\begin{equation*}\label{eq: asymptotic behavior of final state}
    \lim_{r\rightarrow \infty} P_{\epsilon, l}(r) = \pi^{-1/2} \epsilon^{-1/4} ~\sin \left[\epsilon^{1/2}-\frac{1}{2}l\pi - \epsilon^{-1/2}\log (2\epsilon^{1/2}r) + \delta_l(\epsilon)\right],
\end{equation*}
\end{widetext}
where $\delta_l(\epsilon)$ is the phase shift.

We express the $\hat{\boldsymbol{\varepsilon}}\cdot \boldsymbol{r}$ in the spherical coordinate system by
\begin{equation*}\label{eq: dipole operator 1}
    \begin{cases}
    \varepsilon_x = \sin\theta_\varepsilon \cos\phi_\varepsilon\\
    \varepsilon_y = \sin\theta_\varepsilon \sin\phi_\varepsilon\\
    \varepsilon_z = \cos\theta_\varepsilon
    \end{cases}
\end{equation*}
where $\theta_\epsilon$ and $\phi_\epsilon$ are defined in fig. \ref{Fig: matrix_element}(a) and 
\begin{equation}\label{eq: dipole operator 2}
    \begin{cases}
    x/r = \sin\theta \cos\phi = (2\pi/3)^{1/2} (-Y_{1,1} + Y_{1,-1})\\
    y/r = \sin\theta \sin\phi = i(2\pi/3)^{1/2} (Y_{1,1} + Y_{1,-1})\\
    z/r = \cos\theta = (4\pi/3)^{1/2}Y_{1,0}
    \end{cases}
\end{equation}
Combining eq. \ref{eq: matrix element definition}, \ref{eq: scenario 1 initial state}, \ref{eq: scenario 1 final state}, and \ref{eq: dipole operator 2}, we have
\begin{widetext}
\begin{equation}\label{eq: scenario 1 overall expression 1}
    \begin{split}
    \bra{\Phi_{E_{\text{kin}, \boldsymbol{k}}}}\hat{\boldsymbol{\varepsilon}}\cdot \boldsymbol{r}\ket{\Phi_{nlx}} = 4(2/3)^{1/2}\pi^{3/2} \sum_{l',m',m} n(m) (-i)^{l'} \exp(i\delta_{l'}) Y_{l',m'}(\theta_k, \phi_k)R_{n,l'}(E_\text{kin})\\
    \cdot \Big[\varepsilon_x \big(-\bra{l',m'}1,1\ket{l,m} + \bra{l'm'}1,-1\ket{l,m}\big) + i\varepsilon_y \big(\bra{l',m'}1,1\ket{l,m} + \bra{l',m'}1,-1\ket{l,m}\big)\\ + \sqrt{2}\varepsilon_z \bra{l',m'}1,0\ket{l,m}\Big]
\end{split}
\end{equation}
\end{widetext}
where 
\begin{equation}\label{eq: def of radial integral}
    R_{n, l\pm1}(\epsilon) = \int_0^\infty P_{n,l}(r)rP_{\epsilon,l\pm1}(r)dr
\end{equation}
and we introduce the notation
\begin{equation*}
\begin{split}
    \bra{l,m}\hat{l}, \hat{m}\ket{l',m'} := \int Y^*_{l,m}(\theta, \phi)Y_{\hat{l},\hat{m}}(\theta, \phi)Y_{l',m'}(\theta, \phi)d\Omega\\
    = \int (-1)^m Y_{l, -m}(\theta, \phi) Y_{\hat{l}, \hat{m}}(\theta, \phi) Y_{l', m'}(\theta, \phi) d\Omega \\
    =: (-1)^m \sqrt{\frac{3(2l+1)(2l_f+1)}{4\pi}}
     \begin{pmatrix}l & \hat{l} & l'\\0 & 0 & 0\end{pmatrix}
     \begin{pmatrix}l & \hat{l} & l'\\-m & \hat{m} & m'\end{pmatrix},
\end{split}
\end{equation*}
where in the last line we use the definition of Wigner 3-jm symbols.

The matrix element for the P and As $p$-electrons with $p$-polarized photons ($\theta_\epsilon=2\pi/9,~ \phi_\epsilon=\pi/2$) at Stanford Synchrotron Radiation Lightsource beamline 5-2 are shown in Fig. \ref{Fig: matrix_element} (b, c). The matrix elements for Eu $4f$ states cannot be calculated under the framework above due to their complicated indirect photoemission process\cite{suga2021photoelectron} and are phenomenologically chosen to be twice the matrix elements of pnictogen $p$ electrons based on the experimental spectral weight.

\subsection{Spectral function calculation and $k_z$ resolution}
In the weak interaction case, the interactions can be treated perturbatively in Green's function formalism with proper self-energy,
\begin{equation}
    A(\boldsymbol{k}, \omega) = -\frac{1}{\pi}\frac{\Sigma''(\boldsymbol{k},\omega)}{[\omega-\epsilon_{\boldsymbol{k}}-\Sigma'(\boldsymbol{k},\omega)]^2 + [\Sigma''(\boldsymbol{k}, \omega)]^2}
\end{equation}
with the eigenenergy $\epsilon_{\boldsymbol{k}}$ and the self-energy $\Sigma(\boldsymbol{k}, \omega) = \Sigma'(\boldsymbol{k}, \omega) + i\Sigma''(\boldsymbol{k}, \omega)$. The real and imaginary parts of the self-energy are related by the Kramers-Kronig condition.

We model the e-ph interactions phenomenologically by assuming that the imaginary part of self-energy takes a simple form:
\begin{equation}
    \Sigma''(\omega) = g\times H(\Omega_p-\omega),
\end{equation}
where $H$ is the Heaviside function and $\Omega_p$ is the phonon frequency. $g$ is a phenomenological measure of the e-ph coupling strength. While multiple phonon frequencies have been reported in the optical experiments\cite{homes2023optical}, here we choose $\Omega_p = 0$ for simplicity, since (i) our discussion on simulated spectra focuses on the high energy part ($|\omega| \gg |\Omega_p|$) and (ii) the dominant phonon mode that electrons couple to remains an ongoing research project. The argument in the main text remains qualitatively the same by choosing a more realistic phonon energy in the literature.

In the interpretation of ARPES spectra, it is crucial to account for broadening effects beyond the intrinsic interactions described by self-energy. One such effect is the finite resolution in the $k_z$ dimension, which can be particularly significant in materials with pronounced $k_z$ dispersion. The band structures of EuCd\textsubscript{2}P\textsubscript{2} and EuCd\textsubscript{2}As\textsubscript{2} along the $\Gamma$-$K$ trajectory, presented in Fig. \ref{Fig: band kz dependence}, exhibit substantial $k_z$ dependence. Consequently, our simulations incorporate the broadening effected attributable to $k_z$ to ensure a more accurate representation of the experimental observations by rewriting eq. \ref{eq: ARPES intensity} as
\begin{equation}
\begin{aligned}
    I(k_\parallel, \omega) & = \int |M_{f,i}^{\boldsymbol{k}}|^2A_f(k_\parallel, k_z, \omega+h\nu) A_i(k_\parallel, k_z, h\nu) dk_\perp\\
    & \propto \int \frac{\sigma A_i(k_\parallel, k_z, \omega)}{(k_z-k_0)^2+\sigma^2}dk_z
\end{aligned}
\end{equation}
where $\sigma=\Sigma''/v_z$ with $v_z=(\partial \omega_{k_\parallel, k_z}/\partial k_z)|_{k_0}$.

In our simulation, we choose an inner potential $V_0$ of 13.5 eV and 15.5 eV for \ECP\ and \ECA, respectively, to give the best fit to experimental spectra in terms of in-plane band dispersion.

\subsection{Numerical results}
\begin{figure*}[h]
    \centering
    \includegraphics[width=\textwidth]{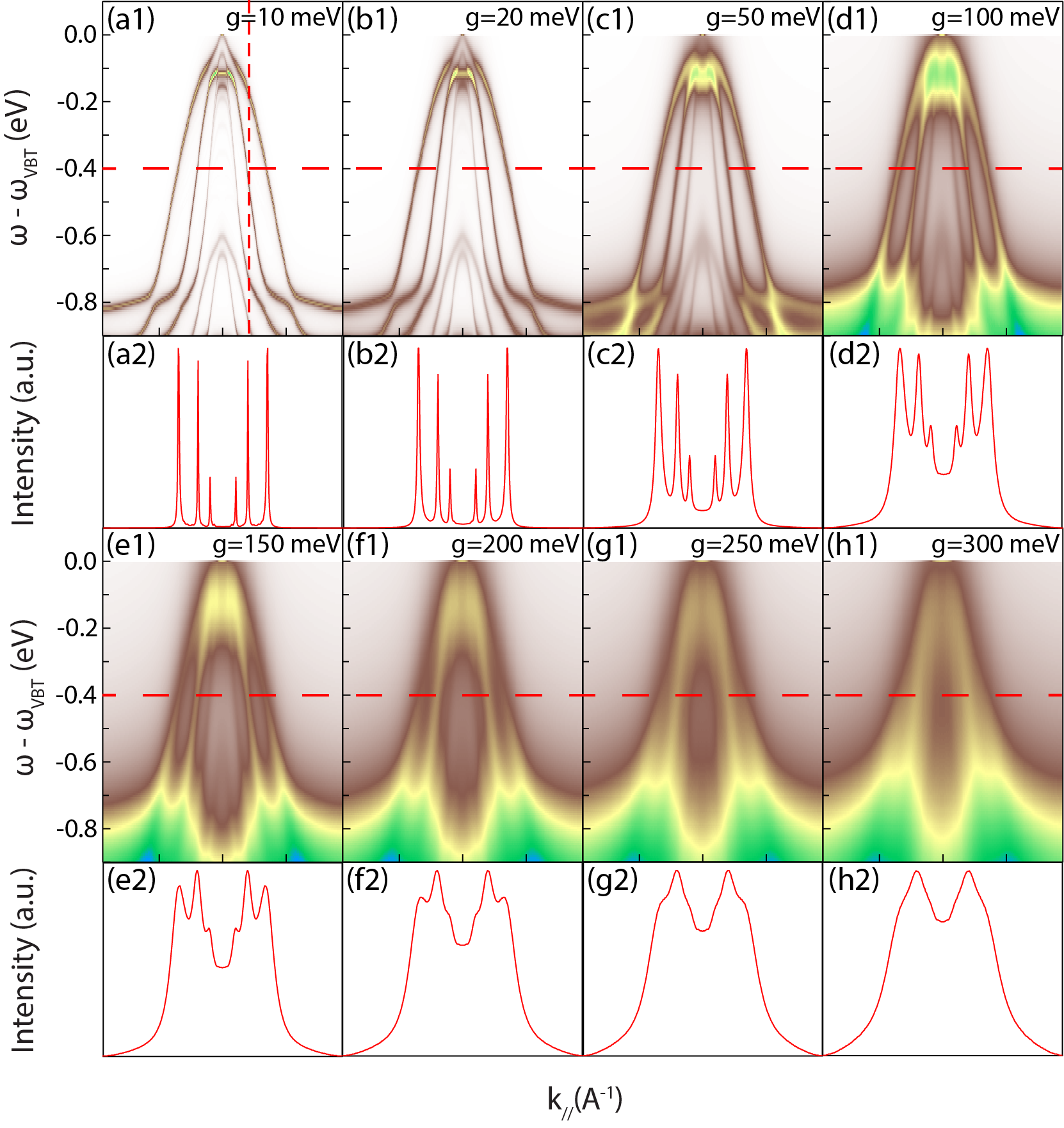}
    \caption{\textbf{Simulated ARPES spectra based on DFT calculations for \ECP.}  \textbf{(a1-h1)} Simulated ARPES Spectra for \ECP\ with $p$-polarized photons under different electron-phonon (e-ph) coupling strength. \textbf{(a2-h2)} The corresponding MDCs at a binding energy of -0.4 eV as marked by horizontal red dashed lines.}
    \label{Fig: DFT_simulation_ECP}
\end{figure*}

\begin{figure*}[h]
    \centering
    \includegraphics[width=\textwidth]{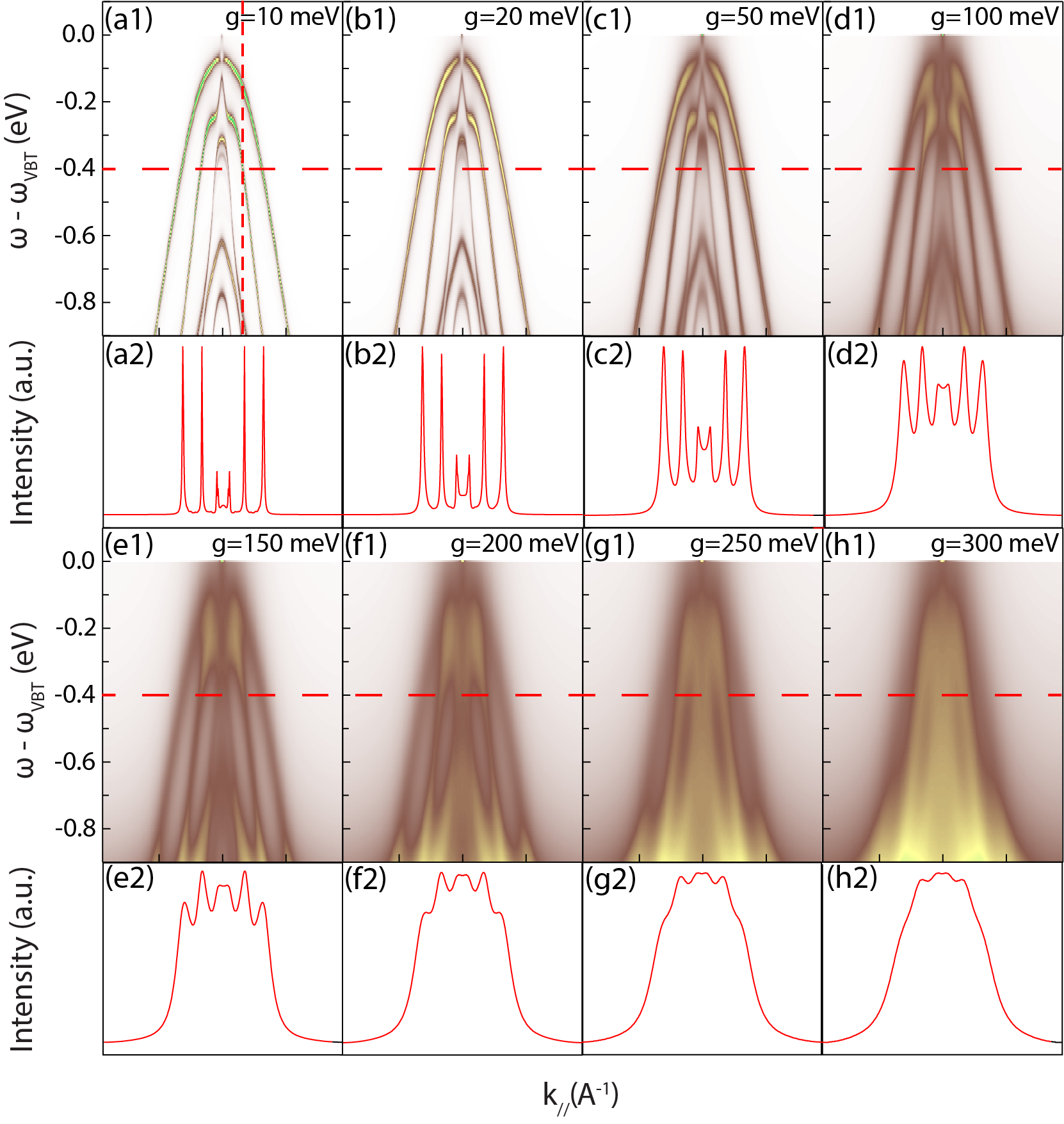}
    \caption{\textbf{Simulated ARPES spectra based on DFT calculations for \ECA. (a1-h1)} Simulated ARPES Spectra for \ECA\ with $p$-polarized photons under different electron-phonon (e-ph) coupling strength. \textbf{(a2-h2)} The corresponding MDCs at a binding energy of -0.4 eV as marked by horizontal red dashed lines.}
    \label{Fig: DFT_simulation_ECA}
\end{figure*}

The simulated ARPES spectra are shown in Fig. \ref{Fig: DFT_simulation_ECP} and Fig. \ref{Fig: DFT_simulation_ECA} for \ECP\ and \ECA, respectively. We assign $g=150$ (250) meV here for small (large) e-ph coupling in the main text. The single-peak-like (multi-peak-like) scenario in MDCs corresponds to a strong (weak) e-ph coupling strength. This correspondence remains consistent under reasonable variations in $\Omega_p$, $\sigma$, $V_0$, and $\Sigma''$.

\begin{figure*}[h]
    \centering
    \includegraphics[width=\textwidth]{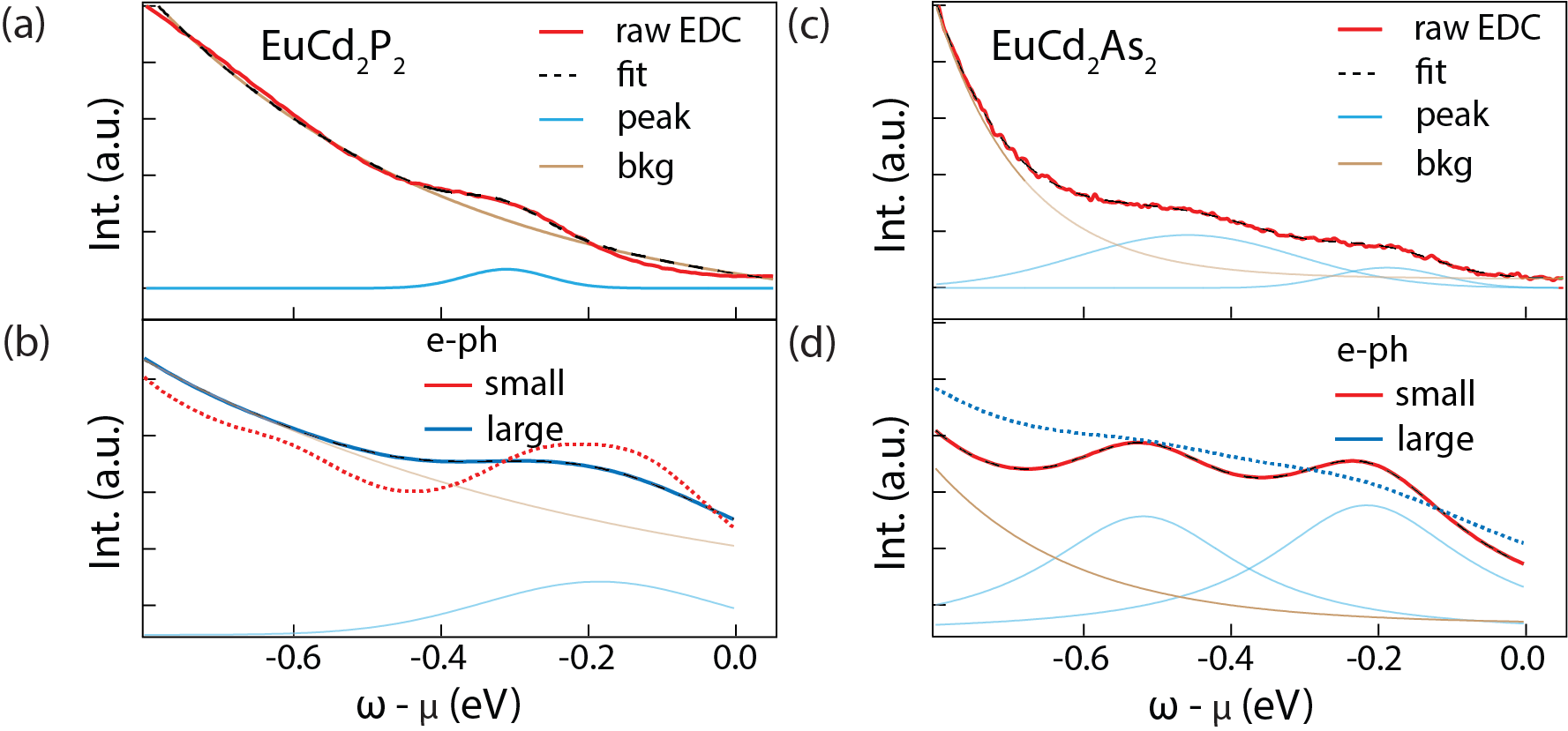}
    \caption{\textbf{The EDC analysis of the experimental and simulated spectra.  (a)} The EDC from the experimental spectrum for \ECP. The red line shows the raw EDC with cut position marked by the vertical red dashed line in Fig. \ref{Fig: DFT_simulation_ECP}(a1). The black dashed line shows the fitted EDC with a Gaussian curve (blue) and an exponential background (brown). \textbf{(b)} The EDCs from the simulated spectra. The red dashed (blue solid) line shows the EDC with small (large) e-ph coupling strength at the same cut position in (a). The black dashed line shows the fit to the EDC with a large e-ph coupling with a Gaussian curve (sky blue) and an exponential background (brown). \textbf{(c)} Similar to (a) for \ECA. The cut position is marked by the vertical red dashed line in Fig. \ref{Fig: DFT_simulation_ECA}(a1). \textbf{(d)} Similar to (b) for \ECA. The black dashed line shows the fit to the EDC with a small e-ph coupling.}
    \label{Fig: DFT_simulation_EDC}
\end{figure*}

Similar to the correspondence between the single- (multi-) peak scenario and the large (small) e-ph coupling in MDCs, the e-ph coupling strength also leads to qualitatively different EDCs. We show the EDCs from the experimental and simulated spectra with different e-ph coupling for \ECP\ in Fig. \ref{Fig: DFT_simulation_EDC}(a) and (b) and for \ECA\ in Fig. \ref{Fig: DFT_simulation_EDC}(c) and (d), respectively. We further fit the EDCs with a Gaussian envelope arising from polaronic pnictogen $p$ bands under the Frank-Condon broadening\cite{shen2004missing} and an exponential background arising from the mixing of Eu $4f$ states. In the \ECP\ (\ECA) case, the experimental EDC shows a similar lineshape to the EDC from large (small) e-ph coupling, with the exponential background dominating over the Gaussian envelope (the Gaussian envelope dominating over the exponential background). Therefore, the enhanced e-ph coupling in \ECP\ compared with \ECA\ is further supported by the EDC analysis here.  

\section{Sample characterization}
\begin{figure*}[h]
    \centering
    \includegraphics[width=\textwidth]{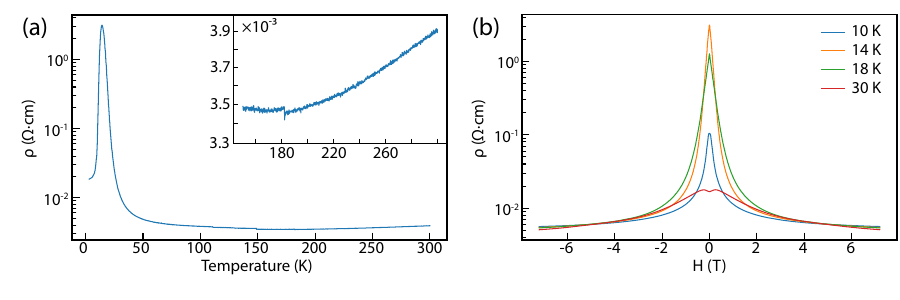}
    \caption{\textbf{Supplementary transport properties. (a)} The temperature dependence of in-plane resistivity. The inset shows a zoom-in of the resistivity under zero external fields at high temperatures and shows a slowly increasing resistivity with increasing temperatures, a hint of bad metal behavior. \textbf{(b)} The magnetic field dependence of in-plane resistivity at different temperatures. The resistivity shows a consistent decrease in resistivity with increasing magnetic field.}
    \label{Fig: transport_detail}
\end{figure*}

We present detailed transport measurements on our samples in Fig. \ref{Fig: transport_detail}. Our sample features a magnetoresistance peak at 14 K which is further suppressed rapidly with an external magnetic field. More importantly, the resistivity shows bad metal behavior with a slight increase in resistivity with increasing temperatures above 180 K.

\bibliography{supplementary_bib}